\documentclass[reprint, preprintnumbers, superscriptaddress, prx, aps, amsmath,amssymb, twocolumn, a4paper, floatfix, noeprint]{revtex4-2}

\usepackage{dcolumn}% Align table columns on decimal point
\usepackage{bm}
\usepackage[utf8]{inputenc}
\usepackage[T1]{fontenc}
\usepackage{graphicx,psfrag}
\usepackage{mathrsfs}
\usepackage{amsmath,amsfonts,amssymb}
\usepackage{multirow}
\usepackage{diagbox}
\usepackage{comment}
\usepackage[table]{xcolor}
\usepackage{enumerate}
\usepackage{booktabs}
\usepackage[normalem]{ulem}
\usepackage{mathtools}
\usepackage{siunitx}
\usepackage{textgreek}
\usepackage[section]{placeins}
\usepackage{orcidlink}

\usepackage{color}
\definecolor{cyan}{rgb}{0,0.9,0.9}
\definecolor{orange}{rgb}{0.9,0.5,0}
\definecolor{magenta}{rgb}{1,0,1}
\definecolor{purple}{rgb}{0.8,0.4,0.8}
\definecolor{gray}{rgb}{0.8242,0.8242,0.8242}
\definecolor{green}{rgb}{0.,0.8,0.}

\usepackage{hyperref}
\hypersetup{
    colorlinks = true,
    linkcolor = {blue},
    citecolor = {blue},
    urlcolor = {blue},
    linkbordercolor = {white},
    breaklinks=true,
    }
\newcommand*\aap{A\&A}

\newcommand*\aaps{A\&AS}

\renewcommand*\apj{ApJ}
\newcommand*\apjl{ApJ}

\newcommand*\apjs{ApJS}

\newcommand*\araa{ARA\&A}

\newcommand*\mnras{MNRAS}

\renewcommand*\nat{Nature}
\newcommand*\nphysa{Nucl.~Phys.~A}
\newcommand*\pasa{PASA}
\newcommand*\pasj{PASJ}

\newcommand*\physrep{Phys.~Rep.}
\newcommand*\physscr{Phys.~Scr}

\renewcommand*\prc{Phys.~Rev.~C}
\renewcommand*\prd{Phys.~Rev.~D}
\renewcommand*\pre{Phys.~Rev.~E}
\renewcommand*\prl{Phys.~Rev.~Lett.}

\newcommand*\ssr{Space~Sci.~Rev.}

\newcommand{\msun}{\text{M}_{\odot}}
\newcommand{\nsat}{\textit{n}_{\text{sat}}}
\newcommand{\chiEFT}{$\chi$EFT}
\newcommand{\mtov}{M_{\text{TOV}}}
\newcommand{\fmiq}{\text{fm}^{-3}}

\newcommand{\sat}{\mathrm{sat}}
\newcommand{\sym}{\mathrm{sym}}

\begin{document}
\preprint{LA-UR-24-20420}

\title{From Existing and New Nuclear and Astrophysical Constraints to Stringent Limits on the Equation of State of Neutron-Rich Dense Matter}
%An overview of existing and new nuclear and astrophysical constraints on the equation of state of neutron-rich dense matter

\author{Hauke \surname{Koehn}\,\orcidlink{0009-0001-5350-7468}}
\email{hauke.koehn@uni-potsdam.de}
\affiliation{Institut f\"ur Physik und Astronomie, Universit\"at Potsdam, Haus 28, Karl-Liebknecht-Straße 24/25, 14476, Potsdam, Germany}

\author{Henrik \surname{Rose}\,\orcidlink{0009-0009-2025-8256}}
\affiliation{Institut f\"ur Physik und Astronomie, Universit\"at Potsdam, Haus 28, Karl-Liebknecht-Straße 24/25, 14476, Potsdam, Germany}

\author{Peter T. H. \surname{Pang}\,\orcidlink{0000-0001-7041-3239}}
\affiliation{Nikhef, Science Park 105, 1098 XG Amsterdam, The Netherlands}
\affiliation{Institute for Gravitational and Subatomic Physics (GRASP), Utrecht University, Princetonplein 1, 3584 CC Utrecht, The Netherlands}

\author{Rahul \surname{Somasundaram}\,\orcidlink{0000-0003-0427-3893}}
\affiliation{Department of Physics, Syracuse University, Syracuse, New York 13244, USA}

\author{Brendan T. \surname{Reed}\,\orcidlink{0000-0002-7775-5423}}
\affiliation{Theoretical Division, Los Alamos National Laboratory, Los Alamos, New Mexico 87545, USA}

\author{Ingo \surname{Tews}\,\orcidlink{0000-0003-2656-6355}}
\affiliation{Theoretical Division, Los Alamos National Laboratory, Los Alamos, New Mexico 87545, USA}

\author{Adrian Abac\,\orcidlink{0000-0003-4786-2698}} 
\affiliation{Max Planck Institute for Gravitational Physics (Albert Einstein Institute), Am M{\"u}hlenberg 1, Potsdam 14476, Germany}
\affiliation{Institut f\"ur Physik und Astronomie, Universit\"at Potsdam, Haus 28, Karl-Liebknecht-Straße 24/25, 14476, Potsdam, Germany}

\author{Oleg \surname{Komoltsev}\,\orcidlink{0000-0002-2188-3549}}
\affiliation{Faculty of Science and Technology, University of Stavanger, 4036 Stavanger, Norway}

\author{Nina Kunert\,\orcidlink{0000-0002-1275-530X}}
\affiliation{Institut f\"ur Physik und Astronomie, Universit\"at Potsdam, Haus 28, Karl-Liebknecht-Straße 24/25, 14476, Potsdam, Germany}

\author{Aleksi \surname{Kurkela}\,\orcidlink{0000-0001-7991-3096}}
\affiliation{Faculty of Science and Technology, University of Stavanger, 4036 Stavanger, Norway}

\author{Michael W. Coughlin\,\orcidlink{0000-0002-8262-2924}}
\affiliation{School of Physics and Astronomy, University of Minnesota, Minneapolis, Minnesota 55455, USA}

\author{Brian F. Healy\,\orcidlink{0000-0002-7718-7884}}
\affiliation{School of Physics and Astronomy, University of Minnesota, Minneapolis, Minnesota 55455, USA}

\author{Tim \surname{Dietrich}\,\orcidlink{0000-0003-2374-307X}}
\affiliation{Institut f\"ur Physik und Astronomie, Universit\"at Potsdam, Haus 28, Karl-Liebknecht-Straße 24/25, 14476, Potsdam, Germany}
\affiliation{Max Planck Institute for Gravitational Physics (Albert Einstein Institute), Am M{\"u}hlenberg 1, Potsdam 14476, Germany}

\begin{abstract}
Through continuous progress in nuclear theory and experiment and an increasing number of neutron-star (NS) observations, a multitude of information about the equation of state (EOS) for matter at extreme densities is available.
To constrain the EOS across its entire density range, this information needs to be combined consistently. 
However, the impact and model dependency of individual observations vary.
Given their growing number, assessing the various methods is crucial to compare the respective effects on the EOS and discover potential biases.  
For this purpose, we present a broad compendium of different constraints and apply them individually to a large set of EOS candidates within a Bayesian framework.
Specifically, we explore different ways of how chiral effective field theory and perturbative quantum chromodynamics can be used to place a likelihood on EOS candidates. 
We also investigate the impact of nuclear experimental constraints, as well as different radio and X-ray observations of NS masses and radii. 
This is augmented by reanalyses of the existing data from binary neutron star coalescences, in particular of GW170817, with improved models for the tidal waveform and kilonova light curves, which we also utilize to construct a tight upper limit of \qty{2.39}{\msun} on the TOV mass based on GW170817's remnant.
Our diverse set of constraints is eventually combined to obtain stringent limits on NS properties. We organize the combination in a way to distinguish between constraints where the systematic uncertainties are deemed small and those that rely on less conservative assumptions.
For the former, we find the radius of the canonical \qty{1.4}{\msun} neutron star to be $R_{1.4}= 12.26_{-0.91}^{+0.80}$\,km and the TOV mass at $M_{\rm TOV}= 2.25_{-0.22}^{+0.42}$\,$\msun$ (95\% credibility).
Including all the presented constraints yields $R_{1.4}= 12.20_{-0.48}^{+0.50}$\,km and $M_{\rm TOV}= 2.30_{-0.20}^{+0.07}$\,$\msun$. 
When comparing these limits to individual data points, we find that the quoted radius of \mbox{HESS J1731-347} displays noticeable tension with other constraints.
Constraining microphysical properties of the EOS proves more challenging. For instance, the symmetry energy slope is restricted to $L_\sym = {48}^{+21}_{-25}$\,MeV, where this constraint is mainly dominated by our reanalysis of the PREX-II and CREX experiment.
\end{abstract}

\maketitle

\date{\today}

%\tableofcontents

\section{Introduction}
\begin{figure}[t]
    \centering
    \includegraphics[width = \linewidth]{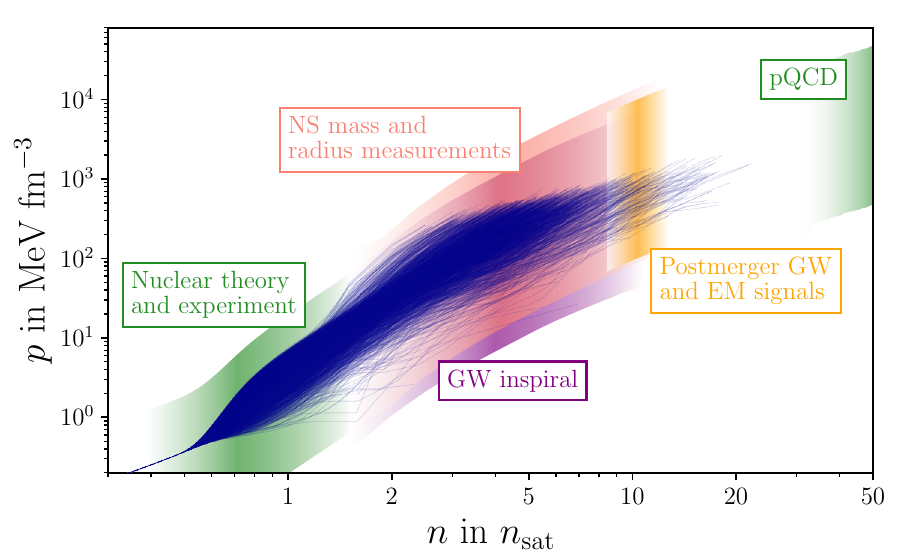}
    \caption{Schematic overview of different sources of information about the dense matter EOS. The set of possible EOS candidates (see Sec.~\ref{sec:EOS_construction}) is shown by dark blue lines up to the respective maximum-mass configurations (TOV points). 
    The colored bands roughly indicate the density regime where the different inputs constrain the EOS. 
    We point out that the postmerger physics in BNS coalescences (orange band) depends also on finite-temperature effects of the EOS.
    }
    \label{fig:overview}    
\end{figure}

Matter compressed to densities around and above the nuclear saturation density occurs throughout the Universe in neutron stars (NSs), atomic nuclei, and during core-collapse supernovae of massive stars.
Although the properties of dense strongly interacting matter are governed by quantum chromodynamics (QCD), confinement and the sign problem of lattice QCD make direct analytical and numerical QCD calculations infeasible~\citep{Wilson_1974, Forcrand_2010, Nagata_2022}. 
Thus, determining the properties of matter in the phase diagram of QCD remains an open problem of physics that has to be addressed with effective theoretical approaches and empirical observations. 
Of particular interest is the thermodynamic relationship between density and pressure of dense matter far below its Fermi temperature, i.e., the equation of state (EOS) for cold dense matter. 
This cold EOS plays a fundamental role for the interior composition of isolated NSs~\citep{Oppenheimer_1939, Baym_2018, Lattimer_2021, Ascenzi_2024, Chatziioannou_2024}, the tidal imprint on gravitational-wave (GW) signals from binary neutron stars (BNSs) or neutron-star--black-hole binaries (NSBH)~\citep{Hinderer_2010, Vines_2011, Lackey_2012}, and is intimately connected to the bulk properties of atomic nuclei~\citep{Roca_2018, Hu_2022}. 
Although we restrict ourselves to the investigation of cold dense matter, we mention that finite-temperature effects of the EOS are relevant during the formation of proto-NSs during the first milliseconds after bounce in core-collapse supernovae~\citep{Yasin_2020, Andersen_2021, Boccioli_2022}, nuclear collision experiments~\citep{Amsden_1975, Hofmann_1976, Chandra_2007}, the (post)merger dynamics of compact binary coalescences involving NSs~\citep{Lattimer_2016, Raithel_2019}, and studies of the Universe immediately after the big bang~\citep{Aoki_2006, Boeckel_2011}. 
The dense matter EOS is therefore highly relevant for many applications in nuclear and astrophysical research, with additional deep theoretical implications for the nature of QCD matter~\citep{Angel_2020, Kojo_2020}.

Conversely, it is possible to place constraints on the EOS by using information from the aforementioned phenomena, i.e., from astrophysical observations, nuclear experiments, and nuclear theory. 
The various types of constraints rely on different physical processes, and hence, affect the EOS at different densities~\citep{Oezel_2016, Lattimer_2021, Huth_2022, Gosh_2022}.
Fig.~\ref{fig:overview} shows a schematic overview indicating the regime in which a specific input constrains the EOS. 
Naturally, any information based on nuclear theory and experiment will be relevant in the regime around the nuclear saturation density $\nsat = 0.16$\,$\fmiq$~\citep{Hebeler_2013, Margueron:2017eqc}. 
The global properties of NSs, i.e., their masses, radii, and tidal deformabilities, are mainly determined by the star's core where densities of several $\nsat$ are reached. Therefore, measuring NS properties constrains the EOS up to this density regime~\citep{Oezel_2016, Burgio_2021, Lattimer_2021}. Around \qty{40}{\nsat}, QCD becomes perturbative and direct analytical calculations of quark matter restrict the relation between density and pressure~\citep{Freedman_1977, Kurkela_2010}, which provides information on the EOS also at NS densities~\cite{Gorda_2023a, Komoltsev_2023}. 

Given this wide density range, different information sources need to be combined to constrain the EOS across all scales.
Several reviews on this topic are readily available; for instance, Refs.~\citep{Sorensen_2024, Tsang_2012, Tsang_2023, Lattimer_2023} present numerous data points from nuclear physics and their implication for the EOS. 
References~\citep{Oezel_2016, Miller_2020} summarize multiple ways to extract masses and radii of NS from radio and X-ray observations.
Many studies also investigated the consequences of the multimessenger data of GW170817 for the EOS~\citep{LIGO_GW170817, LIGO_GW170817_EM, LIGO_GW170817_EOS, Coughlin_2018, Radice_2019, Raithel_2019A, Bauswein_2017}. 
In Ref.~\citep{Capano_2020} this data was supplemented with constraints from nuclear theory, and Refs.~\citep{Dietrich_2020, Raaijmakers_2021, Huth_2022, Essick_2021, Biswas_2021, Annala_2022, Breschi_2024} further added additional constraints from electromagnetic X-ray observations of NS or nuclear experimental data. 
For reviews on general theoretical approaches to the EOS see, e.g., Refs.~\citep{Burgio_2021, Lattimer_2021, Chatziioannou_2024}, and for a review that also addresses constraints on the finite-temperature EOS see Ref.~\citep{Kumar_2024}. 

Yet, so far no contiguous resource for combining and comparing these various constraints exists. 
However, through improved radio and X-ray measurements~\citep{Sieniawska_2018, Greif_2020, Kramer_2009, Kramer_2021}, further GW detections in the current and upcoming observing runs~\citep{Bandopadhyay_2024, Kiendrebeogo_2023}, or improved measurements of nuclear properties~\citep{Sorensen_2024, Durante_2019, Russotto_2021, Ivanov_2024, Becker_2018}, the available information on the EOS is expected to grow.
Therefore, juxtaposing and cross-checking different data points becomes increasingly relevant, since with more data points, mutual tensions may arise that would introduce inconsistencies in the inferred results. 
Moreover, the constraining power and model dependency of different constraints vary substantially. 
Hence, it is necessary to reexamine existing data points, quantify their impact on specific parts of the EOS, and assess potential systematic biases.

In the present article, we create an extensive compendium for different constraints on the EOS. 
This compendium utilizes a diverse set of data points, which we apply step by step to a wide, general prior for the dense matter EOS. 
In this way, we can compare the impact each input asserts individually on the space of possible EOS candidates. 
We employ a Bayesian approach; i.e., we determine the likelihood across all our EOS candidates for each given constraint.
This Bayesian analysis also necessitates engagement with the question of how each constraint on the EOS is constructed from the observed data, as various assumptions could introduce biases. 
For this reason, we comment on potential sources of systematic errors while applying a particular constraint, and further discuss how different (statistical) interpretations of the data may affect the inferred results.

For example, there is no specific {\it a priori} choice of how to implement theoretical constraints from chiral effective field theory (\chiEFT) or from perturbative quantum chromodynamics (pQCD) in a Bayesian likelihood function~\citep{Brandes_2023, Drischler_2021, Gorda_2023a, Somasundaram_2023}.
We compare different possibilities in Secs.~\ref{sec:ChiralEFT} and \ref{sec:pQCD} and find that \chiEFT\ softens the EOS and restricts the radii of NS, while the pQCD constraint also softens the EOS but mostly impacts the maximum NS mass (TOV mass).

Similarly, it is difficult to translate nuclear experiments to information about properties of the EOS. 
In Sec.~\ref{Sec:PREX} we discuss the measurements of the parity violating asymmetry from the lead radius experiment (PREX-II) and the calcium radius experiment (CREX)~\citep{PREX2, Adhikari_2022}, as well as the $^{208}$Pb dipole polarizability measurement~\citep{Tamii_2011} and reanalyze these experimental results with energy density functionals (EDFs) to obtain fully Bayesian posteriors on the symmetry energy parameters at saturation density. 
These posteriors allow us to restrict our EOS set in a statistically consistent fashion and when combined deliver a limit on the slope of the symmetry energy of  $L_\sym = {48}^{+21}_{-25}$\,MeV.
%, however the previously noted tension~\citep{Miyatsu_2023, Reed_2023} between CREX and PREX-II for the symmetry energy slope remains.

For electromagnetic measurements of NS masses and radii, there is less ambiguity in translating observables to the EOS. 
However, especially for analyses that rely on modeling thermal X-ray emission, systematic errors might be comparable to the quoted statistical uncertainties. 
In particular, the mass-radius measurement of \mbox{HESS J1731-347}~\citep{Doroshenko_2022} could be an outlier in the dataset, based on the uncertain assumptions about the compact object's atmosphere and distance~\citep{Alford_2023}. 
We discuss such potential error sources in Sec.~\ref{sec:Thermal} and compare this measurement to the expectation from the other constraints on our EOS set in Sec.~\ref{sec:outlier}. 
We find substantial evidence that the results for \mbox{HESS J1731-347} do not agree with predictions about the EOS from other constraints, though the uncertainties are still too large to reject \mbox{HESS J1731-347} decisively as an outlier.

The GW and electromagnetic data from BNS coalescences are reanalyzed in this work with current state-of-the-art models for the waveform and kilonova (KN) emission. 
This is done employing the nuclear multimessenger-analysis framework NMMA~\citep{NMMA_2023} that directly incorporates our EOS set in the inference. 
In particular, the joint inference of the electromagnetic and GW data for GW170817 provides stricter limits on the EOS than the GW data alone, as was shown in previous works~\citep{Dietrich_2020, NMMA_2023}. 
In Sec.~\ref{subsec:postmerger}, we supplement these joint inferences by further incorporating the black-hole (BH) collapse hypothesis for the postmerger of GW170817, thereby using all the information from GW170817 in a practical, self-consistent manner and obtaining a tight upper limit of \qty{2.39}{\msun} on the TOV mass.

Our Bayesian approach also allows us to combine the different constraints in a flexible and consistent fashion. 
Combining the distinct inputs yields stringent limits on the EOS, consistent with previous works~\citep{Capano_2020, Annala_2022, Dietrich_2020, Huth_2022, LIGO_GW170817_EOS, Raaijmakers_2021, Gueven_2023, Breschi_2024, Brandes_2023}. 
However, because of the flexibility in our approach, we are able to group the constraints into different sets and study how the final estimates change, when we include tighter, but more model-dependent constraints. 
We quote the results for these constraint sets in Sec.~\ref{subsec:combined_constraints}. 
We also offer a resource for custom constraint combinations via a Web interface~\citep{Rose_2024}. 

The present article is organized as follows.
In Sec.~\ref{sec:EOS_construction}, we provide a description of how our EOS candidate set was constructed. 
We describe the constraints from theoretical and experimental nuclear physics in Sec.~\ref{sec:Nuclear}, continue with the electromagnetic mass and mass-radius measurements of NSs in Sec.~\ref{sec:Astro}, and finally address the observables from BNS coalescences in Sec.~\ref{sec:Multimessenger}. 
The detailed discussion of the constraints is arranged in subsections, in which we initially illustrate how this particular data point relates to the EOS, then describe the method and the result of our Bayesian inference, and afterward conclude with discussions about potential systematic error sources. 
The comparison and combination of different constraints are then eventually performed in Sec.~\ref{sec:Comparing_and_combining}.
Though we try to keep our compendium as comprehensive as possible, we do not include all constraints discussed in the literature, and we comment on occasion about why we omit specific sources.

Throughout this article, we denote the pressure of dense matter by $p$, the energy density by $\epsilon$, the number density by $n$, the mass of an NS by $M$, and its radius by $R$. 
Similarly, $R_{1.4}$ is the radius of a canonical \qty{1.4}{\msun} NS and $p_{3\nsat}$ the pressure at 3 times the saturation density. 
The maximum mass of a nonrotating NS in equilibrium is the TOV mass $\mtov$.
We denote the density at the center of such a star as $n_{\text{TOV}}$. 
Likelihoods are written as $\mathcal{L}$ and probabilities or probability density functions as $P$. 
Credible intervals are usually quoted at the 95\% credibility level if not stated otherwise.

\section{Constructing the EOS prior}
\label{sec:EOS_construction}

To construct the set with 100\,000 EOS candidates that is employed throughout the present work, we follow a procedure similar to the one outlined in our previous works~\cite{Dietrich_2019}. 
At low densities, we assume that the EOS can be described in terms of nucleonic matter, while we attach a model-agnostic extrapolation scheme at high densities to account for possible new and exotic phases. 
With these well-motivated assumptions about the EOS, we adopt conservative choices to construct a sufficiently general EOS prior. 
In the following, we describe our construction in detail.

At the lowest densities in NSs, matter forms a solid crust where atomic nuclei are arranged in a Coulomb lattice. 
Here, we use the crust model of Ref.~\cite{Douchin:2001sv} for all EOSs. 
We keep the crust EOS fixed, i.e., we do not explore uncertainties in the crust EOS; see, however, Refs.~\cite{Grams:2021lzx, Gamba_2019} for the potential impact on global NS properties.
The crust EOS of Ref.~\cite{Douchin:2001sv} is used up to the crust-core transition density predicted by this model, which is \qty{0.076}{\fmiq}. 

For the EOS of the outer NS core, we assume that matter is composed of only nucleonic degrees of freedom in beta equilibrium up to a certain density $n_{\text{break}}$. 
We randomly draw the density $n_{\text{break}}$ from a uniform distribution in the range \qtyrange{1}{2}{\nsat} to account for the possibility that non-nucleonic degrees of freedom appear at higher densities~\cite{Somasundaram:2021clp}. 
To model the homogeneous matter below $n_{\text{break}}$, we employ the metamodel (MM) introduced in Refs.~\cite{Margueron:2017eqc, Margueron:2017lup}; see also Refs.~\cite{Somasundaram:2020chb,Grams:2021lzx}. 
The MM is a density functional approach, similar to the Skyrme model~\cite{Chabanat:1997un} that allows one to directly incorporate nuclear-physics knowledge encoded in terms of nuclear empirical parameters (NEPs). 
These parameters are defined via a Taylor expansion of the energy per particle in symmetric matter $e_{\sat}$ and the symmetry energy $e_{\sym}$ about saturation density $n_\sat$:
\begin{eqnarray}
\label{eq:sat}
&&\hspace{-1cm}e_{\sat}(n) = E_\sat + K_\sat \frac{x^2}{2} + Q_\sat \frac{x^3}{3!} \nonumber \\
&&\hspace{2cm} + Z_\sat \frac{x^4}{4!} + \dots, \\
\label{eq:sym}
&&\hspace{-1cm}e_{\sym}(n) = E_\sym + L_\sym x + K_\sym \frac{x^2}{2}+ Q_\sym \frac{x^3}{3!} \nonumber \\ 
&&\hspace{2cm} + Z_\sym \frac{x^4}{4!} + \dots,
\end{eqnarray}
where 
\begin{align}
    x \coloneqq \frac{n-n_\sat}{3n_\sat}
\end{align}
is the expansion parameter. 
For a given set of NEPs, the MM provides an EDF that can be used to calculate the EOS of nuclear matter in beta equilibrium. 
The MM is able to reproduce the EOSs predicted by a large number of nucleonic models that exist in the literature~\cite{Margueron:2017eqc, Margueron:2017lup}, including those from involved microscopic calculations such as in the framework of \chiEFT~\cite{Somasundaram:2020chb}. 
To account for nuclear-physics uncertainties and to generate a wide EOS prior for the analysis presented here, we vary the NEPs uniformly in the ranges specified in Table~\ref{tab:NEP_range}. 
In this manner, we construct the EOS in the density range $\qty{0.12}{\fmiq}~<~n~<~n_{\text{break}}$. 
The lower limit of \qty{0.12}{\fmiq} is chosen arbitrarily, but we verified that our choice has negligible impact on the construction of our EOS prior. 
To combine the crust and the core EOSs, we use a cubic spline in the speed of sound $c^2_s$ versus density plane between the crust-core transition density \qty{0.076}{\fmiq} and the onset of the MM at \qty{0.12}{\fmiq}. 
For a few samples, this construction of the EOS below $n_{\text{break}}$ leads to mechanically unstable behavior, i.e., $c^2_s<0$ in certain density intervals. 
For these EOSs, we replace the negative sound speed with $c^2_s = 0$.

\begin{table}[t]
\centering
\tabcolsep=0.3cm
\def\arraystretch{1.5}
\caption{The prior distributions from which the NEPs are drawn to generate the EOS below $n_{\text{break}}$. The parameters $E_\sat$ and $\nsat$ are fixed at \qty{-16}{MeV} and \qty{0.16}{\fmiq}, respectively. We denote uniform priors by $\mathcal{U}$.}
\begin{tabular}{>{\centering\arraybackslash} p {3.5cm} >{\centering\arraybackslash} p {3.5 cm}}
 \toprule
 \toprule
 Parameter & Prior\\
 \midrule
 $n_{\text{break}}$ [$\nsat$]& $\mathcal{U}(1,2)$ \\
 $K_{\sat}$ [\unit{MeV}] & $\mathcal{U}(150,300)$\\
 $Q_{\sat}$ [\unit{MeV}]& $\mathcal{U}(-500,1100)$\\
 $Z_{\sat}$ [\unit{MeV}]& $\mathcal{U}(-2500,1500)$\\ 
 $E_{\sym}$ [\unit{MeV}]& $\mathcal{U}(28,45)$ \\
 $L_{\sym}$ [\unit{MeV}]& $\mathcal{U}(10,200)$ \\
 $K_{\sym}$ [\unit{MeV}]& $\mathcal{U}(-300,100)$ \\
 $Q_{\sym}$ [\unit{MeV}]& $\mathcal{U}(-800,800)$ \\
 $Z_{\sym}$ [\unit{MeV}]& $\mathcal{U}(-2500, 1500)$\\
\bottomrule
\end{tabular}
\label{tab:NEP_range}
\end{table}

Above $n_{\text{break}}$, we need to take into account the possibility that non-nucleonic degrees of freedom might appear~\cite{Somasundaram:2021ljr}, while simultaneously allowing for an EOS prior that remains as conservative and broad as possible. 
Hence, we employ a speed-of-sound approach which is a modified version of the scheme in Ref.~\cite{Tews:2018iwm}. 
In this approach, we create a nonuniform grid in density between $n_{\text{break}}$ and \qty{25}{\nsat} with $9$ grid points, where the ninth point is fixed at \qty{25}{\nsat} and the other points are randomly distributed. 
Then, at each density grid point, the squared sound speed is varied uniformly between $0$ and $c^2$, where $c$ is the speed of light. 
Finally, to create the full density-dependent sound speed profile $c^2_s(n)$, we interpolate linearly between the grid points in the $c^2_s$-$n$ plane.
The density-dependent speed of sound $c^2_s(n)$ can be integrated to give the EOS, i.e., the pressure $p(n)$, the energy density $\epsilon(n)$, and the baryon chemical potential $\mu(n)$; see Refs.~\cite{Tews:2018iwm,Somasundaram:2021clp} for more details.
Note that first-order phase transitions are not explicitly included in our EOS set. 
However, the manner in which we sample the $c_s^2$ and the density grid allows for arbitrarily soft EOSs which strongly resemble such phase transitions.
Following this approach, we have created a set of 100\,000 individual candidate EOSs that was recently also used to study the impact of perturbative QCD on the inference of the NS EOS~\citep{Komoltsev_2023}.

Our set of EOS candidates gives rise to a natural prior for EOS-related quantities like the canonical NS radius $R_{1.4}$ or the TOV mass $\mtov$.
They are simply given by taking the corresponding values from our EOS candidates as samples.
We show these prior distributions, e.g., in Fig.~\ref{fig:chiEFT}. 
The samples also allow us to determine the posterior distribution for these quantities when a certain constraint is applied. 
Since we determine $\mathcal{L}(\text{EOS}|d)$ for each constraint $d$ individually, we obtain the associated posteriors by simply weighing the samples with the likelihoods of the associated EOSs. 
Because our priors are informative and nonuniform, they also impact the quoted posterior credible intervals to a non-negligible extent. 
Nevertheless, the prior on the actual parameter space, i.e., the EOS candidate set, is flat.

\section{Information from nuclear physics and perturbative QCD}
\label{sec:Nuclear}

\begin{figure}
    \centering
    \includegraphics[width = \linewidth]{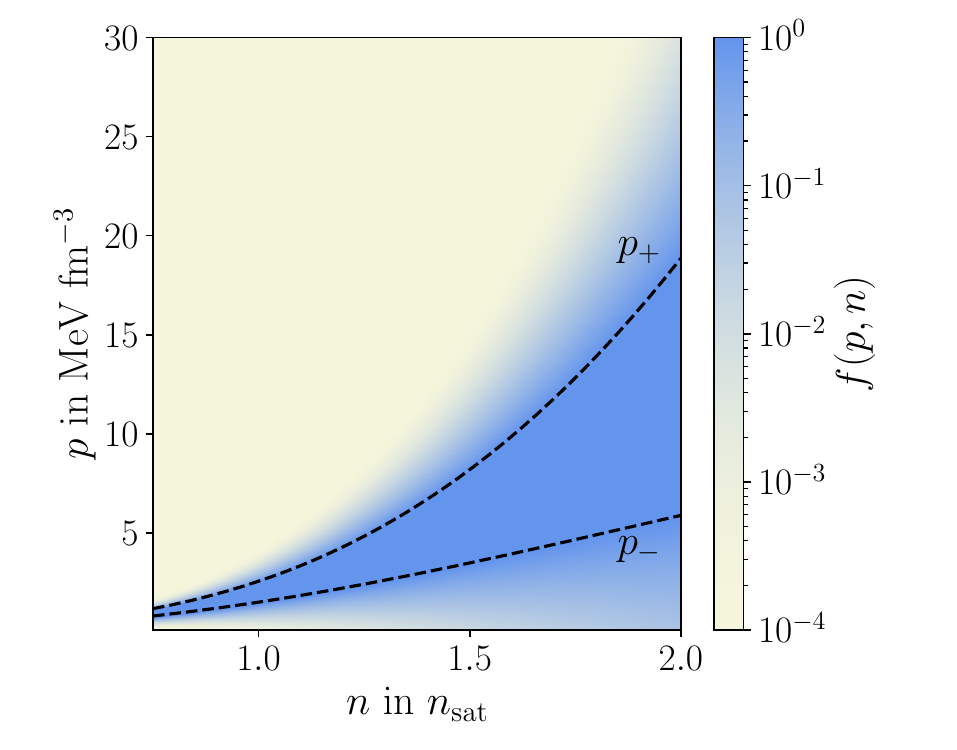}
    \caption{Score function $f(p,n)$ from Eq.~(\ref{eq:score_function}) used in Eq.~(\ref{eq:chiEFT_likelihood}) to calculate the likelihood of an EOS given \chiEFT\ constraints.
    The black dashed lines show the band obtained by \chiEFT\ calculations in Ref.~\citep{Tews:2018kmu}.}
    \label{fig:score_function}
\end{figure}

\begin{figure}
    \centering
    \includegraphics[width = \linewidth]{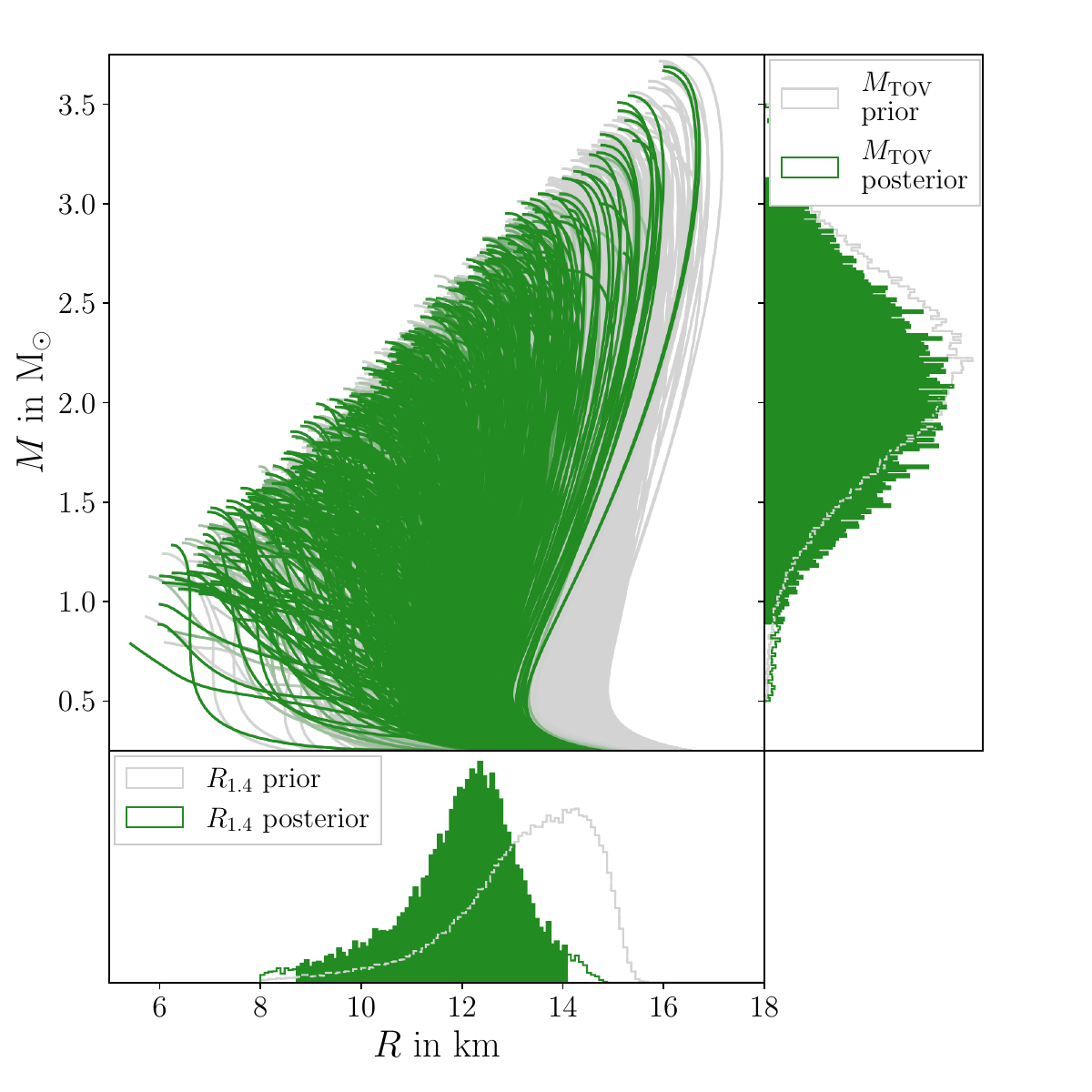}
    \includegraphics[width = \linewidth]{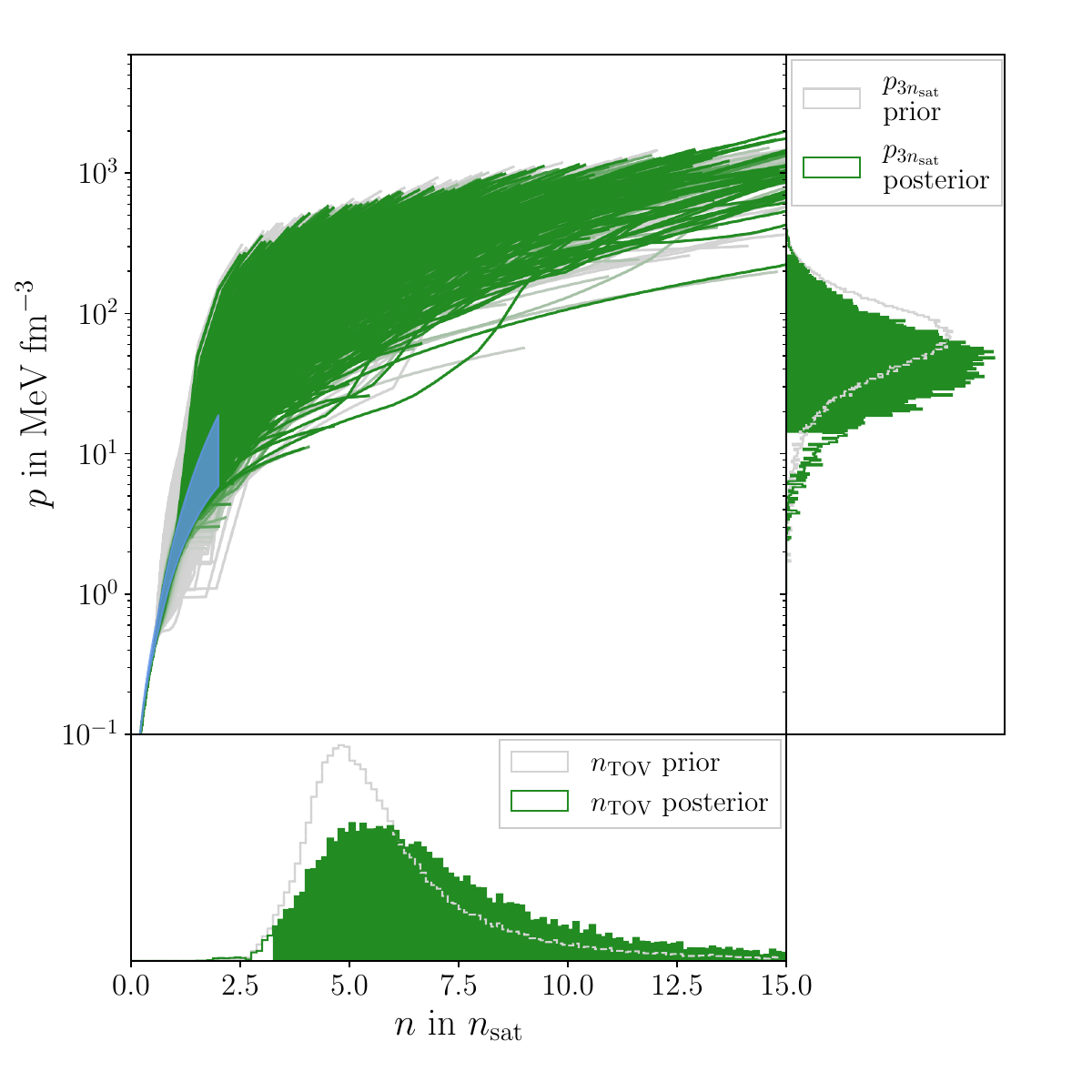}
    \caption{EOS inference based on the constraints set by \chiEFT. 
    Top: The central panel shows the relationships between NS mass $M$ and radius $R$ from the EOS candidate set. 
    The color coding marks EOS posterior probability on a scale from gray (zero) to the maximum value (green). 
    Additionally, the prior and posterior on $\mtov$ (right-hand panel), and on $R_{1.4}$ (bottom panel) are shown.
    The bottom figure's central panel displays the relationship between number density $n$ and pressure $p$ for the EOS, drawn up to the individual $n_{\text{TOV}}$. The EOS candidates are again colored according to posterior likelihood. The attached panels feature the prior and posterior on $p_{3n_{\text{sat}}}$ and $n_{\text{TOV}}$. 
    The blue band shows the range of pressure values for the \chiEFT\ calculation of Ref.~\citep{Tews:2018kmu}. 
    High-likelihood EOS may deviate from this band beyond their individual breakdown density.
  }
    \label{fig:chiEFT}
\end{figure}

Theoretical first-principles calculations regarding the properties of dense nucleonic matter at neutron-star conditions are currently impossible to obtain directly from the QCD Lagrangian \citep{Nagata_2022, Kojo_2020}. 
Nevertheless, in certain density regimes, theoretical approaches connected to QCD can be used to constrain the EOS. 
Specifically, we employ \chiEFT, which is valid at densities up to about \qty{2}{\nsat}, and pQCD, which is applicable to perturbative quark matter and valid at $n \gtrsim \qty{40}{\nsat}$. 
While both of these theoretical approaches break down at intermediate densities, they nevertheless provide valuable EOS information in this regime because any EOS has to match their predictions while respecting thermodynamic consistency and causality. 
We first explore how these two theoretical approaches can be used for Bayesian inference on our EOS candidate set and find that they already rule out the most extreme EOS candidates in our prior.
Then, we shift to complementary information from nuclear experiments. 
The symmetry energy at saturation density can be inferred from the Lead Radius Experiment (PREX) measurement of the $^{208}$Pb neutron-skin thickness~\citep{Reed_2021}, the Calcium Radius Experiment (CREX) measurement of the $^{48}$Ca neutron-skin thickness~\citep{Adhikari_2022}, and the electric dipole response of $^{208}$Pb. 
Reanalyzing these measurements with our ensemble of EDFs, we find, respectively $110^{+50}_{-72}$, $27^{+65}_{-26}$, and $81^{+59}_{-61}$\,MeV for $L_\sym$. 
Correspondingly, larger or smaller radii for NSs are preferred while the TOV mass is mostly unaffected.
Finally, we discuss the information on the symmetry energy at higher densities obtained from heavy-ion collision (HIC) experiments using $^{197}$Au ~\citep{Huth_2022, Russotto_2016, Fevre_2016}.
Because of large uncertainties, these have only marginal influence on neutron-star properties.

\subsection{Chiral effective field theory}
\label{sec:ChiralEFT}

At low energies and momenta, quarks are confined to baryons, such as nucleons or mesons.
Nuclear interactions can be described in terms of these effective degrees of freedom while keeping an intimate connection to QCD, by obeying all symmetries posed by the fundamental theory, in particular the approximate chiral symmetry of QCD.
While phenomenological models have used these effective degrees of freedom for a long time, the 1990s saw the introduction of \chiEFT~\cite{Epelbaum:2008ga,Machleidt:2011zz}.
\chiEFT\ provides a systematic expansion in terms of nucleon momenta over a breakdown scale $\Lambda_b$ and expands the effective nucleon-nucleon (NN), three-nucleon (3N), and multinucleon interactions in terms of explicitly resolved pion exchanges and short-range contact interactions that absorb physical processes at momenta above $\Lambda_b$.
The \chiEFT\ expansion can then be truncated at a desired order, providing a nuclear Hamiltonian for which the many-body Schrödinger equation can be solved numerically.
For nuclear matter, these approaches usually determine the energy per particle $E/A(n)$ as a function of baryon number density, which in turn allows one to determine the energy density $\epsilon$ and the pressure $p$ at a given number density $n$.
The missing terms in the truncated Hamiltonian introduce systematic uncertainties that can be estimated from the order-by-order convergence of a given calculation~\cite{Epelbaum:2014efa,Drischler_2020}.
Hence, \chiEFT\ enables us to quantify uncertainty bands for the possible range of energies and pressures at $n$. 
The band employed in the present work is taken from Ref.~\cite{Tews:2018kmu} and was computed using the auxiliary-field diffusion Monte Carlo (AFDMC) algorithm~\cite{Schmidt:1999lik} with local \chiEFT\ interactions from Refs.~\cite{Gezerlis:2013ipa,Gezerlis:2014zia,Tews:2015ufa,Lynn:2015jua}.
This AFDMC band was calculated at N$^2$LO, i.e., at third order in the EFT expansion.
Other EFT calculations using a variety of many-body methods are available in the literature~\cite{Tews:2012fj,Hebeler_2013,Lynn:2015jua,Drischler:2017wtt,Keller:2022crb} but lead to comparable results~\cite{Huth:2020ozf}.

To infer the likelihood of an EOS given \chiEFT\ constraints, we do not interpret the uncertainty band imposed by the \chiEFT\ calculation as a strict boundary, but instead use it to construct a score function $f(p, n)$ that grades the conformity of a given pressure value $p$ at density $n$ with the \chiEFT\ prediction. 
We set $f$ constant on the interval proposed by the AFDMC band and model the likelihood outside its boundaries with an exponential decay:
\begin{align}
    %f(p, n) = \frac{(2/\beta+1)^{-1}}{p_{+}-p_{-}} \begin{cases} \exp\left(-\beta \frac{p-p_{+}}{p_{+} - p_{-}}\right) \qquad \text{if $p > p_{+}$}\\
    f(p, n) = \begin{cases} \exp\left(-\beta \frac{p-p_{+}}{p_{+} - p_{-}}\right) \qquad \text{if $p > p_{+}$}\\
         \exp\left(-\beta \frac{p_{-}-p}{p_{+} - p_{-}}\right) \qquad \text{if $p < p_{-}$}\\
        1 \qquad \qquad \qquad \qquad \qquad \text{else\,.}
    \end{cases}
\label{eq:score_function}
\end{align}
Here, $p_{+}$ and $p_{-}$ are the $n$-dependent upper and lower bounds set by the AFDMC uncertainty band, shown as dashed black lines in Fig.~\ref{fig:score_function}.
To be sufficiently conservative, we set $\beta=6$, so that 75\% of the weight is contained within the interval $(p_{-},p_{+})$. 
This interpretation follows from Ref.~\citep{Furnstahl_2015} where it is shown that for the truncation errors proposed in Ref.~\cite{Epelbaum:2014efa} that are used for our AFDMC band, the error band from the calculation at $j$th order can be interpreted to contain a credibility of \mbox{$j/(j+1)$} when assuming a uniform prior distribution on the unknown higher order coefficients. 
Figure~\ref{fig:score_function} shows $f(p,n)$ across the range of densities for which we impose the constraint. 

The function $f$ measures for each pressure-density point $(p,n)$ its agreement with the \chiEFT\ calculations. We then identify the total likelihood of an EOS as the product of all values of $f$ along its $p(n)$ curve~\citep{Brandes_2023}
\begin{equation}
    \mathcal{L}(\text{EOS}|\chi\text{EFT}) \propto \prod_j f(p(n_j, \text{EOS}), n_j)\,.
\end{equation}
This may also be expressed by an integral over $\log (f)$ along the curve: 
\begin{align}
\begin{split}
    &\mathcal{L}(\text{EOS}|\chi\text{EFT}) \propto \\
   & \exp\left(\int_{\qty{0.75}{\nsat}}^{n_{\text{break}}}\ \frac{\log f(p(n, \text{EOS}), n)}{n_{\text{break}} - \qty{0.75}{\nsat}}\ dn \right)\,.
   \label{eq:chiEFT_likelihood}
   %  &\mathcal{L}(\text{EOS}|\chi\text{EFT}) \propto \\
   % & \exp\left(\int_{\qty{0.75}{\nsat}}^{n_{\text{break}}}\ \frac{\ln\{f[p(n, \text{EOS}), n]\}}{n_{\text{break}} - \qty{0.75}{\nsat}}\ dn \right).
   % \label{eq:chiEFT_likelihood}
\end{split}
\end{align}
The limit of the integral is given by the range in which our EOSs are considered to follow the nucleonic description of the metamodel, i.e., from \qty{0.75}{\nsat} to $n_{\text{break}}$. 
We emphasize that the breakdown densities are EOS dependent, and hence certain EOSs can deviate strongly from the \chiEFT\ prediction above their individual $n_{\text{break}}$.

In Fig.~\ref{fig:chiEFT}, we show our EOS candidates in the $p$-$n$ plane as well as in the equivalent macroscopic $M$-$R$ plane, color coded by their respective posterior probability according to Eq.~(\ref{eq:chiEFT_likelihood}). 
The constraints from \chiEFT\ imply a canonical NS radius of $R_{1.4} = 12.11^{+1.96}_{-3.39}$\,\unit{km} (95\% credibility).
Even though the predictions of \chiEFT\ require the EOS to be relatively soft at lower densities, stiff EOSs with high TOV masses are not ruled out since our extrapolation scheme allows for significant increases in stiffness after the breakdown density.
Hence, the estimated value of the TOV mass differs only mildly between prior and posterior, although the tail of the distribution on $n_{\text{TOV}}$ gets shifted up to $\sim \qty{17}{\nsat}$.

As mentioned above, the truncation of \chiEFT\ at a finite order causes a systematic uncertainty that is expressed as a band of possible pressure values. 
Interpreting such a systematic uncertainty band for Bayesian inference raises ambiguity related to the form of the likelihood in Eq.~(\ref{eq:chiEFT_likelihood}). 
We discuss possible alternatives to our choice of $f$ and the likelihood of Eq.~(\ref{eq:chiEFT_likelihood}) and their impact on the posterior distribution in Appendix~\ref{app:chiEFT}. 

\subsection{Perturbative quantum chromodynamics}
\label{sec:pQCD}

\begin{figure}
    \centering
    \includegraphics[width = \linewidth]{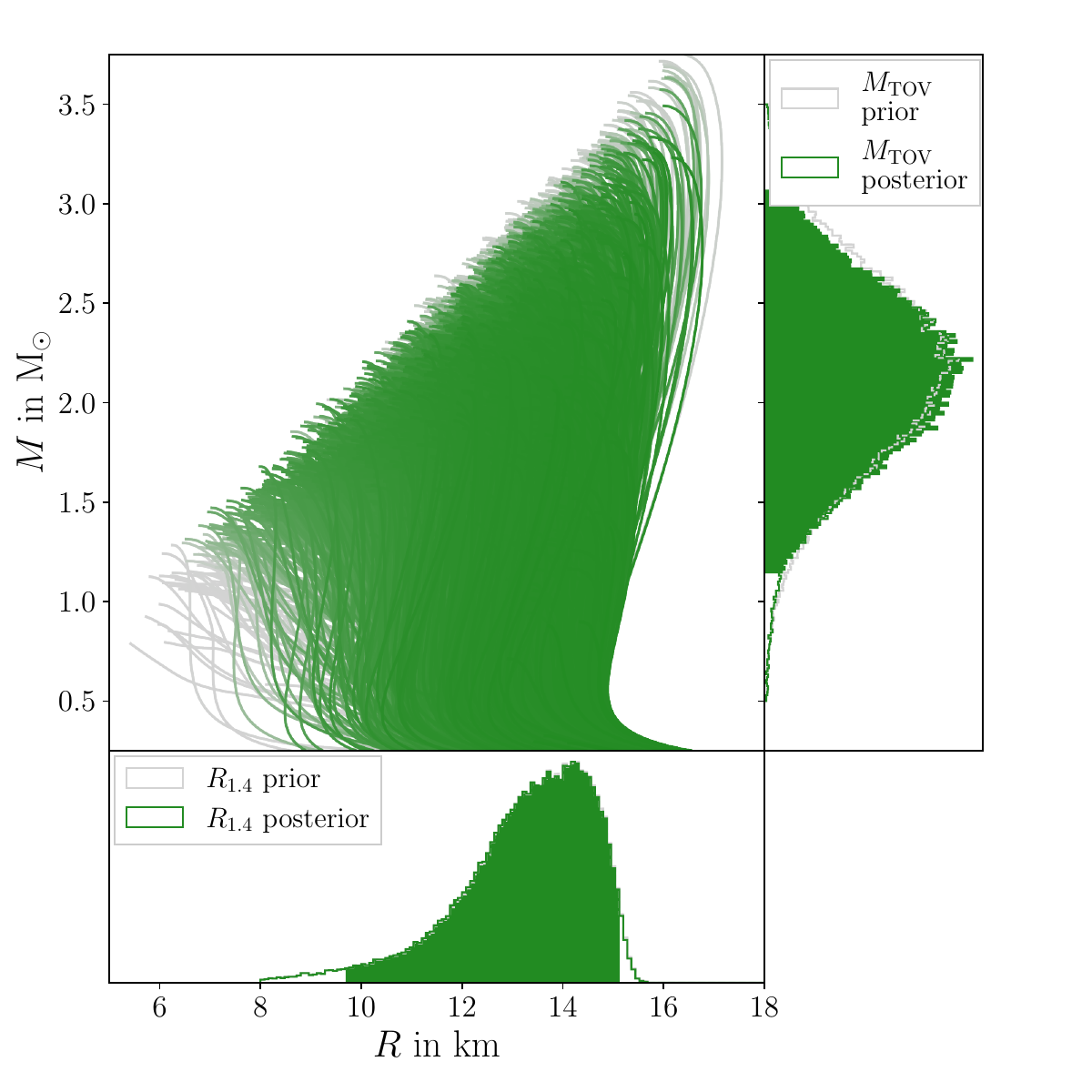}
    \includegraphics[width = \linewidth]{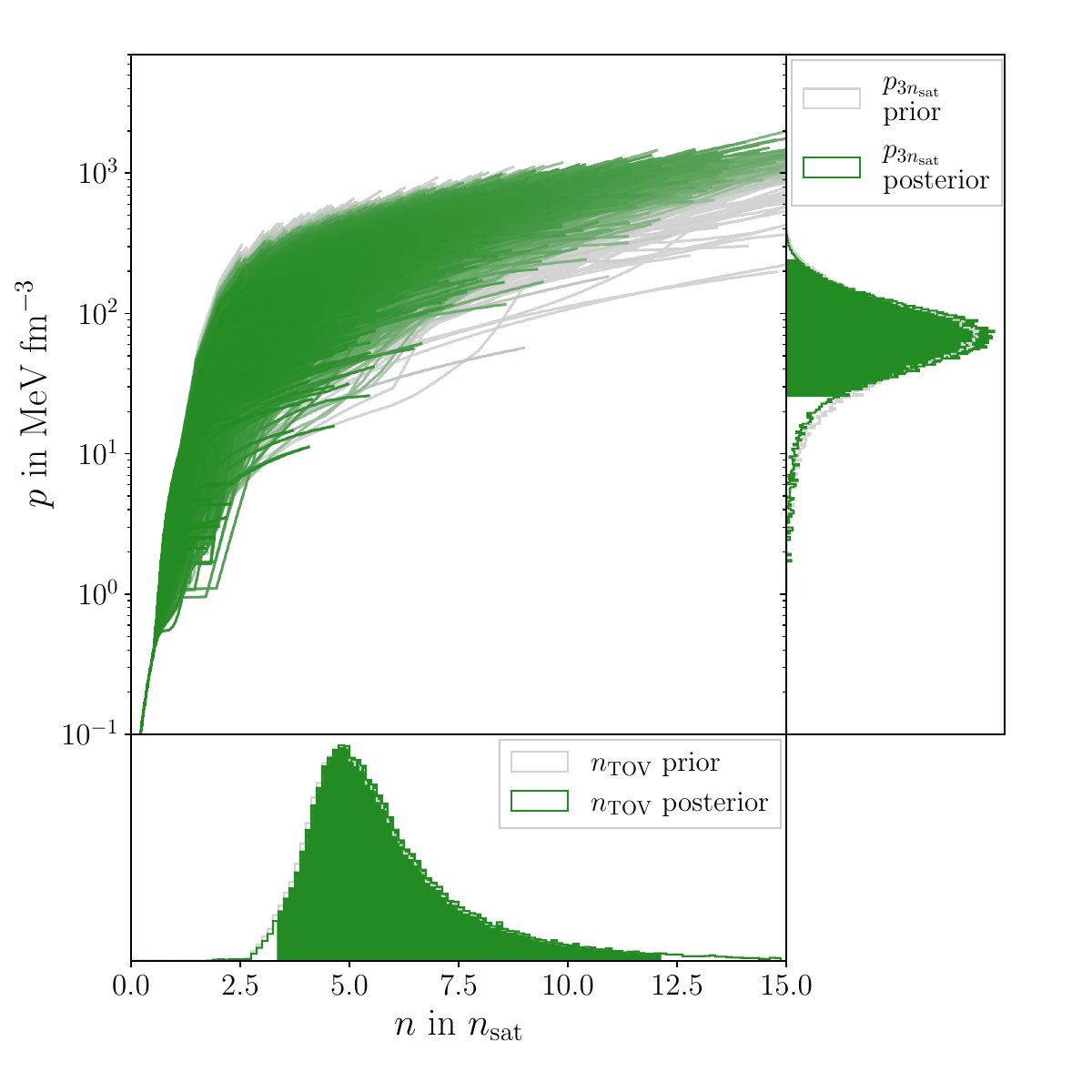}
    \caption{EOS inference based on the constraints set by pQCD when the matching density is set to the TOV density $n_L = n_{\text{TOV}}$. 
    Arrangement and color coding as in Fig.~\ref{fig:chiEFT}.}
    \label{fig:pQCD}
\end{figure}

\begin{figure}
    \centering
    \includegraphics[width = \linewidth]{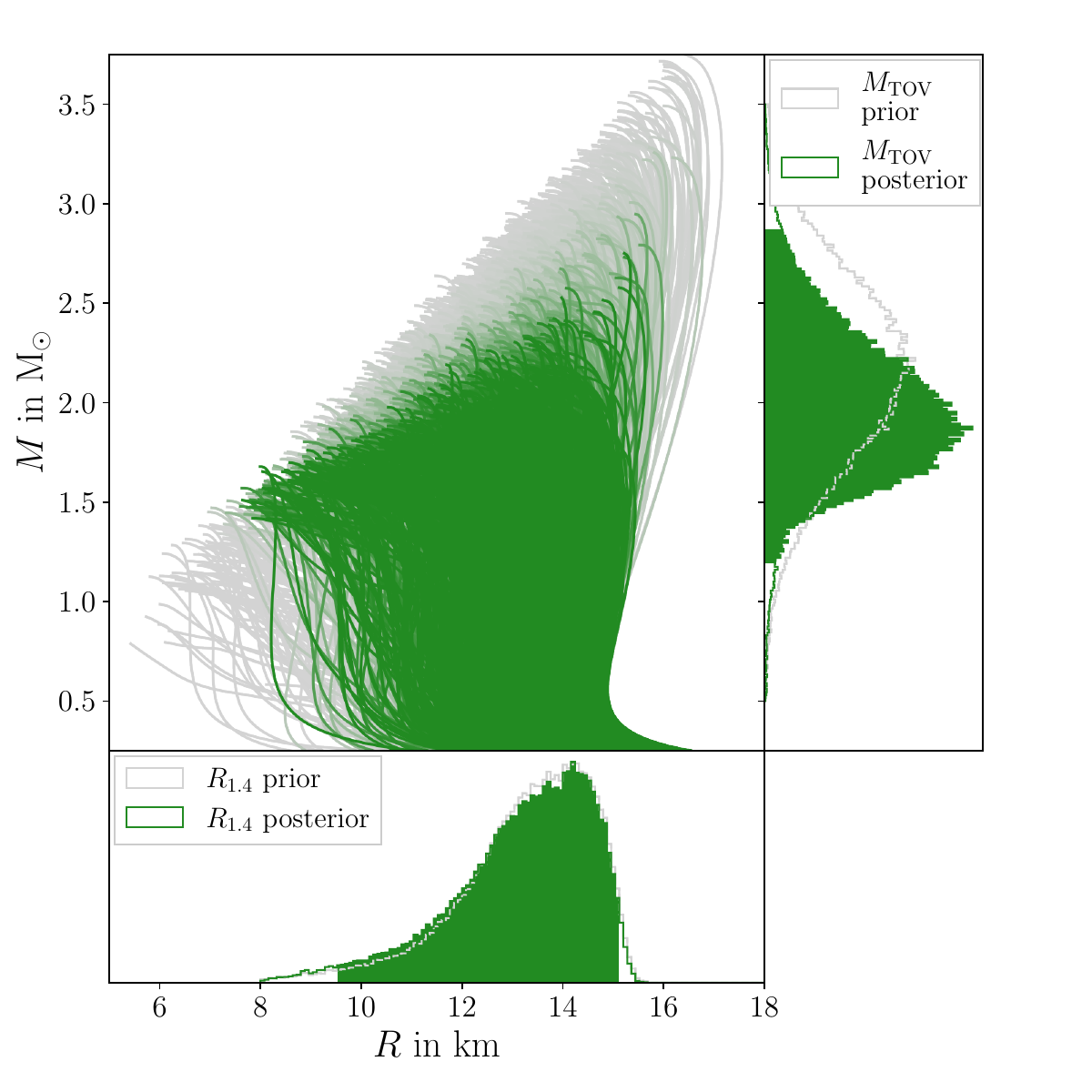}
    \includegraphics[width = \linewidth]{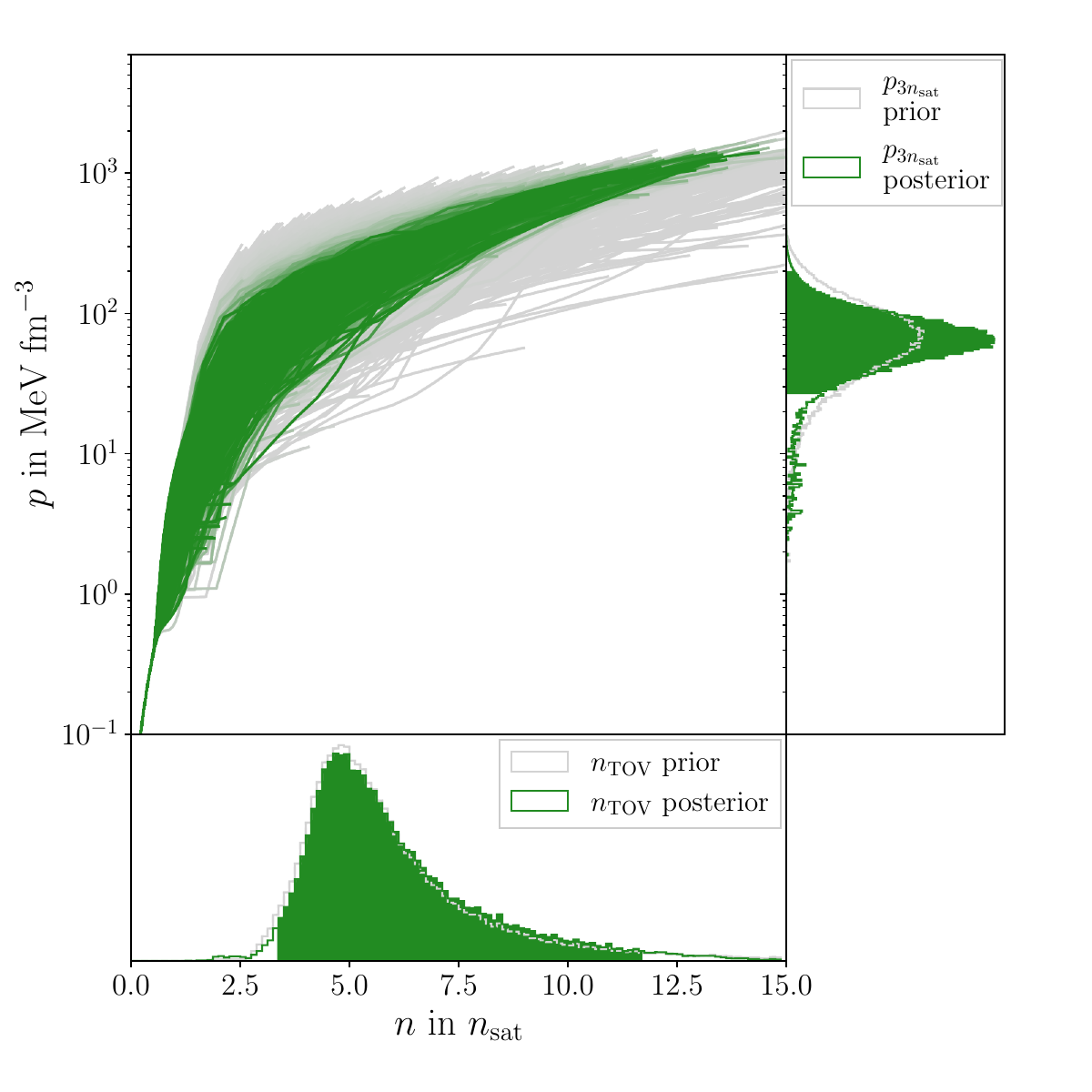}
    \caption{EOS inference based on the constraints set by pQCD$*$, where the likelihood function is determined from a high-density EOS ensemble that is conditioned to the pQCD speed-of-sound prediction between \qtyrange{25}{40}{\nsat}. Arrangement and color coding as in Fig.~\ref{fig:chiEFT}.}
    \label{fig:pQCD_GPR}
\end{figure}

For $n\approx\qty{40}{\nsat}$, it is expected that matter has made the transition to a quark-matter phase and the EOS can be determined through pQCD~\citep{Freedman_1977, Kurkela_2010, Gorda_2021}. 
Naturally, this density regime is far beyond the scope of any terrestrial or astrophysical laboratory, but given the EOS from \textit{ab initio} pQCD calculations, one can work backward to exclude certain pressure regions at lower densities. 
This method has been proposed recently in Refs.~\citep{Komoltsev_2021,Gorda_2023a}. 
It checks whether a point on an EOS candidate $(\mu_L, n_L, p_L)$ with baryon chemical potential $\mu_L$ and pressure $p_L$ at a low density $n_L$ can be connected to the pQCD EOS $(\mu_H, n_H, p_H)$ at a higher density $n_H$ with at least some interpolation which is mechanically stable and causal (i.e., $0 < c_s < c$) at all densities. 
By constructing the most extreme EOS interpolations that satisfy these conditions, it can be shown that if the pressure difference does not lie within the interval
\begin{align}
  \Delta p_{\text{min}} \leq p_H - p_L \leq \Delta p_{\text{max}}\,,
  \label{eq:pQCD_condition}
\end{align}
with 
\begin{align}
    \Delta p_{\text{min}} = \frac{\mu_H^2 - \mu_L^2}{2 \mu_L}n_L\,,\\
    \Delta p_{\text{max}} = \frac{\mu_H^2 - \mu_L^2}{2 \mu_H}n_H \,,
\end{align}
no causal and stable interpolation between the two points exists and the candidate EOS is inconsistent with the pQCD EOS. 

For the inference on the EOS, we first follow the approach of Refs.~\citep{Gorda_2023a, Gorda_2023b}, in which the posterior probability for the EOS is determined by choosing a matching point $(n_L, \epsilon_L, p_L)$ on the EOS and checking whether it can be matched to the prediction from pQCD:
\begin{align}
    P(\epsilon_L, p_L |n_L, \mu_H, n_H, p_H) = \begin{cases} 
        1\quad \text{if Eq.~(\ref{eq:pQCD_condition}) is fulfilled} \\
        0\quad \text{otherwise.}
    \end{cases}
    \label{eq:pQCD_stepfunction}
\end{align}
Reference~\citep{Gorda_2023b} quantifies several uncertainties affecting the pQCD calculations for $p_H$ and $n_H$, specifically with regards to missing higher-order (MHO) contributions and the renormalization scale. 
The errors from missing higher-order contributions are estimated through the Bayesian machine-learning algorithm MiHO~\citep{Duhr_2021} and can be incorporated in the form of a posterior \footnote{As in \citep{Gorda_2023b}, we here estimate the missing higher-order errors only of the pressure and not for the density which has much smaller MHO errors than the pressure.},
\begin{align}
    P_{\text{MHO}}(p_H, n_H|\vec{p}^{(j)}(\mu_H), \vec{n}^{(j)}(\mu_H))\,,
\end{align}
where $\vec{p}^{(j)}(\mu_H)$, $\vec{n}^{(j)}(\mu_H)$ represent the pQCD series expansion for the pressure and number density at chemical potential $\mu_H$. 
The individual terms in the pQCD series depend on an unphysical renormalization scale $\bar{\Lambda}$, which arises due to the truncation of the series at a finite order. 
To avoid artificial preference to any specific scale, we marginalize over the dimensionless renormalization scale 
\begin{align}
    X = \frac{3}{2} \frac{\bar{\Lambda}}{\mu_H}\,,
\end{align}
with a log-uniform distribution between $1/2$ and $2$. %Moreover, also the high density matching scale $\mu_H$ can be marginalized over in the range \qtyrange{2.2}{3}{GeV}. 
%We summarize this below as $P_{\text{ren}}(X, \mu_H|\vec{p}^{(j)})$. 
The likelihood function can then be written as (see Eq.~(2.9) in Ref.~\citep{Gorda_2023b})
%Under the simplifying assumption that the series $\vec{n}^{(j)}$ immediately converges to the true value $n_H$, the final likelihood can be written as (Eq.~2.9 in~\citep{Gorda_2023b})
\begin{align}
    \mathcal{L}(\text{EOS}|&\text{pQCD})  = \int dX dp_H dn_H \nonumber \\
     &P(\epsilon_L, p_L|n_L, \mu_H, n_H, p_H)\\
    \times &P_{\text{MHO}}(p_H, n_H|\vec{p}^{(j)}(\mu_H,X), \vec{n}^{(j)}(\mu_H,X)) \nonumber
    \\
    \times&P_{\text{SM}}(X|\vec{p}^{(j)})\,,\nonumber
\end{align}
where $P_{\text{SM}}(X|\vec{p}^{(j)})$ is an integration weight proportional to the marginalized likelihood of $X$ within the statistical model used in the MiHO algorithm. 
The chemical potential $\mu_H$ is chosen to be \qty{2.6}{GeV} corresponding to $\approx$\qty{40}{\nsat}.

All points on the candidate EOS have to fulfill the condition of Eq.~\eqref{eq:pQCD_condition}, but as long as the candidate EOS itself is causal and stable, it is enough to check only the highest density point for each candidate EOS.
Because the condition is by construction more conservative than any prior on the candidate EOS, the conclusions drawn from pQCD will depend on the value up to which the candidate EOS is extrapolated, i.e., what value for $n_L$ is chosen~\citep{Somasundaram_2023, Gorda_2023a}. 
For a lower $n_L$, the region in the $\epsilon$-$p$ plane from which a stable, causal connection to the pQCD regime is possible grows. 
The pQCD constraint becomes therefore less restrictive at lower matching densities. 
For our purposes, we follow Ref.~\citep{Somasundaram_2023} and terminate our EOSs at the TOV density $n_{\text{TOV}}$.
By doing so, we discard the unstable branch for our EOS candidates, i.e., the part above their TOV densities. 
This is because although a candidate might be compatible with pQCD at its TOV point, the extrapolation above it could still violate the pQCD constraint. 
Matching to the TOV density is a conservative choice that avoids overemphasizing the model for speed-of-sound extrapolation at higher densities. Likewise, the unstable branch of the EOS is not accessible with any other constraint, since all astrophysical and nuclear information is applicalbe only up to the TOV density. 
Densities $n>n_{\rm TOV}$ may be reached in BNS mergers; however, as shown recently this part of the EOS will leave no observable imprint given current and next-generation astrophysical observatories \cite{Ujevic_2023}, making the pQCD the only source of information in this density range in the near future.
A downside of matching the EOS to pQCD at $n_{\text{TOV}}$ is that it treats the EOS differently below and above that density, e.g. allowing for very strong phase transitions with density jumps of several times the saturation density right above $n_{\rm TOV}$ but not below~\cite{Gorda_2023a}. 

In Fig.~\ref{fig:pQCD}, we show the result when applying the pQCD constraint. 
Very stiff and very soft EOS are disfavored by pQCD, though the overall shift in NS radii and masses is slight. 
We note that by matching at the TOV density, very soft EOSs with higher $n_{\text{TOV}}$ are more affected by the pQCD constraint, because they reach densities closer to the actual pQCD regime.

Using the information from pQCD as just described is a conservative approach. More stringent, but more model-dependent, constraints can be obtained by extending the candidate EOS either to a fixed higher density, e.g., \qty{10}{\nsat} as in Ref.~\citep{Gorda_2023a}, or by additionally marginalizing over the possible extensions an EOS can have above $n_{\text{TOV}}$ to reach the correct pQCD limit. 
By relying on the condition in Eq.~(\ref{eq:pQCD_stepfunction}), we assign equal likelihood to an EOS that matches the pQCD constraint only through one of the most extreme extrapolations and an EOS that allows for a variety of plausible extensions beyond $n_{\text{TOV}}$.
Reference~\citep{Komoltsev_2023} addresses this imbalance by introducing a method to determine the QCD likelihood function from an ensemble of Gaussian-process-generated EOS segments at high densities conditioned to the well-convergent pQCD speed-of-sound band between \qtyrange{25}{40}{\nsat}. 
The QCD likelihood at a given matching point $(n_L, \epsilon_L, p_L)$ is then simply given by the kernel density estimate of the samples provided through this conditioned high-density EOS ensemble; for further details see Ref.~\citep{Komoltsev_2023}. 
We call this implementation of the pQCD constraint pQCD$*$. 
In Fig.~\ref{fig:pQCD_GPR}, we show the resulting posterior distribution when this method is applied to our EOS candidates, using the publicly available code in Ref.~\citep{Komoltsev_2024}. 
Since now the pQCD information from the speed of sound is used at lower densities, more EOSs are rejected and the posterior on $\mtov$ and $p_{3\nsat}$ becomes more informative, yet NS radii and the TOV number density remain relatively unaffected.
This illustrates that more potent constraints from pQCD are achievable, though the constraining power of the pQCD input depends on how exactly the pQCD prediction at high densities is backpropagated to the TOV density at several $\nsat$. 
Unless stated otherwise, we refer to the more conservative approach of Eq.~\eqref{eq:pQCD_stepfunction} when we speak of the pQCD constraint on our EOS set in following sections.

\subsection{Symmetry energy constraints from atomic nuclei}
\label{Sec:PREX}
\begin{figure}
    \centering
    \includegraphics[width = \linewidth]{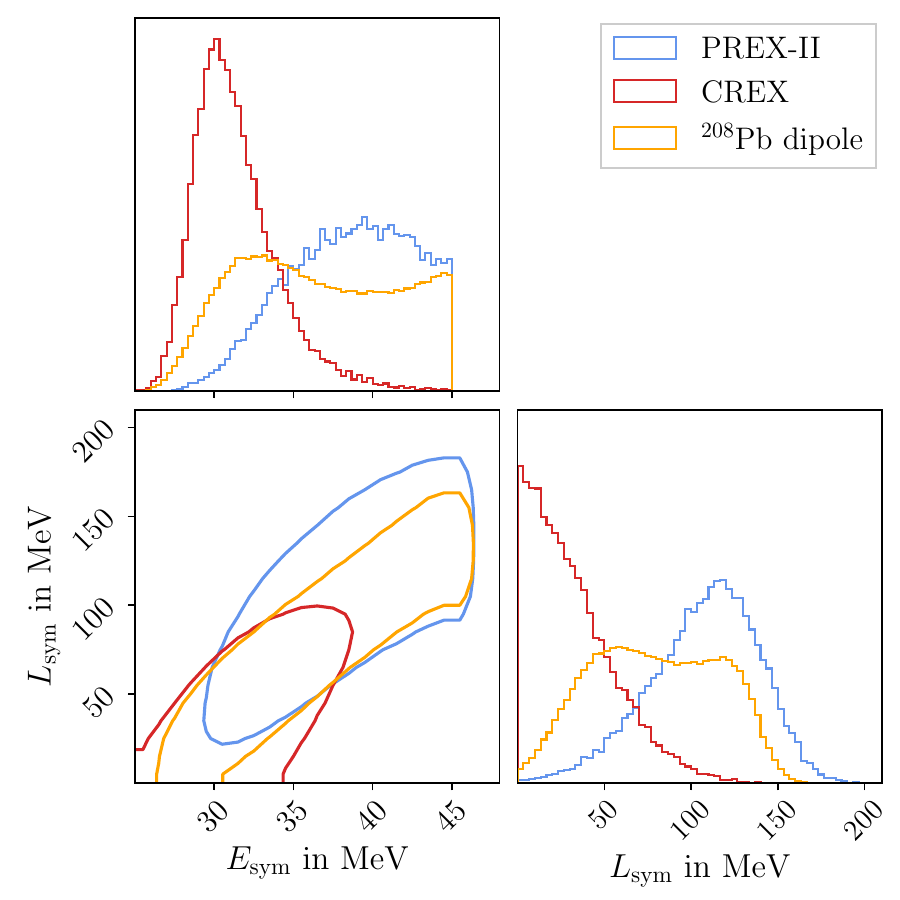}
    \caption{Posterior of the symmetry energy at saturation density $E_{\rm sym}$ and its slope $L_{\rm sym}$ as inferred from the PREX-II (blue line), CREX (red line), and $^{208}$Pb dipole measurements (orange line). Contours are shown at 95\% credibility.}
    \label{fig:ENP}
\end{figure}

\begin{figure}
    \centering
    \includegraphics[width = \linewidth]{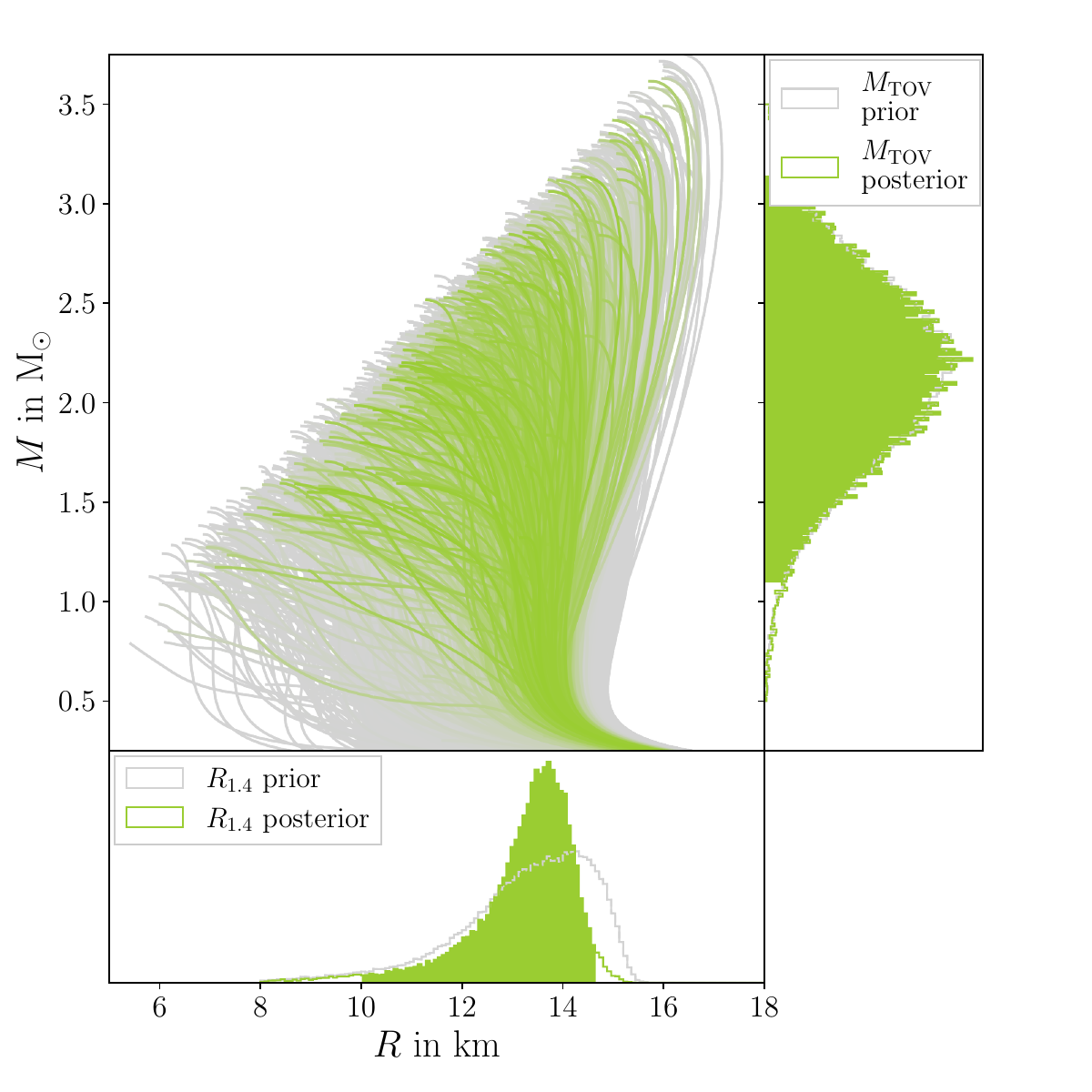}
    \includegraphics[width = \linewidth]{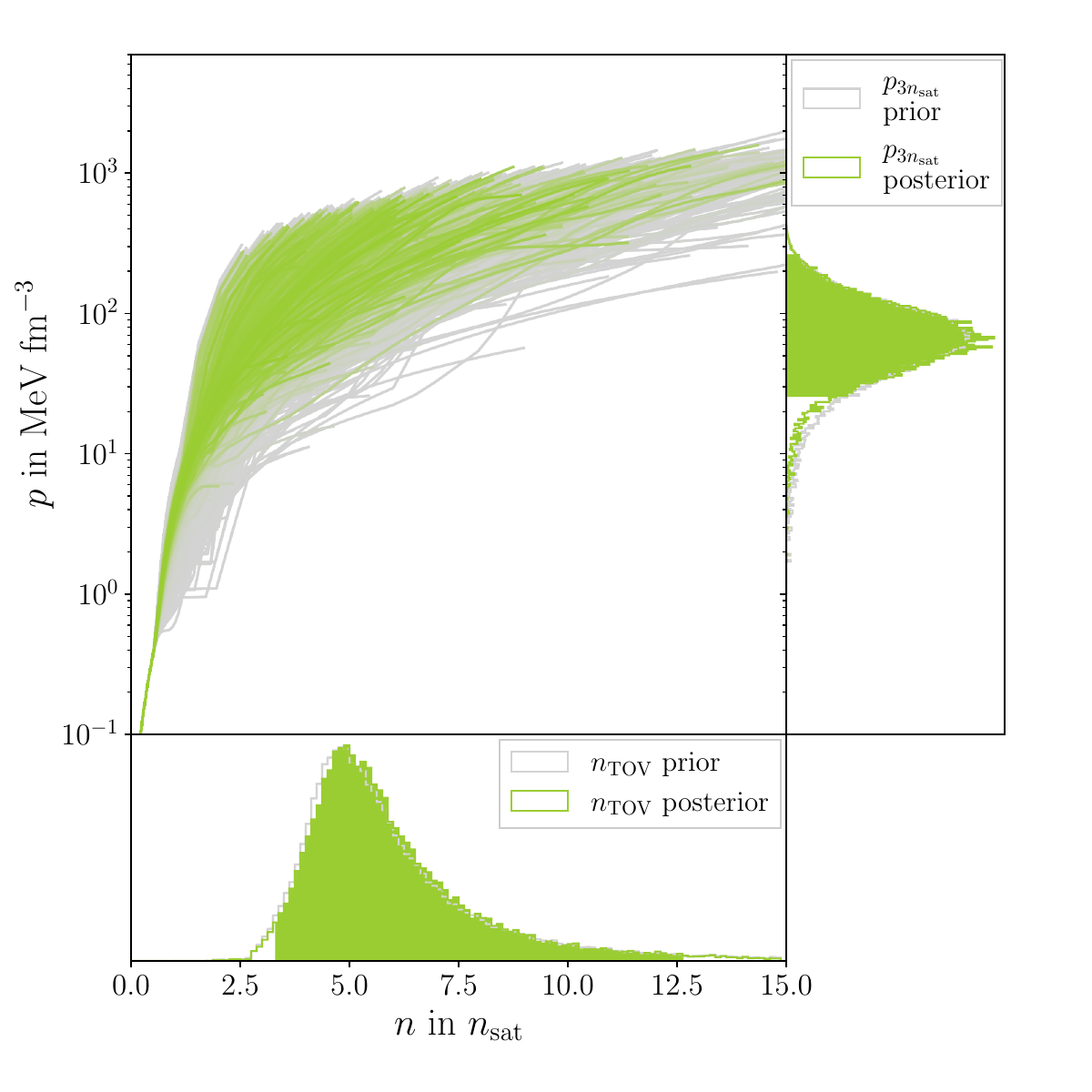}
    \caption{EOS inference based on the information of the PREX-II measurement. Arrangement and color coding as in Fig.~\ref{fig:chiEFT}.}
    \label{fig:PREX}
\end{figure}

\begin{figure}
    \centering
    \includegraphics[width = \linewidth]{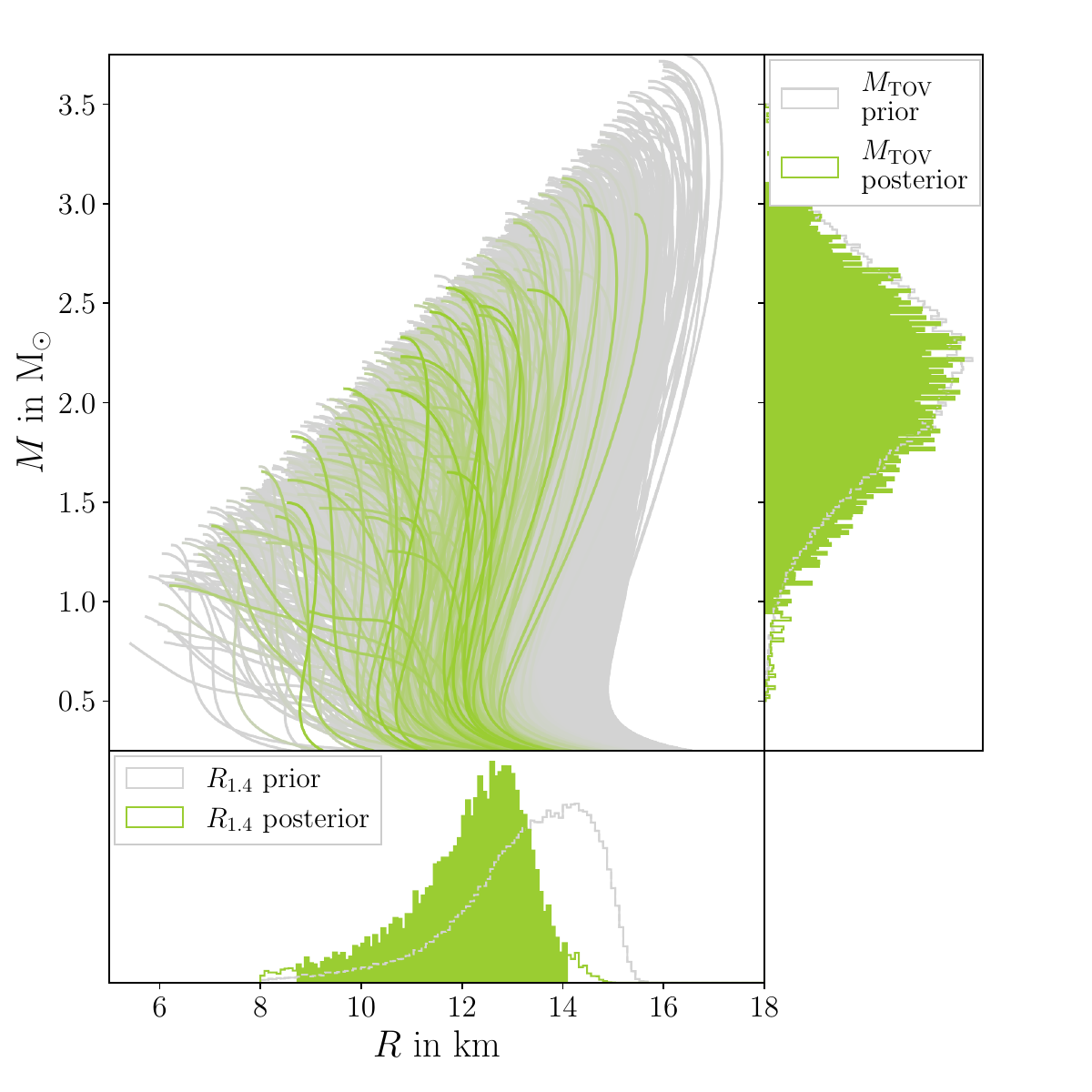}
    \includegraphics[width = \linewidth]{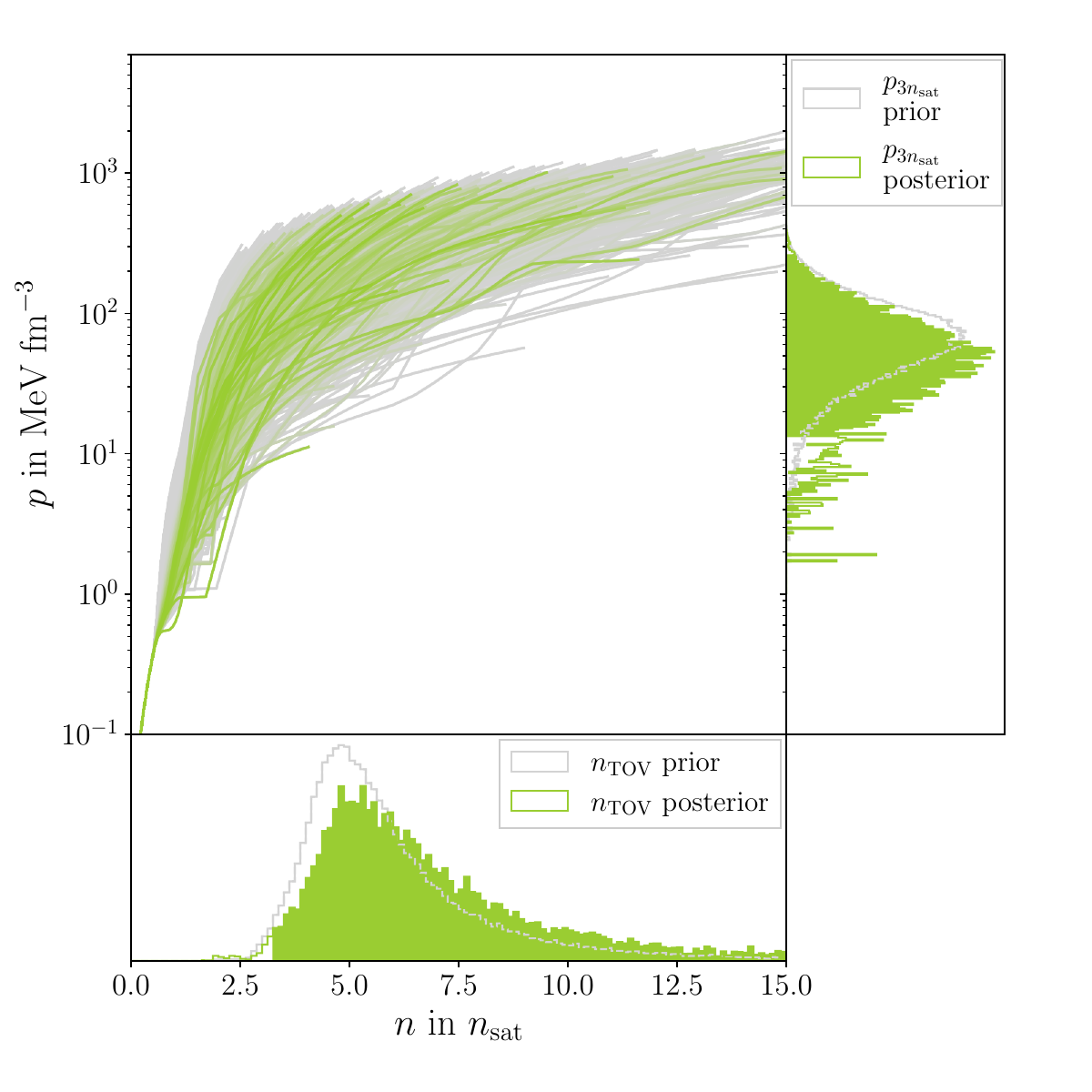}
    \caption{EOS inference based on the information of the CREX measurement. Arrangement and color coding as in Fig.~\ref{fig:chiEFT}.}
    \label{fig:CREX}
\end{figure}
\begin{figure}
    \centering
    \includegraphics[width = \linewidth]{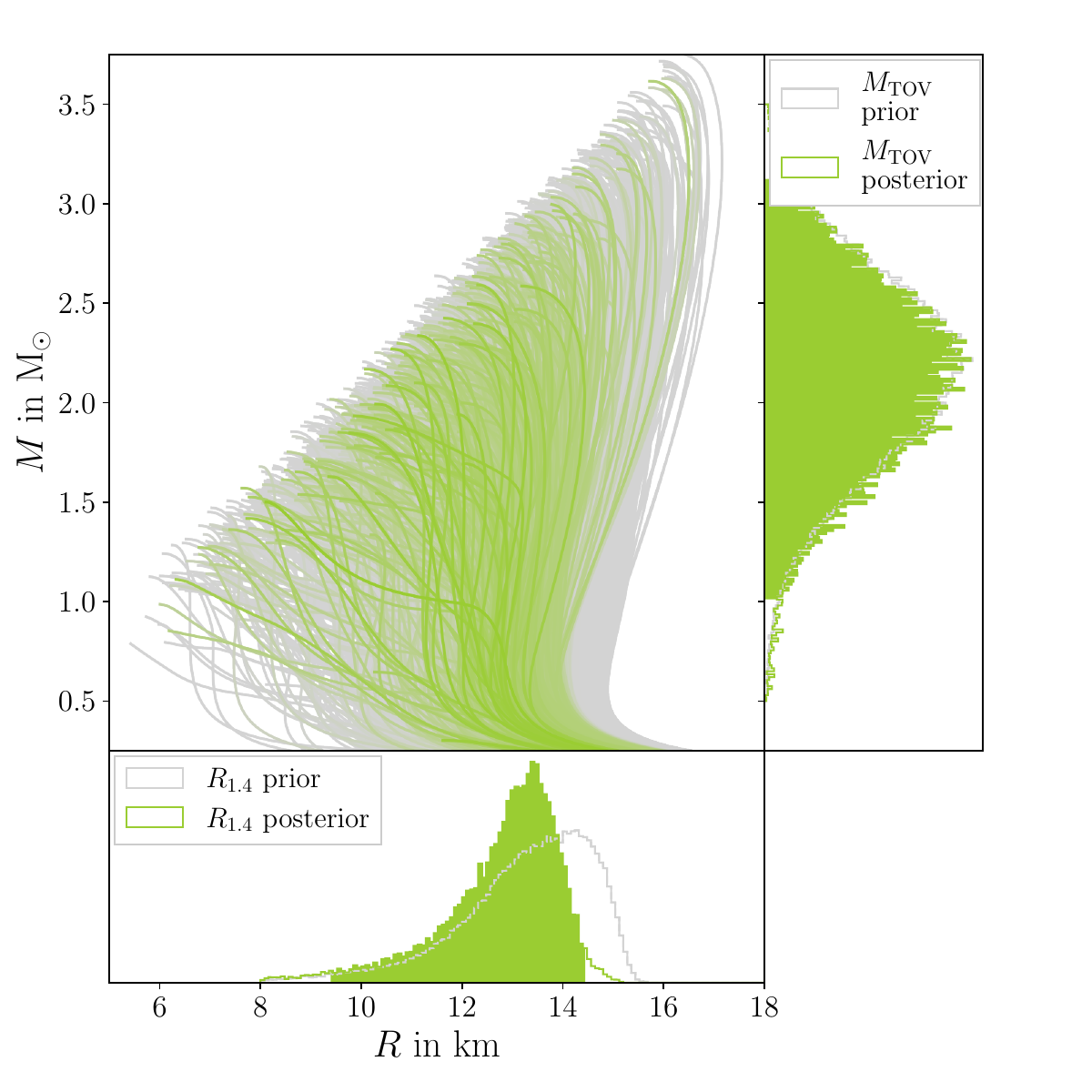}
    \includegraphics[width = \linewidth]{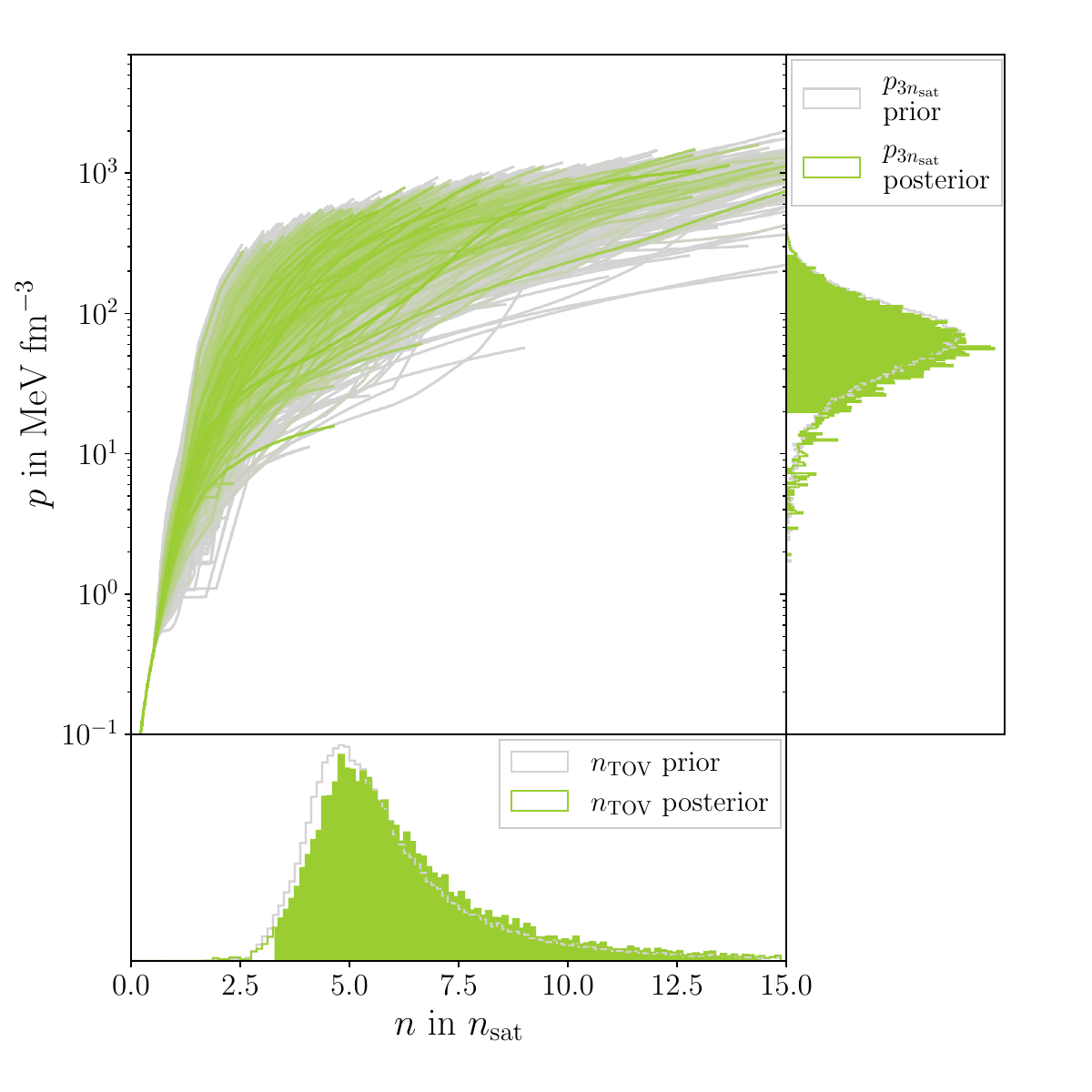}
    \caption{EOS inference based on the measurement of the electric dipole polarizability in $^{208}$Pb. Arrangement and color coding as in Fig.~\ref{fig:chiEFT}.}
    \label{fig:dipole}
\end{figure}
A significant part of the uncertainty about the EOS at nuclear densities can be attributed to the lack of knowledge about nuclear interactions at large isospin asymmetries.
This can be mapped into the uncertainty of the symmetry energy and specifically its slope parameter $L_{\rm sym}$. 
The parameter $L_{\rm sym}$ is proportional to the pressure of pure neutron matter at saturation density $p_{\rm pnm}(\nsat)$,
\begin{align}
    p_{\rm pnm}(\nsat) = \frac{1}{3} \nsat L_{\rm sym}\,,
\end{align}
which emphasizes the strong connection between $L_{\rm sym}$ and the EOS, showcasing its importance in constraining the EOS from experiments at subsaturation densities~\cite{fattoyev:2012}. 
Notably, the neutron-skin thickness of an atomic nucleus, defined as the difference in the point neutron radius $R_n$ and point proton radius $R_p$ \citep{Thiel_2019}, obeys a strong correlation with $L_{\rm sym}$~\cite{horowitz:2001,carriere:2003,klupfel:2009}, which is one of the motivations to experimentally measure the neutron-skin thickness of atomic nuclei. 
The neutron skin has been inferred from weak form factors~\citep{PREX2,Adhikari_2022}, from the electric dipole polarizability~\citep{Birkhan_2017, Tamii_2011, Roca-Maza_2015}, and from strong probes~\citep{Giacalone_2023, Zenihiro_2018, Friedman_2012, Tarbert_2014, Brown_2007}. 
Here, we focus on the first two methods since matching the observables from strong probes to the symmetry energy introduces significant model dependencies~\citep{Lattimer_2023, Tsang_2012}. 
We discuss extracting symmetry-energy constraints from heavy-ion collisions in Sec.~\ref{Sec:HIC}.

\textit{Weak form factor.} To determine the neutron radius $R_n$ for the doubly closed-shell isotope $^{208}$Pb, the PREX-II experiment measured the parity-violating asymmetry from scattered electrons and translated this to an estimate for $R_n$ via theoretical mean-field model calculations. 
Similary, the neutron radius of $^{48}$Ca was determined in the CREX experiment in an analogous manner.
Since the proton radius $R_p$ of these nuclei is well known from electron-scattering experiments, these experiments directly yield a neutron-skin thickness $R_{\rm skin} \coloneqq R_n-R_p$ of 0.28$^{+0.14}_{-0.14}$\,fm~\citep{PREX2} for $^{208}$Pb and 0.12$^{+0.07}_{-0.07}$\,fm~\citep{Adhikari_2022} for $^{48}$Ca.

To translate the neutron-skin thickness to a constraint on the EOS, we first determine the posterior on $E_{\rm sym}$ and $L_{\rm sym}$ from the PREX-II and CREX results, following the approach laid out in Ref.~\citep{Reed_2021}.
For this purpose, we collect several model predictions that relate $L_{\rm sym}$ and $R_{\rm skin}$.
The collection of models employed encompasses a wide range of both covariant and nonrelativistic EDFs that have been optimized with respect to nuclear binding energies, charge radii, and giant monopole resonances as described in Refs.~\citep{Chen_2014, Reed_2021}.
While our set is by no means complete, our collection constitutes a representative sample of EDFs that vary widely in predictions of properties of both atomic nuclei and nuclear matter.
The resulting correlation of the predicted $R_{\rm skin}$ with $L_{\rm sym}$ is very pronounced~\cite{horowitz:2001,carriere:2003,Roca-Maza:2011,fattoyev:2012}; hence we perform a simple least-squares linear fit to calculate a relationship:
\begin{align}
    R^{\text{(fit, 208)}}_{\rm skin}(L_{\rm sym})&= 1.69\times10^{-3}\ L_{\rm sym}+0.09\, \rm{fm} \label{eq:Rskin208}\,,\\    
    R^{\text{(fit, 48)}}_{\rm skin}(L_{\rm sym})&= 8.90\times10^{-4}\ L_{\rm sym}+ 0.13\, \rm{fm} \label{eq:Rskin48}\,.
\end{align}

\textit{Electric dipole polarizability.} The dipole response of an atomic nucleus carries information about the symmetry energy and neutron-skin thickness, because the neutrons and protons react differently to the external electromagnetic excitation. 
Hence, the nuclear dipole polarizability $\alpha_D$ is strongly related to the symmetry energy parameters \citep{Piekarewicz_2014}. 
This value has been experimentally measured for $^{208}$Pb~\citep{Tamii_2011}, $^{120}$Sn~\citep{Hashimoto_2015}, $^{48}$Ca~\cite{Birkhan_2017}, and $^{40}$Ca~\citep{Fearick_2023} through inelastic proton scattering at forward angles, as well as for $^{68}$Ni~\cite{Rossi_2013} through measuring the decay neutron energy. 
Because the measurement for $^{208}$Pb has the smallest experimental uncertainty, we here focus only on this instance. 
To relate this measurement back to $E_{\sym}$ and $L_\sym$ we follow Refs.~\citep{Essick_2021, Essick_2021A}, where, based on a collection of EDFs~\citep{Roca-Maza_2013}, a relation between the symmetry energy and $\alpha_D$ was derived:
\begin{align}
    \alpha_D E_{\sym} = 93.5\,\text{fm}^3\,\text{MeV} + 3.08\,\text{fm}^3\,L_{\sym}\,,
    \label{eq:dipole_relation}
\end{align}
with a Gaussian error of $\sigma=27.6\,\text{fm}^{3}\,\text{MeV}$ for Eq.~\eqref{eq:dipole_relation}. 

To get a posterior on $E_{\rm sym}$ and $L_{\rm sym}$ from the aforementioned measurements, we sample on these parameters with a Markov chain Monte Carlo (MCMC) algorithm. 
The likelihood compares the experimentally determined values with the result predicted from Eqs.~(\ref{eq:Rskin208}, \ref{eq:Rskin48}, \ref{eq:dipole_relation}). For instance, in case of PREX-II the likelihood reads
\begin{align}
\begin{split}
   \log \mathcal{L}(E_{\rm sym}, L_{\rm sym}|&\text{PREX-II}) = \\
   -\frac{1}{2} \Biggl(&\biggl(\frac{R_{\rm skin} -R^{(\text{fit}, 208)}_{\rm skin}(L_{\rm sym})}{\sigma_{\rm expt}}\biggr)^2 + \\
   &\biggl(\frac{E_{\rm sym}-E^{\text{(fit)}}_{\rm sym}(L_{\rm sym})}{\sigma_{\rm fit}(L_{\sym})} \biggr)^2 \Biggr)\,.
\label{eq:sym_sampling_likelihood}
\end{split}
\end{align}
Here, $R_{\rm skin}$ and $\sigma_{\rm expt}$ are the experimental value and uncertainty from PREX-II.
The expression is analogous for the CREX and $^{208}$Pb dipole measurements. 
The second summand in Eq.~\eqref{eq:sym_sampling_likelihood} accounts for the correlation between $E_{\rm sym}$ and $L_{\rm sym}$ in our EDF set. It utilizes a fit relation that we obtained from the same set EDFs used to deduce Eqs.~(\ref{eq:Rskin208})~and~(\ref{eq:Rskin48}):
\begin{align}
\begin{split}
    E^{\rm (fit)}_{\rm sym}(L_{\rm sym})& = a L_{\rm sym}^3 + b L_{\rm sym}^2 + c L_{\rm sym} + d\,,\\
    \qquad &a = 3.54\cdot10^{-6}, \ \  b = -1.81\cdot10^{-4}, \\ \qquad &c = 0.063, \ \  d = 29.01.
\end{split}
\end{align}
We denote the 68\% fitting error to this relation as $\sigma_{\rm fit}(L_{\rm sym})$.

The priors for the MCMC sampling are uniform in the ranges
\begin{align}
\begin{split}
    \qty{25}{MeV}\leq E_{\rm sym} \leq \qty{45}{MeV},\\
    \qty{0}{MeV} \leq L_{\rm sym} \leq \qty{200}{MeV},
\end{split}
\end{align}
which are conservative theoretical limits obtained from models that reproduce nuclear-physics data~\cite{carriere:2003,horowitz:2001,fattoyev:2012,Roca-Maza:2011,Dutra:2012} and roughly coincide with the prior ranges for the metamodel part of our EOS set in Table~\ref{tab:NEP_range}.

Figure~\ref{fig:ENP} compares the resulting posterior distributions on $E_{\rm sym}$ and $L_{\rm sym}$ from PREX-II, CREX, and the $^{208}$Pb dipole measurement. 
For $E_\sym$ we find from PREX-II, CREX, and the $^{208}$Pb dipole, respectively ${39}_{-8}^{+6}$, ${31}_{-3}^{+7}$, and ${36}_{-8}^{+9}$\,MeV. 
Likewise, the slope $L_\sym$ is constrained at $110^{+50}_{-72}$, $27^{+65}_{-26}$, and $79^{+62}_{-64}$\,MeV. 
These posteriors serve as a likelihood on our EOS candidate set, since each EOS has corresponding values for $E_{\rm sym}$ and $L_{\rm sym}$ from its generation with the metamodel. 
Hence, we simply set
\begin{align}
    \mathcal{L}(\text{EOS}|\text{PREX-II}) = P(E_{\rm sym}, L_{\rm sym}|\text{PREX-II})\,.
\end{align}
The expressions for CREX and the electric dipole measurement are again completely analogous.

The impact of these measurements on our EOS set is shown in Figs.~\ref{fig:PREX},~\ref{fig:CREX}, and \ref{fig:dipole}. 
While the constraints of CREX are fairly identical to the ones by \chiEFT\ presented in Sec.~\ref{sec:ChiralEFT}, consistent with the calculations of Ref.~\cite{Hagen:2015yea}, the PREX-II analysis leads to larger values for $L_{\rm sym}$, and therefore prefers EOSs that are very stiff around saturation density.
The measurement of $\alpha_D$ for $^{208}$Pb only constrains the relationship between $E_\sym$ and $L_\sym$, and hence, the posterior still covers almost the entire prior range for $E_\sym$. Yet, this measurement still excludes the highest values for $L_\sym$ along the correlation line with $E_\sym$.
All measurements affect $R_{1.4}$ with PREX-II leading to a larger radius of $13.44_{-3.43}^{+1.18}$\,km than CREX with ${12.33}_{-3.59}^{+1.74}$\,km. 
The dipole measurement lies just in between with ${13.03}_{-3.59}^{+1.36}$\,km.
However, none of them has a strong impact on the inferred EOS properties at several times saturation density such as $M_{\text{TOV}}$ or $p_{3\nsat}$.

It has been suggested that the results from both CREX and PREX-II display mutual tension, as the large symmetry energy slope from PREX-II is hard to reconcile with the smaller value recovered from CREX \citep{Miyatsu_2023, Reed_2023}. 
Because the statistical uncertainties are broad, we find posterior overlap for all three measurements in Fig.~\ref{fig:ENP}, similar to other analyses~\citep{Yue_2024, Mondal_2023, Zhang_2023}. 
Yet, the slight disagreement between CREX and PREX-II may hint at a larger systematic error than estimated.
Uncertainties in the experimental setup of CREX and PREX-II are produced mainly from the inherent helicity correlation and polarization in the electron beam. When determining $\alpha_D$, systematic uncertainties can arise from the calibration of excitation energies and the offset for the scattering angle~\citep{Tamii_2009}.
An additional uncertainty arises in the conversion of the measured observables to the symmetry energy parameters. 
When translating the parity-violating asymmetry to the neutron-skin thickness, this theoretical uncertainty is quantified at \qty{0.012}{fm} for $R^{208}_{\rm skin}$~\citep{PREX2} and at \qty{0.024}{fm} for $R^{48}_{\rm skin}$~\citep{Adhikari_2022}.
Extracting symmetry energy information from $\alpha_D$ also relies on EDF predictions that can introduce bias~\citep{Piekarewicz_2012, Roca-Maza_2015}.
This emphasizes the need for enhanced theoretical models in order to reduce the systematic uncertainties currently present and possibly accommodate both the PREX-II and CREX results~\citep{Reinhard_2021}.

\subsection{Heavy-Ion collision experiments}
\label{Sec:HIC}

\begin{figure}
    \centering
    \includegraphics[width = \linewidth]{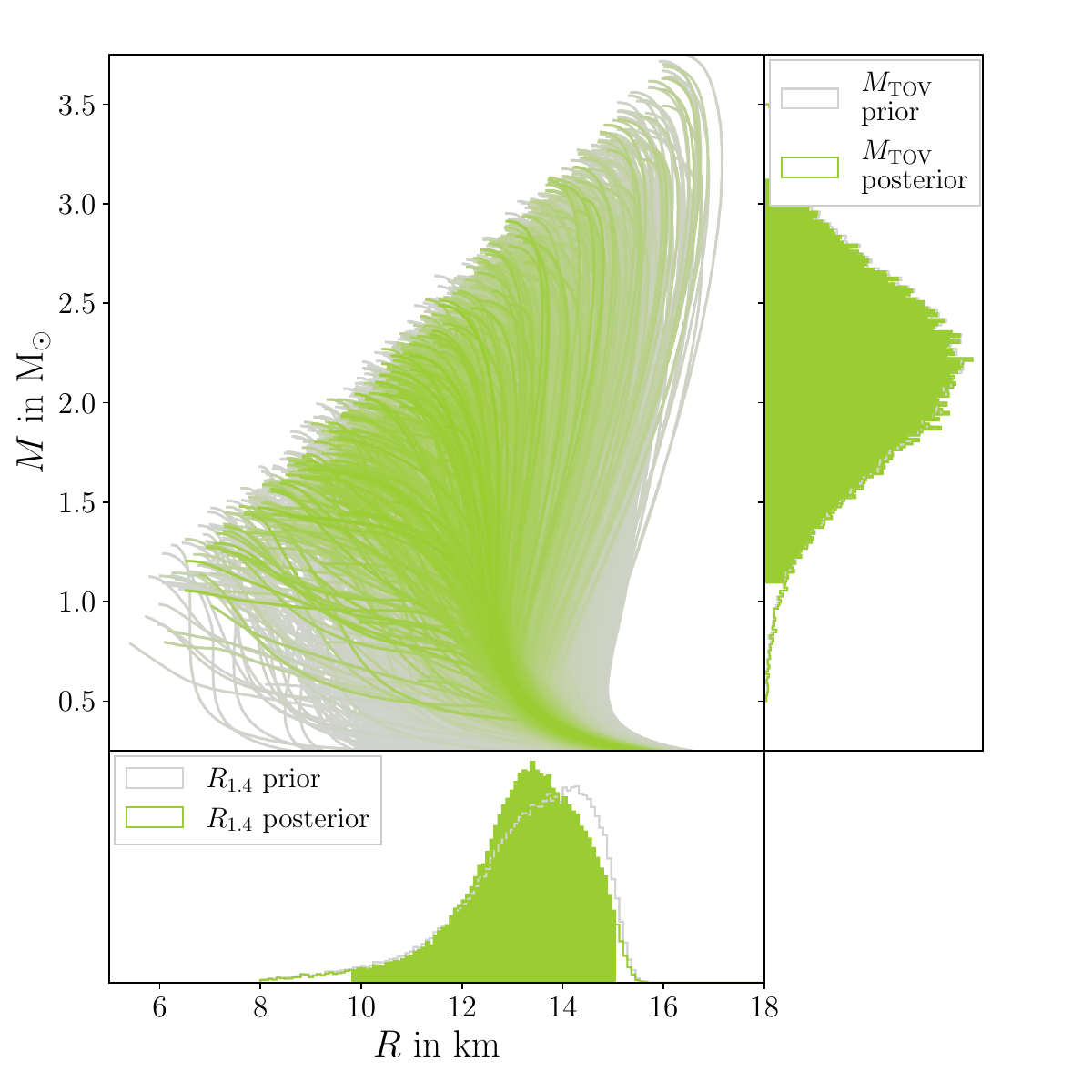}
    \includegraphics[width = \linewidth]{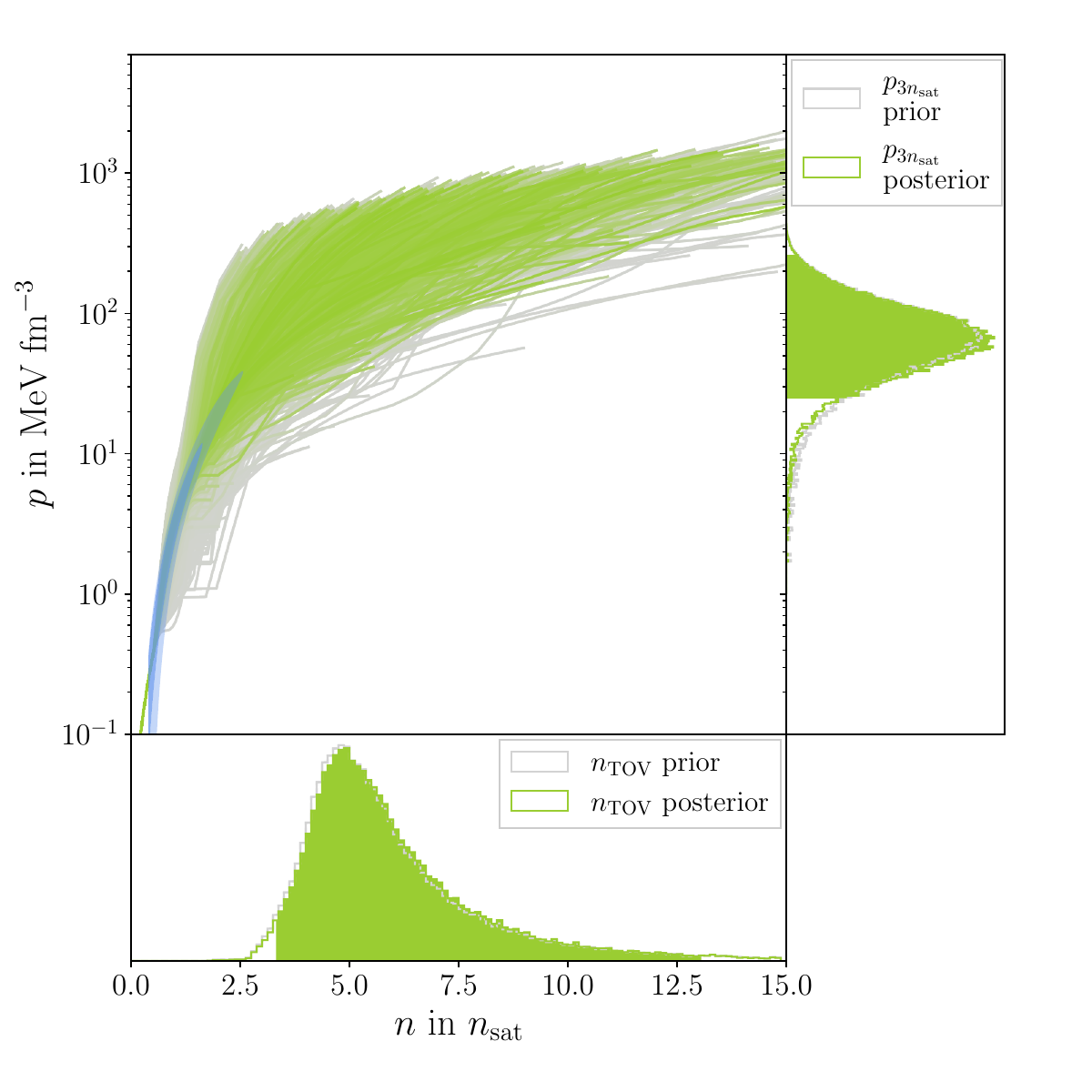}
    \caption{EOS inference based on the information from $^{197}$Au  collisions in Ref.~\citep{Huth_2022}. 
    Arrangement and color coding as in Fig.~\ref{fig:chiEFT}.
    The blue band shows the inferred pressure distribution over the density range times the sensitivity of the ASY-EOS experiment, i.e., the integrand of Eq.~\eqref{eq:HIC_likelihood}.}
    \label{fig:HIC}
\end{figure}

The previously discussed nuclear-physics constraints from \chiEFT\ and the PREX-CREX experiments mostly affect the EOS near nuclear saturation density.
In HIC experiments, heavy atomic nuclei are collided at relativistic energies and their matter gets compressed. Hence, these experiments provide the opportunity to study nuclear matter above saturation density and to constrain the EOS in the density range of \qtyrange{1}{2}{\nsat} for beam energies up to \qty{2}{GeV} per nucleon~\cite{Danielewicz_2002, Fevre_2016, Russotto_2016, Tsang_2019}.

To implement these constraints on our EOS set, we follow the approach of Ref.~\citep{Huth_2022}.
In particular, we employ constraints from the Four-Pi (FOPI)~\cite{Fevre_2016} and the Asymmetric-Matter EOS (ASY-EOS) experimental campaigns~\cite{Russotto_2016} performed at the GSI Helmholtz Centre for Heavy Ion Research, as well as the results of Ref.~\cite{Danielewicz_2002} for symmetric nuclear matter.
The FOPI and ASY-EOS experiments collided gold nuclei at \qtyrange{0.4}{1.5}{GeV} per nucleon and provide constraints both on symmetric nuclear matter as well as the nuclear symmetry energy, due to the initial isospin asymmetry. 
Additionally, the S$\pi$rit experiment recently analyzed the spectral distributions of charged pions created in the collisions of enriched tin isotopes. 
The pion multiplicity depends strongly on the ratio of proton to neutron density in the collision region and thus constitutes a probe of the symmetry energy~\citep{Estee_2021}. 
The results of the S$\pi$rit experiment are so far consistent with the results from ASY-EOS. Because of the similarly large uncertainties, we do not anticipate any additional information on the EOS as of now~\citep{Estee_2021, Yong_2021}. 
Therefore, for the present analysis we consider only the ASY-EOS data.

To extract information on the symmetry energy from collision data, one has to analyze the expansion of the fireball of hadronic matter that forms in the overlapping region of the nuclei. 
Its expansion is dictated by the achieved compression and is thereby sensitive to the EOS.
This sensitivity can be analyzed by investigating the elliptic flow $v_2$~\cite{Fevre_2016,Danielewicz_2002}.
This quantity is determined from the azimuthal distribution of the emitted particles with respect to the reaction plane (RP), i.e., with respect to the angle $\Phi -\Phi_{\rm RP}$. 
The elliptic flow $v_2$ is the second moment in the Fourier expansion of this distribution:
\begin{equation}
\begin{aligned}
    \frac{d\sigma(y,p_{t})}{d\Phi}= &C\ [1+2v_{1}(y,p_t)\cos(\Phi-\Phi_{\rm RP}) \\
    &+2v_{2}(y,p_t)\cos(2(\Phi-\Phi_{\rm RP}))+...].
\end{aligned}
\end{equation}
All Fourier coefficients $v_{n}$ depend on the longitudinal rapidity 
\begin{align}
    y \coloneqq \frac{1}{2}\ln\left(\frac{E+p_z}{E-p_z}\right)\,,
\end{align} 
where $p_z$ is the momentum along the beam axis and $E$ is the total energy perpendicular to the beam axis of the particle. 
Because effects of the initial asymmetry are small, it was suggested to employ the asymmetric flow ratio for neutrons over protons $v_2^{\rm np} = v_2^{\rm n} / v_2^{\rm p}$~\cite{Russotto_2011}. 
Using the flow ratio as measured by the ASY-EOS experiment, simulations using the UrQMD transport model have been used to extract information on the nuclear symmetry energy~\cite{Russotto_2016}.
In Ref.~\cite{Huth_2022}, it was shown that these results are consistent also for other transport models, such as IQMD~\cite{Hartnack_1998} and Tübingen QMD~\cite{Cozma_2018}.

For the UrQMD simulations, the EOS functional is defined as 
\begin{equation}
\dfrac{E}{A}(n,\delta)\approx e_{\rm sat}(n)+e_{\rm sym}(n) \delta^2 + \dots,
\label{eq:expandE}
\end{equation}
where
\begin{align}
    \delta \coloneqq 1-2\ \frac{n_p}{n}
\end{align}
is the asymmetry parameter with $n_p$ denoting the proton number density. 
Note that the truncation of the expansion in Eq.~\eqref{eq:expandE} at second order in $\delta$ is justified since the neglected nonquadratic terms are expected to be small~\citep{Somasundaram:2020chb}.  
The first term in Eq.~\eqref{eq:expandE} denotes the energy per particle for symmetric nuclear matter, and for the analysis here is parametrized as \citep{Fevre_2016}
\begin{equation}
    e_{\rm sat}(n)=\dfrac{3}{5}\left(\dfrac{n}{n_{\rm sat}}\right)^{2/3}E_F + \dfrac{\alpha}{2} \left(\frac{n}{\nsat}\right)+\dfrac{\beta}{\gamma +1}\left(\dfrac{n}{n_{\rm sat}}\right)^{\gamma}\,.
    \label{eq:asy_ESNM}
\end{equation}
Here, the Fermi energy $E_F$ was set to \qty{37}{MeV} and the parameters $\alpha,~\beta,$ and $\gamma$ were fitted to the  binding energy $E_{\rm sat}=-\qty{16}{MeV}$ of symmetric nuclear matter, to the condition of vanishing pressure in symmetric matter at saturation density, and to a (varying) value for $K_{\rm sat}$.
The term $e_{\rm sym}(n)$ in Eq.~\eqref{eq:asy_ESNM} is the nuclear symmetry energy and parametrized as
\begin{equation}
    e_{\rm sym}(n) = E_\mathrm{kin,0}\left(\dfrac{n}{n_{\rm sat}}\right)^{2/3} + E_\mathrm{pot,0}\left(\dfrac{n}{n_{\rm sat}}\right)^{\gamma_{\rm asy}}\,,
    \label{eq:Esym_QMD}
\end{equation}
in the ASY-EOS analysis.
Specifically, $E_\mathrm{kin,0}$ was set to \qty{12}{MeV} and $E_\mathrm{pot,0}=E_{\rm sym}-E_\mathrm{kin,0}$.
Here, $E_{\rm sym}$ is the symmetry energy at saturation density and was also taken as a variable parameter.
Then, the parameter $\gamma_{\rm asy}$ was extracted from fits to data using transport model simulations~\cite{Russotto_2016}, leading to $\gamma_{\rm asy}=0.68\pm 0.19$ when assuming $E_{\rm sym}=\qty{31}{MeV}$ and $\gamma_{\rm asy}=0.72\pm 0.19$ for $E_{\rm sym}=\qty{34}{MeV}$.
To obtain a result for arbitrary $E_{\rm sym}$, $\gamma_{\rm asy}$ was interpolated linearly between these two points, keeping the uncertainty fixed at $0.19$.
With the model matched to the HIC data in this way, one can compute the pressure in beta equilibrium, assuming the electrons form an ultrarelativistic degenerate Fermi gas.
We here use the immediate results of Ref.~\citep{Huth_2022}, where $E_{\rm sym}$ was varied uniformly between \qtyrange{31}{34}{MeV} and $K_{\rm sat}$ is drawn from a Gaussian distribution with mean \qty{200}{MeV} and standard deviation \qty{25}{MeV}.

Accordingly, for each density, the results yield a probability distribution on the pressure at that particular $n$, which we denote by $P(p, n|\text{HIC})$.
The constraint is then applied at those densities for which the experiment is sensitive, which can be determined by the sensitivity of the flow ratio for neutrons over charged particles of the ASY-EOS experiment~\cite{Russotto_2016}.
The likelihood of an EOS with respect to the information from heavy-ion collisions is thus written as
\begin{align}
    \mathcal{L}(\text{EOS}|\text{HIC}) = \int dn\ P(p(n, \text{EOS}), n|\text{HIC})\ C(n)\,,
    \label{eq:HIC_likelihood}
\end{align}
where $C(n)$ denotes the sensitivity curve.

The impact of the HIC data on our candidate EOS set is shown in Fig.~\ref{fig:HIC}.
We find that the data prefers the EOS to be stiff in the region between \qtyrange{1}{2}{\nsat}, requiring high pressures but shifting the canonical NS radius to slightly smaller values compared to the prior.
Because the sensitivity of the experiment declines quickly beyond \qty{2}{\nsat}, we find no impact on the TOV mass or $p_{3\nsat}$, even given our relatively broad prior EOS set.

The complicated extraction of constraints from HIC collision data suffers from several shortcomings, among which is the simplicity of the energy-density functional used to extract $\gamma_{\rm asy}$.
Future theoretical work is needed to improve the extraction of constraints from HIC data. 
Moreover, some bias might arise from the likelihood function that we use to apply the HIC constraint to our EOS candidates. 
For instance, Ref.~\citep{Tsang_2023} uses several Gaussian likelihood functions for symmetry energy parameters, whereas we apply the symmetry energy constraint over the whole density range and include the sensitivity of the experiment.

\section{Neutron-star measurements through radio and X-ray observations}
\label{sec:Astro}

For a given EOS, the Tolman-Oppenheimer-Volkoff equations~\citep{Oppenheimer_1939} uniquely determine the relationship between mass and radius of a nonrotating NS. 
In return, mass, and especially mass-radius measurements of observed NSs can be used to test the EOS and the properties of neutron-rich matter at several times $\nsat$. 
In this section, we first discuss the impact of mass-only measurements of heavy NSs~\citep{Demorest_2010, Antoniadis_2013, Cromartie_2020, Romani_2022}, which require the true EOS to have a TOV mass of $\gtrsim \qty{2}{\msun}$.
Afterward, we focus on simultaneous measurements of NS radii and masses~\citep{Riley_2019, Miller_2019, Riley_2021, Miller_2021, Salmi_2024, Dittmann_2024, Steiner_2018, Doroshenko_2022, Naettilae_2017, Goodwin_2019}.
Such measurements employ various techniques, but they all rely on modeling the system's X-ray emission in one way or another. 
We briefly characterize the methods and models for these observations and comment on remaining systematic uncertainties.
Generally, we find that all these measurements are consistent with a canonical radius of $R_{1.4}$ between \qtyrange{11}{13}{km}.

\subsection{Heavy pulsars and radio timing measurements}
\label{subsec:Shapiro}
\begin{figure}[ht!]
    \centering
    \includegraphics[width = \linewidth]{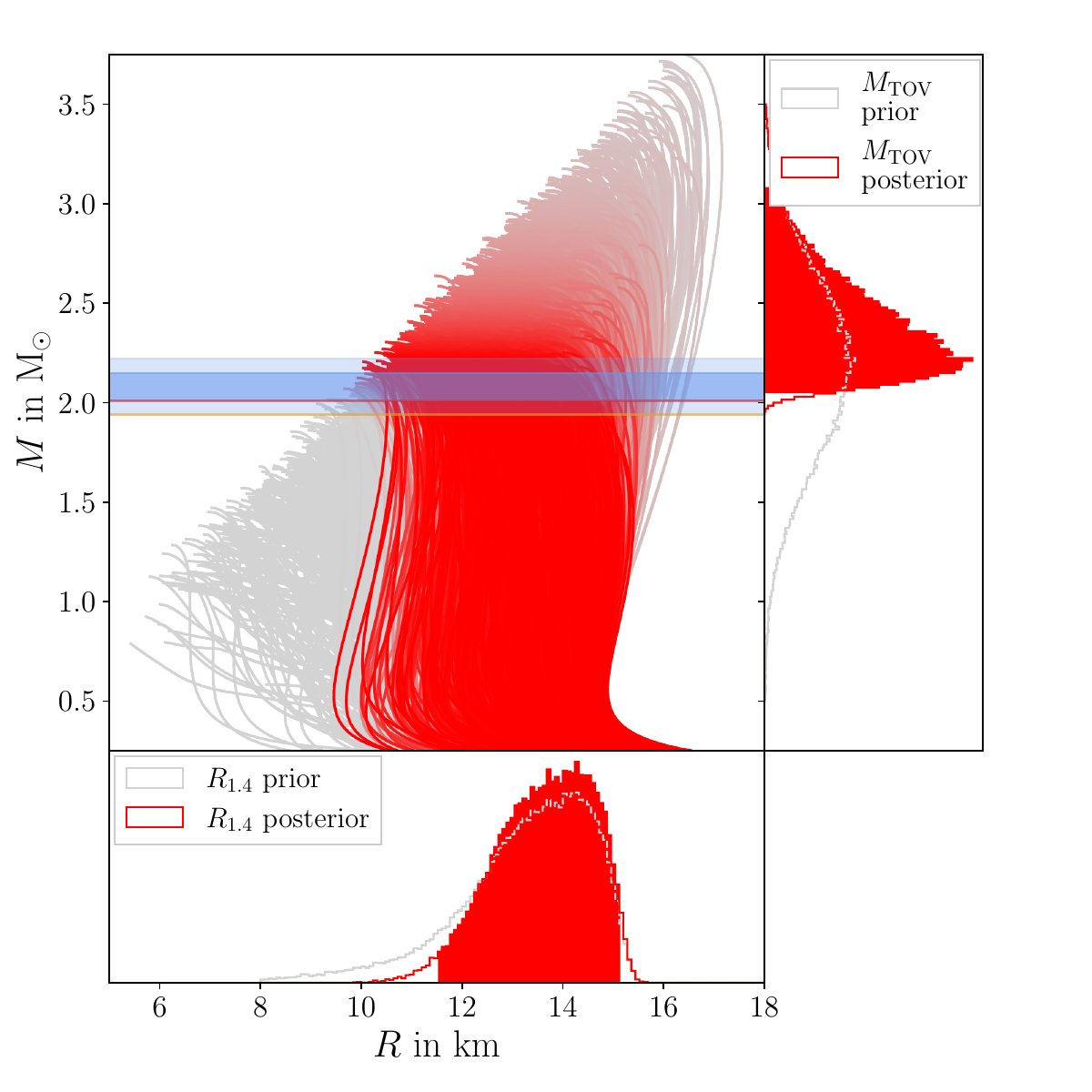}
    \includegraphics[width = \linewidth]{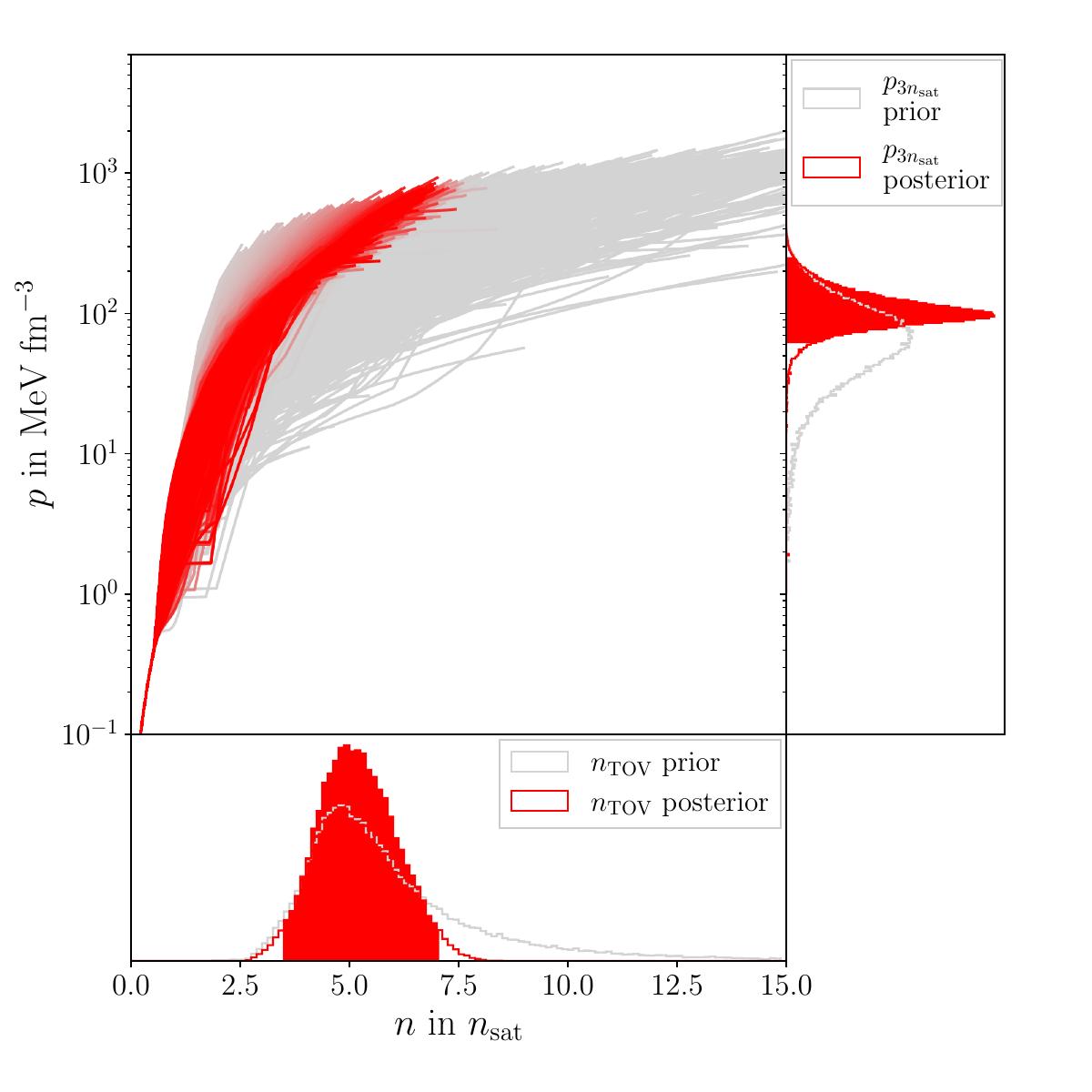}
    \caption{EOS inference based on the radio timing mass measurement of three radio pulsars. Arrangement and color coding as in Fig.~\ref{fig:chiEFT}. The blue band marks the 68\% and 95\% credible interval for the mass measurement of \mbox{PSR J0740+6620}.
    The colored lines indicate median mass values for the measurement of \mbox{PSR J0348+0432} (red) and \mbox{PSR J1614-2230} (orange), coincidentally very close to the lower ends of the credible intervals from \mbox{PSR J0740+6620}.}
    \label{fig:Shapiro}
\end{figure}

For highly inclined (i.e.,~edge-on) binary pulsar systems, the pulsar's radio signal has to pass through the companion's gravitational field to reach Earth.
Thus, it is affected by Shapiro time delay and other relativistic effects~\citep{Demorest_2010}.
Through precise pulsar timing measurements that track the signal along the orbit, this effect can be used to study the companion mass, which allows for the determination of the NS mass via the binary orbital period and radial velocity.
This method was applied first in Ref.~\citep{Demorest_2010} for \mbox{PSR J1614-2230}, finding a mass of $1.97^{+0.08}_{-0.08}\,\unit{\msun}$. We emphasize that the uncertainty quoted originally in Ref.~\citep{Demorest_2010} is at 68\% credibility, while here
we quote the \mbox{95\%} credibility intervals. 
Later, a refined analysis reported the value $1.94^{+0.06}_{-0.06}\,\unit{\msun}$~\citep{Shamohammadi_2023}, which we will adopt here, but see also $1.922^{+0.030}_{-0.030}\,\unit{\msun}$ of Ref.~\citep{Agazie_2023} that was reported during the development of the present work. 
Furthermore, we consider two additional heavy NS masses. The mass of the pulsar \mbox{PSR J0348+0432} was determined to be $2.01^{+0.08}_{-0.08}\,\unit{\msun}$~\citep{Antoniadis_2013} based on measuring its radial velocity from radio timing observations, together with spectral modeling and radial velocity measurements for its white dwarf companion.
The mass of \mbox{PSR J0740+6620} was reported at 2.08$^{+0.14}_{-0.14}\,\unit{\msun}$~\citep{Cromartie_2020, Fonseca_2021} based on the Shapiro time delay technique.
All of the quoted studies use radio timing observations of the pulsar signals and employ the \textsc{TEMPO2} package~\citep{Hobbs_2006} for an analytical model of the pulse time series. 
Binary orbital parameters were either fitted directly from the time-of-arrival model or in the case of Ref.~\citep{Antoniadis_2013} jointly inferred with spectroscopic observations of the optical counterpart. 
These three NSs provide the most relevant constraints on the TOV mass, although the masses of a few dozens of less massive pulsars have been determined by radio timing techniques too; see, e.g., Refs.~\cite{Fonseca_2016, Arzoumanian_2018, Shamohammadi_2023}. 

Any of these observations provides a posterior $p(M|\text{Obs.})$ on the NS mass $M$. 
If an EOS predicts a TOV mass below the observed NS mass, it should be ruled out. 
We express the likelihood of an EOS given such a mass measurement as
\begin{align}
    &\mathcal{L}(\text{EOS}|\text{Obs.}) = \frac{1}{M_{\text{TOV}}}\ \int_0^{M_{\text{TOV}}} dM\ P(M | \text{Obs.})\,,
    \label{eq:l1}
\end{align}
where $M_{\text{TOV}}$ denotes the EOS-specific TOV mass and we assume that the prior of $M$ for a given EOS is flat on the interval $\left[0, M_{\text{TOV}}\right]$.
The mass posteriors for \mbox{PSR J1614-2230}, \mbox{PSR J0348+0432}, and \mbox{PSR J0740+6620} are described well by a normal distribution; hence we adopt the quoted values above for the mean and standard deviation and calculate the EOS likelihood according to Eq.~(\ref{eq:l1}). 
We assume all observations to represent independent events, so we can combine these inferences through multiplication, leading to the result shown in Fig.~\ref{fig:Shapiro}. 

Clearly, the existence of high-mass pulsars requires generally stiff EOSs with higher pressure and lower TOV density. 
Effectively, this places an upper limit of \qty{8}{\nsat} on $n_{\text{TOV}}$. 
At the same time, EOS candidates with very high TOV masses become also slightly disfavored due to the prefactor $1/\mtov$ in Eq.~(\ref{eq:l1}). 
This factor appears in the derivation of the likelihood and expresses the fact that a very high TOV mass becomes less plausible if the highest observed NS mass is repeatedly around \qty{2}{\msun}. However, its impact is small, and hence other studies occasionally omit it \citep{Miller_2020, NMMA_2023}. Since we keep it, we obtain a slightly narrower upper limit of $\mtov < \qty{2.93}{\msun}$ at 95\% credibility, instead of $\mtov < \qty{3.10}{\msun}$ had we not used this factor.
Naturally, a cutoff in the observed neutron-star population at \qty{2}{\msun} could also arise from selection effects and formation channels of millisecond pulsar binaries, although population studies including various binary types find a similar cutoff at $\sim\qty{2}{\msun}$~\citep{Alsing_2018, Shao_2020}.

Pulsar mass measurements through Shapiro delay techniques rely only on the validity of general relativity and require no further model assumptions. 
Additionally, pulse arrival times can be measured with high precision, leading to small statistical uncertainties that are further suppressed by accumulating more data points over repeated observations.
The mass measurement for \mbox{PSR J0348+0432} additionally depends on spectroscopic models for the white dwarf companion.
However, these are known to match observations reliably for low-mass white dwarfs. 
Overall, systematic biases stem predominantly from modeling dispersive effects in the interstellar medium, as well as from noise in the timing detectors and fitting techniques~\citep{Antoniadis_2013, Fonseca_2021}. 
In Ref.~\citep{Miller_2021} the systematic uncertainties for the mass measurement of \mbox{PSR J0740+6620} are quoted at \qty{0.02}{\msun}. 
It has been noted~\citep{Demorest_2010, Cromartie_2020} that the inferred companion masses in the binaries of \mbox{PSR J1614-2230} and \mbox{PSR J0740+6620} are slightly off compared to a well-established connection with the binary orbital period~\citep{Rappaport_1995}, though this might also arise from an atypical evolution history.

\subsection{Black widow pulsar PSR J0952-0607}
\label{subsec:Black_Widow}
\begin{figure}[ht!]
    \centering
    \includegraphics[width = \linewidth]{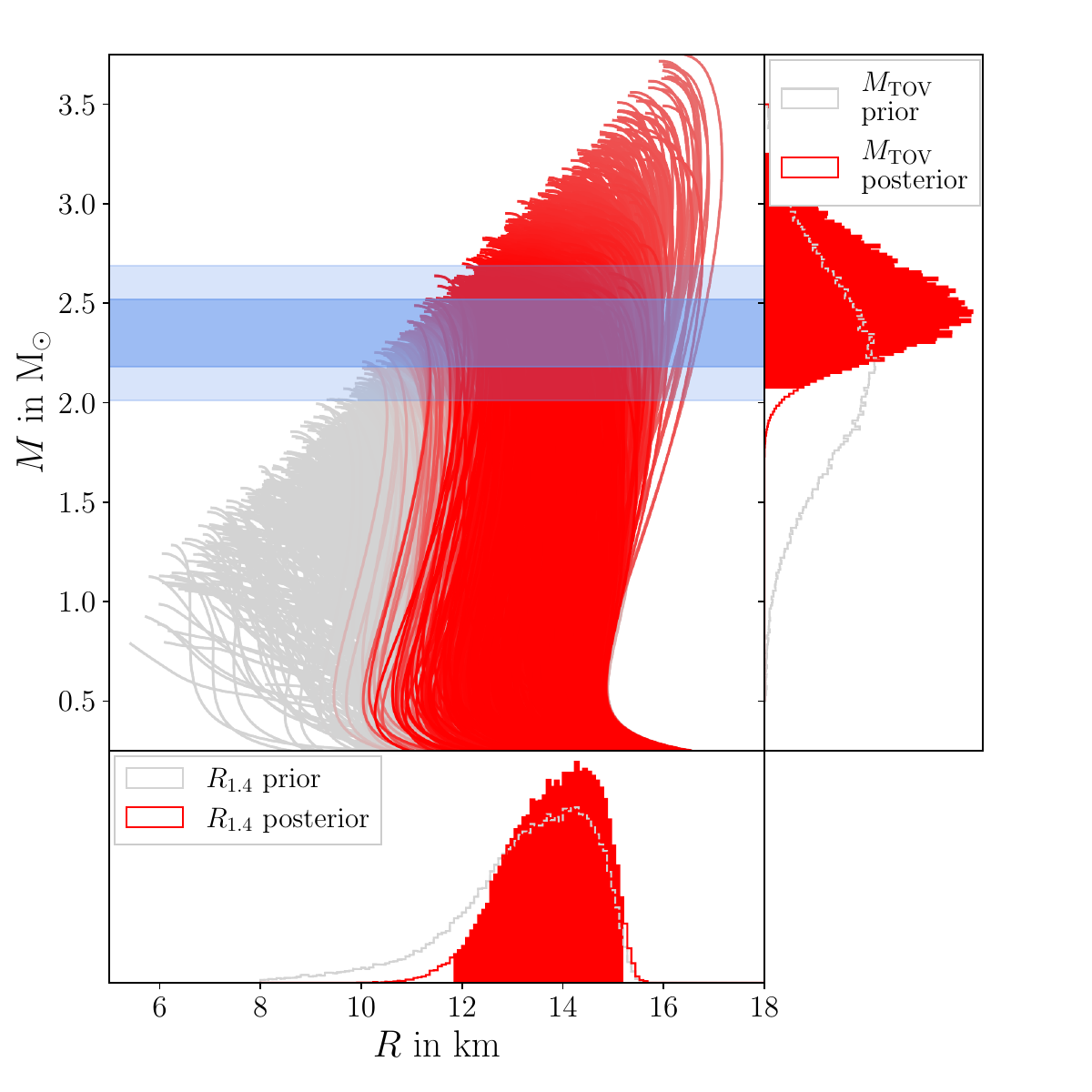}
    \includegraphics[width = \linewidth]{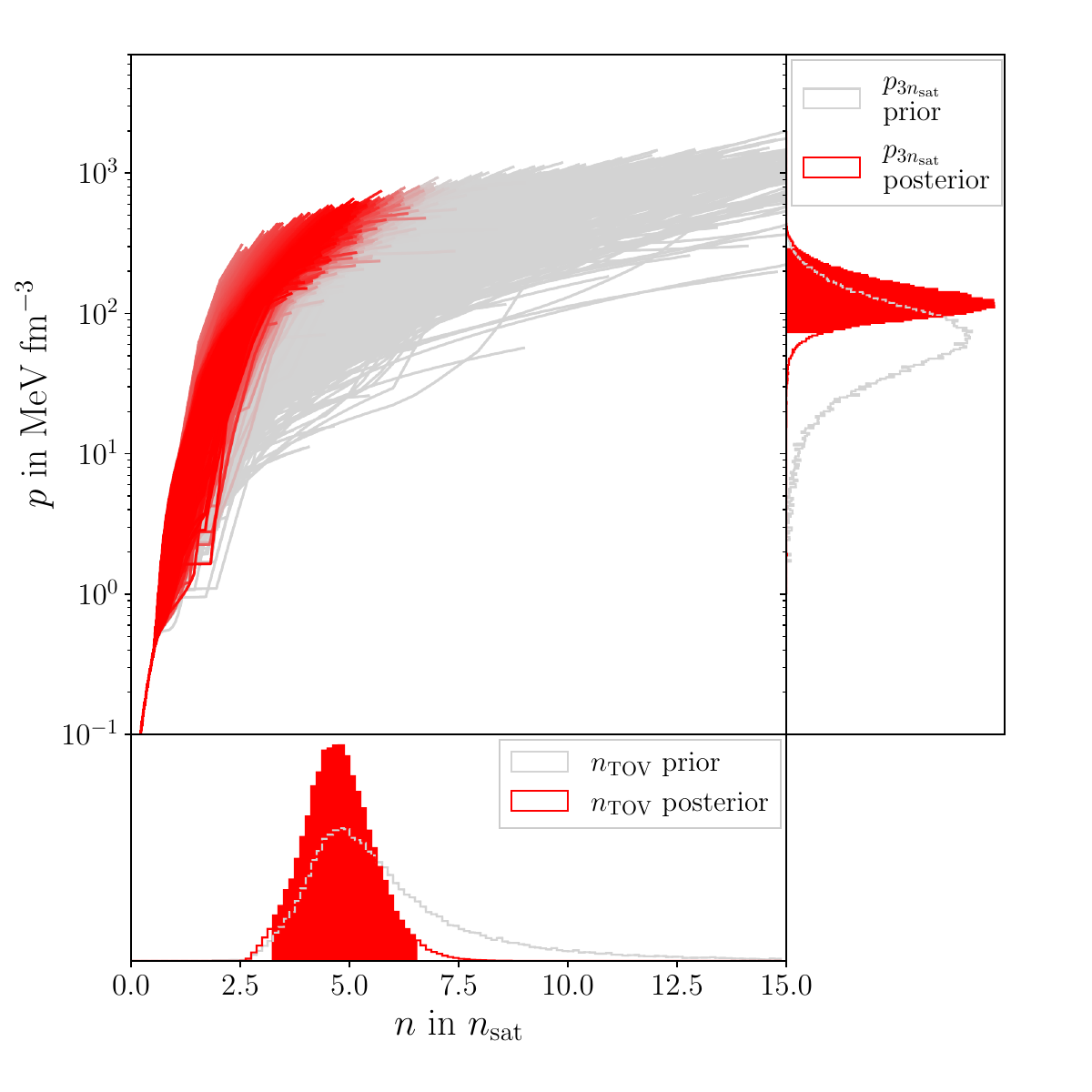}
    \caption{EOS inference based on the mass measurement of PSR J0952-0607. Arrangement and color coding as in Fig. \ref{fig:chiEFT}. The blue bands mark the 68\% and 95\% credible intervals from the mass measurement.}
    \label{fig:BlackW}
\end{figure}

Black widow pulsars constitute a subclass of binary pulsars in which the companion is a low-mass star or brown dwarf whose outer atmosphere is evaporated by the pulsar emission~\citep{Paradijs_1988}. 
Thus, the companion light curve is strongly affected by heating and tidal deformation from the pulsar.
Modeling this emission allows one to infer the binary inclination.
Given additional measurements of other orbital parameters, such as orbital period and radial velocity, the masses of the NS and its companion can be determined. 
Several black widow masses have been assessed through this method, see, e.g., Refs.~\cite{Romani_2021, Romani_2012, Nieder_2020}.
One remarkable example is \mbox{PSR J1810+1744}, as its mass is very well constrained at $2.13^{+0.08}_{-0.08}\,\unit{\msun}$~\cite{Romani_2021}.
Moreover, in Ref.~\citep{Romani_2022}, the authors used the Keck-I 10-m telescope with its Low Resolution Imaging Spectrometer to obtain optical multicolor light curves and spectral radial velocities from the companion of the black widow pulsar \mbox{PSR J0952-0607}. 
They reported an NS mass of $2.35^{+0.34}_{-0.34}\,\unit{\msun}$ at 95\% credibility. 
The analysis relied on a version of the \textsc{icarus} binary light-curve model~\citep{Breton_2012, Kandel_2020} that calculates the thermal emission from surface elements on the companion. 
The model parameters were inferred through Bayesian parameter estimation with MCMC sampling. 
Even though higher masses for some black widow pulsars have been reported, e.g., in Ref.~\citep{Romani_2015}, we here include \mbox{PSR J0952-0607} as the only example of a black widow pulsar in our EOS inference.
This is because the analysis of \mbox{PSR J0952-0607} yields only small fit residuals, indicating it could be a particularly reliable instance of a high-mass black widow pulsar.

To calculate the likelihood of an EOS for this measurement, we integrate \mbox{PSR J0952-0607}'s mass posterior up to the maximum possible mass predicted by that EOS. 
This is similar to Eq.~(\ref{eq:l1}), but because of the pulsar's high spin frequency of \qty{707}{Hz} \citep{Romani_2022, Nieder_2019}, the mass limit in the integral is not given by the TOV mass but instead by $M_{\text{rot}}(\qty{707}{Hz})$, i.e., the maximum possible mass a neutron star with a spin frequency of \qty{707}{Hz} could have. 
When calculating $\mathcal{L}(\text{EOS}|\text{PSR J0952-0607})$, we thus use $M_{\text{rot}}(\qty{707}{Hz})$ in Eq.~\eqref{eq:l1}.

For the 10\,000 EOSs candidates with a TOV mass below \qty{1.5}{\msun}, we set $M_{\text{rot}}(\qty{707}{Hz})$ to 1.1 times their TOV mass, but these candidates anyways have negligible support.
To determine $M_{\text{rot}}(\qty{707}{Hz})$ for the remaining EOS candidates, we use the \textsc{rns} code \citep{Stergioulas_1995, Nozawa_1998} and find that, on average, $M_{\text{rot}}(\qty{707}{Hz})$ is about 3.8\% higher than the TOV mass.
However, for a few EOSs with small compactness $M_{\text{rot}}(\qty{707}{Hz})$ can increase by as much as 15\%. 
Through a least-squares fit, we find the relationship
\begin{align}
\begin{split}
  M_{\text{rot}}(\qty{707}{Hz})=  (1&+ 0.0487\ x - 0.0203\ x^2 \\
  &+ 0.0026\ x^3)\ \mtov,
  \label{eq:Mrot_relation}
\end{split}\\
  x = &\frac{ c^2 R_{\text{TOV}}}{G \mtov} \,, \nonumber
\end{align}
inspired by similar relations in the literature \citep{Konstantinou_2022, Shao_2020a}.
For 139 EOS candidates, numerical issues--likely caused by very stiff segments--prevent the calculation of a mass value with \textsc{rns}. 
In these cases, we set $M_{\text{rot}}(\qty{707}{Hz})$ according to Eq.~\eqref{eq:Mrot_relation}.

The result of the inference is shown in Fig.~\ref{fig:BlackW} and appears similar to the previous inference in Sec.~\ref{subsec:Shapiro} as both require stiff EOS with a TOV mass above \qty{2}{\msun}. 
As before, EOSs with $n_{\text{TOV}} > \qty{8}{\nsat}$ are effectively ruled out. 
For the TOV mass we find $2.52^{+0.73}_{-0.45}$\,$\msun$. Ignoring \mbox{PSR J1810+1744's} rotation and simply using the candidate TOV mass in the likelihood Eq.~\eqref{eq:l1} would have increased the posterior estimate to $2.58^{+0.71}_{-0.44}$\,$\msun$.

As discussed before, the mass value of Ref.~\citep{Romani_2022} is susceptible to systematic biases. 
These may arise mainly from uncertainty about the heat transport across the companion's surface and local temperature peaks. 
This not only affects the estimate for the inclination, but is also needed to accurately link the spectral line widths to the radial velocity~\citep{Kandel_2020}. 
In Ref.~\cite{Romani_2021}, for instance, the authors would have estimated an unreasonably high mass for \mbox{PSR J1810+1744} when ignoring wind heat advection, hot spots, and gravitational darkening. 
For \mbox{PSR J0952-0607} though, a simple direct heating model ignoring all of these effects provided the best fit~\citep{Romani_2022}, and the results were robust when rerunning the analysis with a model incorporating the aforementioned features. 
Simple direct heating may be a good description for \mbox{PSR J0952-0607}, because of its advantageous properties like low heating, small Roche lobe fill factor and large binary period. 
Hence, the possibility of severe systematic uncertainty introduced through modeling appears lower in this system than elsewhere. 
Yet, systematics may also be introduced by instrument noise.
In particular, the complete dataset for the light-curve observations includes some outliers that, when included in the analysis, yield an NS mass of $2.50^{+0.40}_{-0.40}\,\unit{\msun}$ (Table~2 in Ref.~\citep{Romani_2022}), indicating a potential instrument bias. 

\subsection{X-ray pulse profile measurements by the neutron-star interior composition explorer (NICER)}

\begin{figure}
    \centering
    \includegraphics[width = \linewidth]{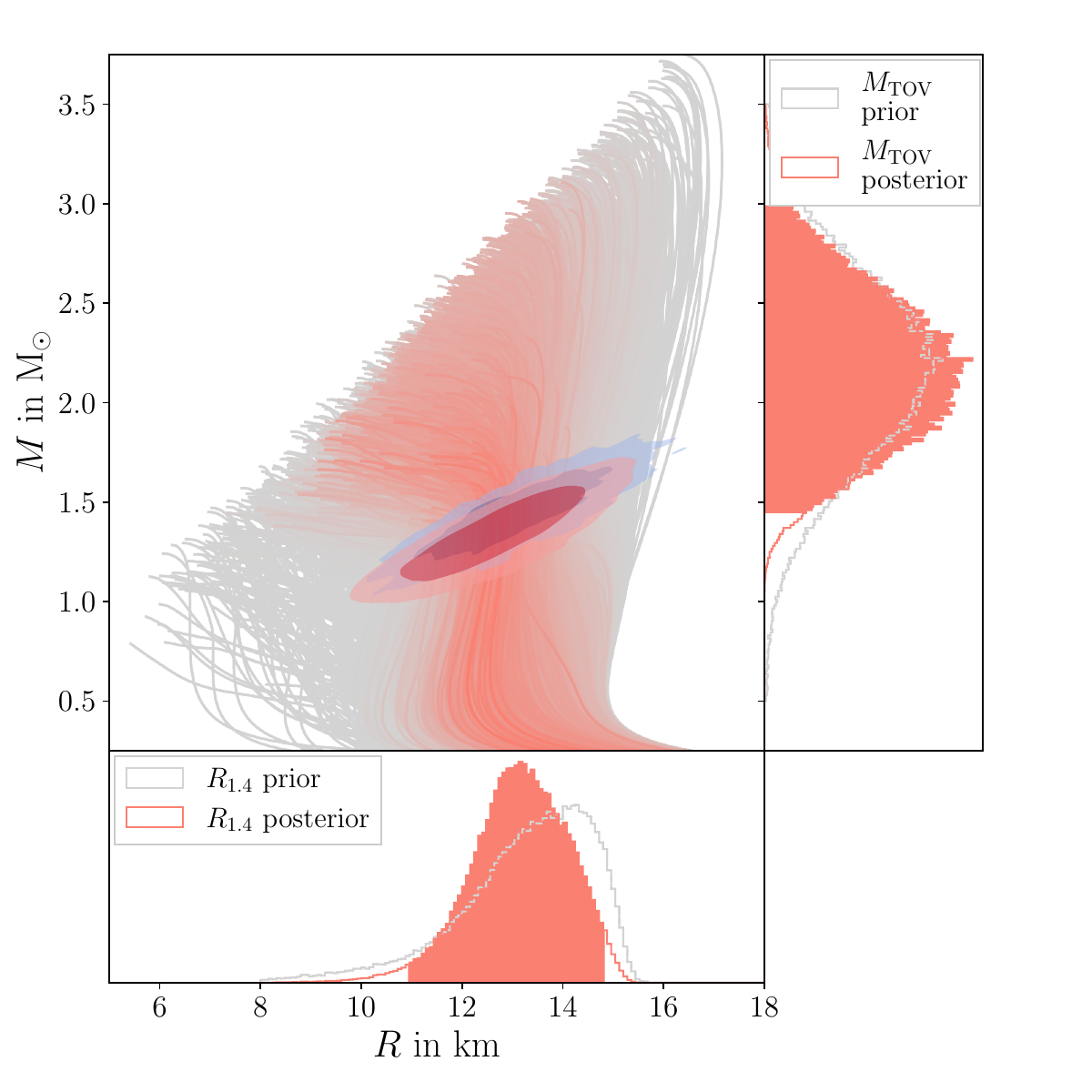}
    \includegraphics[width = \linewidth]{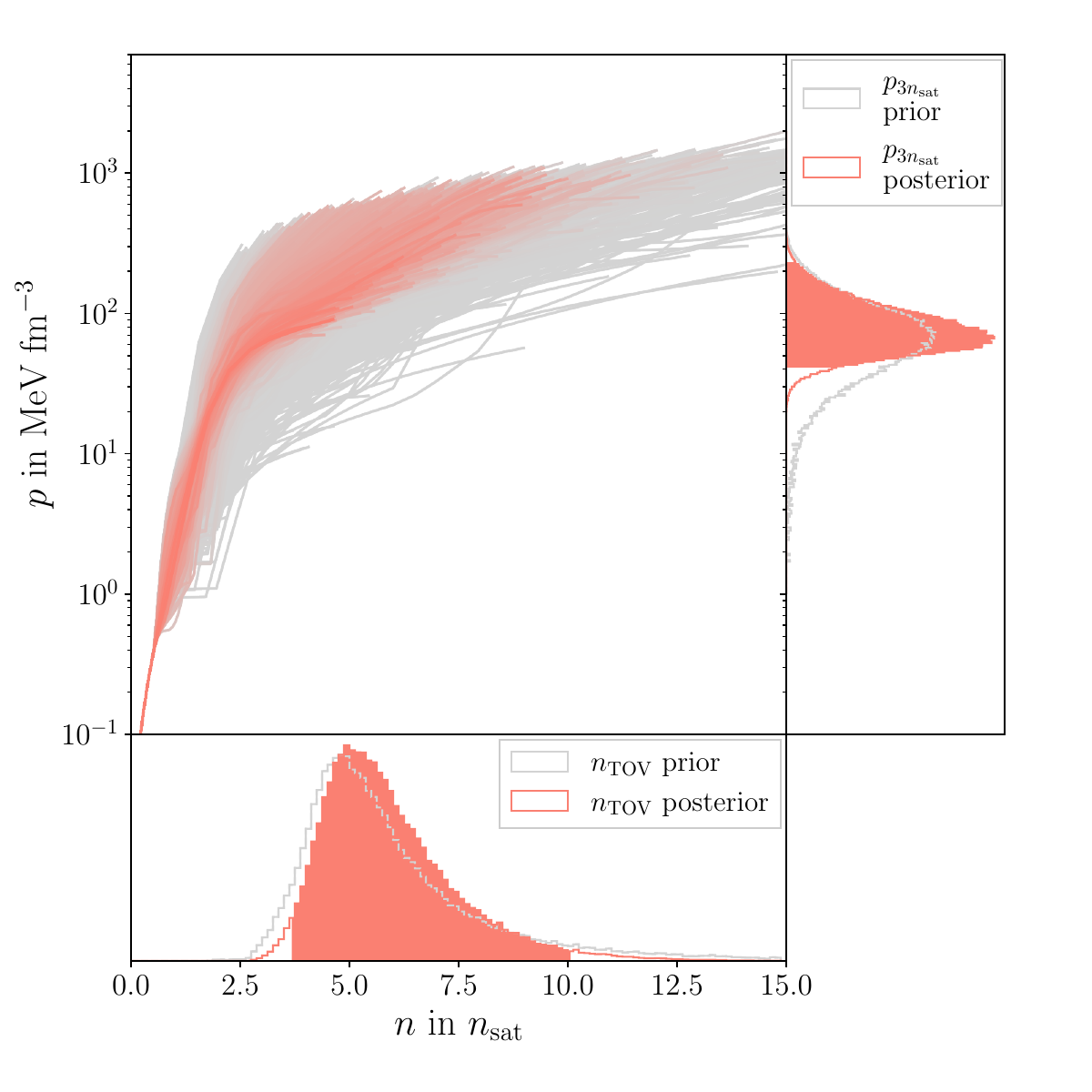}
    \caption{EOS inference based on the NICER measurement of \mbox{PSR J0030+0451}. Arrangement as in Fig.~\ref{fig:chiEFT}. In the middle plot of the top figure, the contours show the 68\% and 95\% credible posterior regions from the $M$-$R$ measurement from Ref.~\cite{Miller_2019} (blue) and Ref.~\cite{Riley_2019} (red).}
    \label{fig:NICER_J0030}
\end{figure}

\begin{figure}
    \centering
    \includegraphics[width = \linewidth]{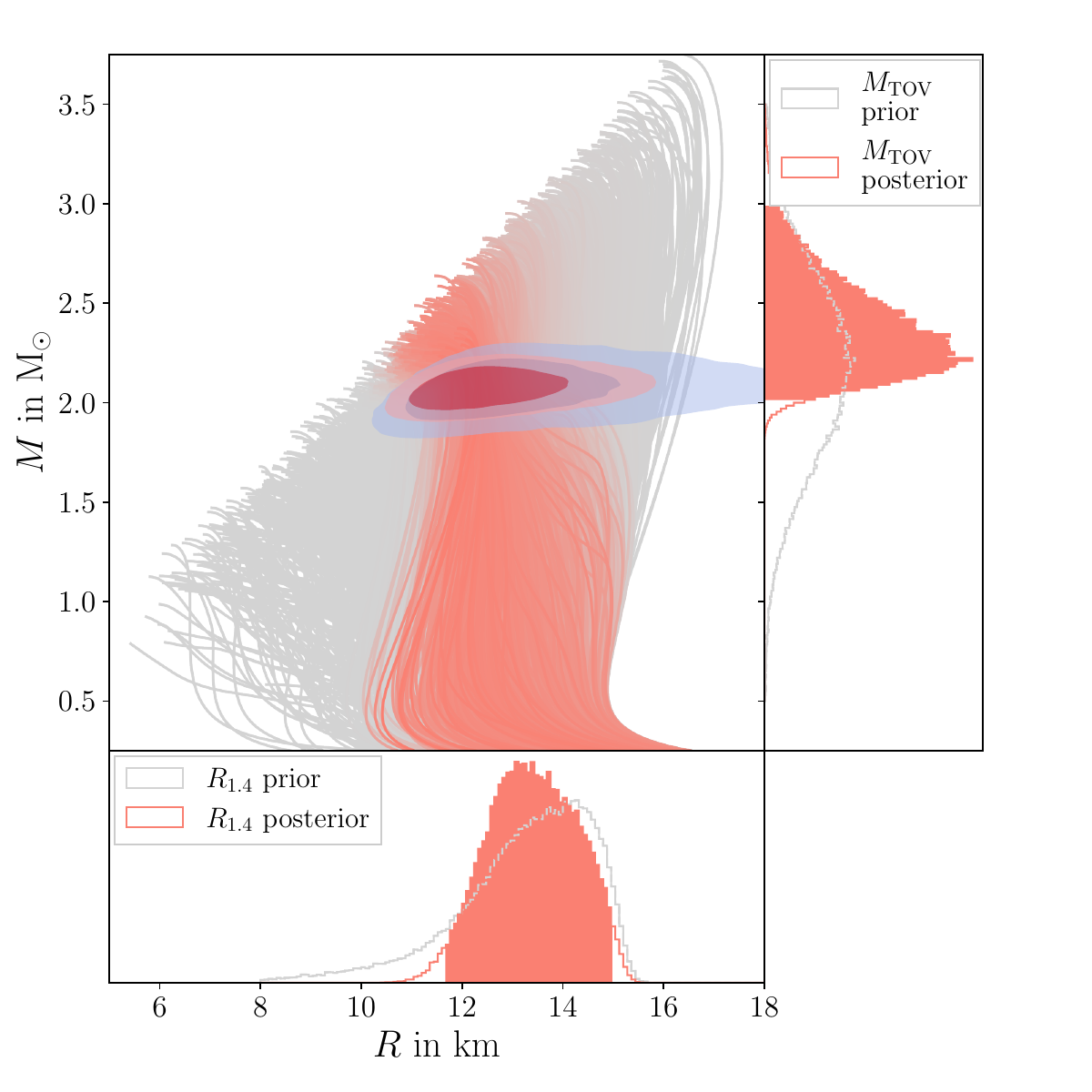}
    \includegraphics[width = \linewidth]{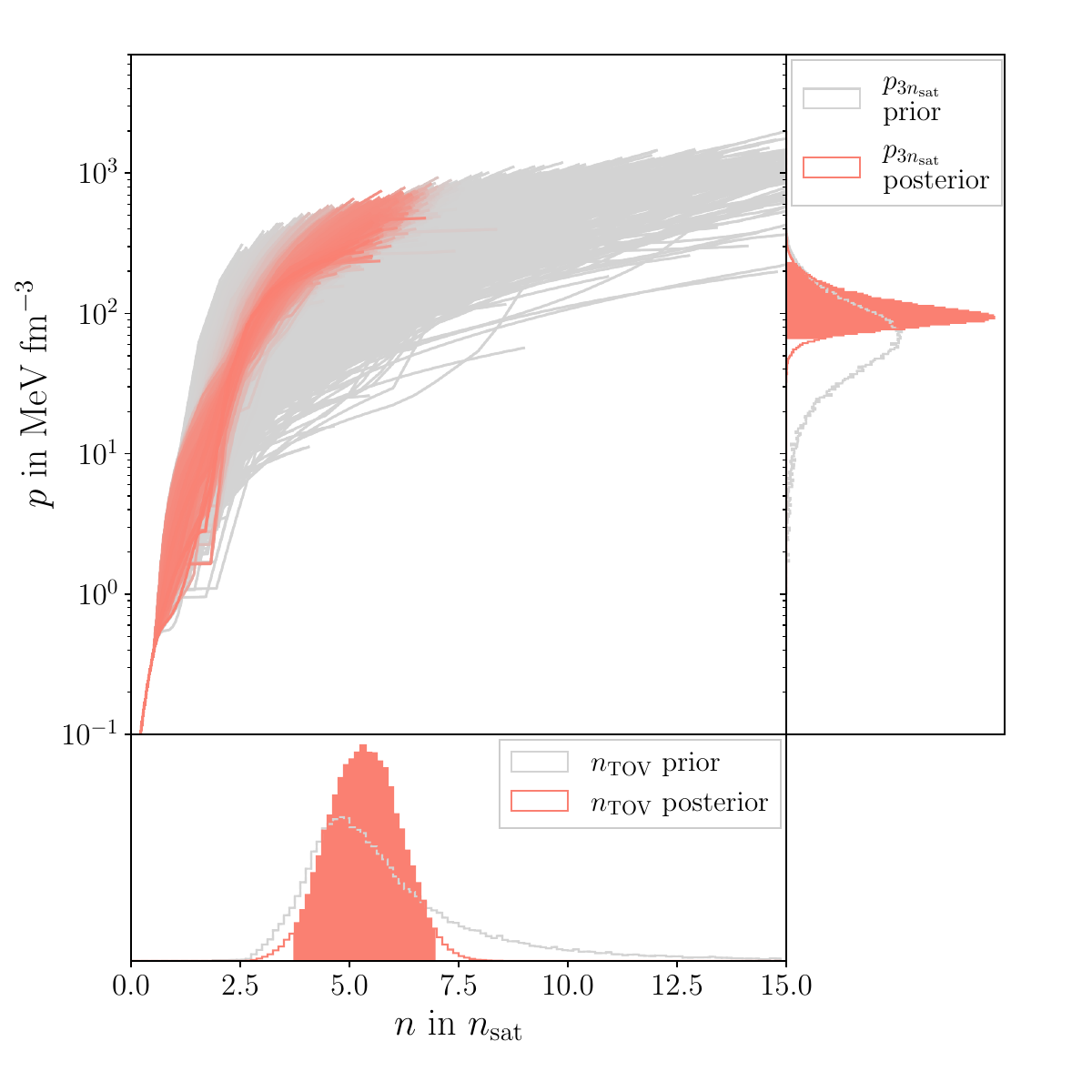}
    \caption{EOS inference based on the NICER measurement of \mbox{PSR J0740+6620}. Arrangement as in Fig.~\ref{fig:chiEFT}. In the middle plot of the top figure, the contours show the 68\% and 95\% credible posterior regions from the $M$-$R$ measurement from Ref.~\cite{Dittmann_2024} (blue) and Ref.~\cite{Salmi_2024} (red).}
    \label{fig:NICER_J0740}
\end{figure}

Shortly after the discovery of pulsars, arguments showed that the temperature distribution on an NS's surface does not need to be uniform~\citep{Sturrock_1971, Ruderman_1975, Tsai_1974, Greenstein_1983}. 
Thermal hot spots on the surface of a pulsar, caused by electron-positron pair cascades heating specific parts~\citep{Harding_2001, Harding_2002, Medin_2010, Potekhin_2014}, lead to repeated fluctuations in the star's X-ray emission as it rotates around its spin axis. 
This effect depends on the chemical composition of the atmosphere and the nature of the hot-spot heating, but also crucially on the compactness of the NS, so its mass $M$ and radius $R$ can be determined when the spin period is known. 
Hence, observing the X-ray pulses offers the potential to constrain the dense matter EOS. 

The Neutron Star Interior Composition Explorer (NICER) can resolve pulsar X-ray emission in the \qtyrange{0.2}{12}{keV} band with a time resolution $<\qty{1}{\micro\second}$~\citep{Gendreau_2016, Bogdanov_2019}. 
Thus, it can track X-ray pulses over the rotation phases of millisecond pulsars. 
Several pulsars have been observed with NICER~\citep{Bogdanov_2019}, but so far inferences of masses and radii have only been carried out in two instances, namely for \mbox{PSR J0030+0451} in Refs.~\citep{Riley_2019, Miller_2019}, and the high-mass pulsar \mbox{PSR J0740+6620} in Refs.~\citep{Riley_2021, Salmi_2024, Miller_2021, Dittmann_2024}. 
The analyses of \mbox{PSR J0740+6620} were supplemented with phase-averaged spectra from the XMM-Newton telescope and used the Shapiro time delay measurement of Ref.~\citep{Fonseca_2021} (see also Sec.~\ref{subsec:Shapiro}) as a prior on the mass and distance. 
The groups in Refs.~\citep{Riley_2019, Riley_2021} and Refs.~\citep{Miller_2019, Miller_2021} both employed Bayesian models to directly predict the expected pulse waveform given a specification of the parameters such as mass, radius, distance, or effective hot-spot temperature.
The \textsc{x-psi} code of Refs.~\citep{Riley_2019, Riley_2021} is publicly available~\citep{Riley_2023}; see also Ref.~\citep{Afle_2023} for a reproducibility study.
Differences in the studies of the two groups include the possible hot-spot geometries, the implementation of the instrument response, and sampling techniques.
In particular, the posteriors of Refs.~\citep{Riley_2019, Riley_2021} rely on the nested sampling algorithm, while final samples for Refs.~\citep{Miller_2019, Miller_2021} are obtained through MCMC sampling.
While different sampling methods can impact the final results, different prior and modeling choices are likely the dominant contribution to differences for the $M$-$R$ estimate. 
In the case of $\mbox{PSR J0740+6620}$ separate choices for the relative effective area of XMM-Newton to NICER also contribute, as well as different prior choices for $R$.

Similarly to Eq.~(\ref{eq:l1}), the likelihood of a certain EOS given an $M$-$R$ posterior $P(M, R| \text{NICER})$ from a NICER measurement can be written as
\begin{align}
\begin{split}
    &\mathcal{L}(\text{EOS}|\text{NICER}) =\\
    &\int_0^{M_{\text{TOV}}} dM\ P(M, R(M, \text{EOS})| \text{NICER})\,, 
\label{eq:l2}
\end{split}
\end{align}
where $M_{\text{TOV}}$ is the EOS-specific TOV mass and $R(M, \text{EOS})$ the mass-radius curve given by the TOV equations. 
Different $M$-$R$ posteriors are available depending on which geometrical hot spot configurations are adopted for the inference. 
Here, we use those $M$-$R$ posteriors from the headline results in the respective publications.
For \mbox{PSR J0030+0451}, these are the samples obtained from the model with one circular and one circular partially concealed hot spot (ST+PST) of Ref.~\citep{Riley_2019} and the model with three oval spots of Ref.~\citep{Miller_2019}. 
For \mbox{PSR J0740+6620}, the recommended model has two circular hot spots (ST-U) both in Refs.~\citep{Salmi_2024, Dittmann_2024}.
For \mbox{PSR J0030+0451} we use the publicly available posterior samples in Refs.~\citep{Riley_2019_samp, Miller_2019_samp}, whereas for \mbox{PSR J0740+6620} we use the samples from the recent release of the 3.6 year dataset~\citep{Salmi_2024_samp, Dittmann_2024_samp}.

In Fig.~\ref{fig:NICER_J0030}, we show the inferred $M$-$R$ contours for \mbox{PSR J0030+0451} together with the resulting posterior likelihood of our EOS set.
Figure~\ref{fig:NICER_J0740} shows analogous results for \mbox{PSR J0740+6620}. 
Since for each pulsar two distinct $M$-$R$ posteriors are available, the combined likelihood for our EOS candidates was calculated as the arithmetic average from both analyses. 
The measurement of \mbox{PSR J0030+0451} mainly places constraints on the equatorial radius of medium-sized NSs, but does affect the posterior estimate for the TOV mass only slightly. 
The inference of the heavier \mbox{PSR J0740+6620} naturally yields overall tighter constraints on $\mtov$ and excludes very soft EOSs, though its uncertainty on the canonical radius $R_{1.4}$ is broader than for \mbox{PSR J0030+451}. 

Systematic effects in the NICER analyses originate from assumptions about the instrument response and hot spot geometries. 
Furthermore, the models typically use fully ionized hydrogen atmospheres, so that the current radius estimates could underestimate the true value~\citep{Salmi_2017, Salmi_2023}. 
At the same time, the general congruence between the two independent analyses indicates that systematic effects have only minor impact.
Recently, a new analysis of the \mbox{PSR J0030+0451} NICER data in Ref.~\citep{Vinciguerra_2023} has reported that, when compared with Ref.~\citep{Riley_2019}, improvements to sampling techniques and instrument response modeling, plus inclusion of XMM-Newton data for background cross-calibration, changes the preferred hot-spot configuration to one with two hot spots with dual temperature emission regions each (PDT-U). 
Using this model, the inferred NS gravitational mass and equatorial circumferential radius shift from the originally reported $1.35^{+0.29}_{-0.27}\,\msun$ and $12.71^{+2.15}_{-2.16}$\,km to $1.70^{+0.34}_{-0.38}\,\msun$ and $14.44^{+1.40}_{-2.06}$\,km.
Because these inferences were conducted as test runs and, given computational limitations, could not be run until convergence was demonstrated, Ref.~\citep{Vinciguerra_2023} emphasizes that the values quoted above are not yet robust. 
The new values do overlap the old ones at the 1$\sigma$ level. Nevertheless, these results might hint at biases in the joint inference of XMM-Newton and NICER data. 
Such systematics could arise from difficulties in sampling over multimodal posterior surfaces and problems in background noise estimation~\citep{Vinciguerra_2023, Vinciguerra_2023A}. 
For example, it could be that, unlike what was assumed in the joint NICER-XMM analysis of Ref.~\citep{Vinciguerra_2023}, the XMM data contain a higher number of background sources than expected from blank sky observations. 

\subsection{Analysis of quiescent thermal X-ray spectra}
\label{sec:Thermal}

\begin{figure}[ht!]
    \centering
    \includegraphics[width = \linewidth]{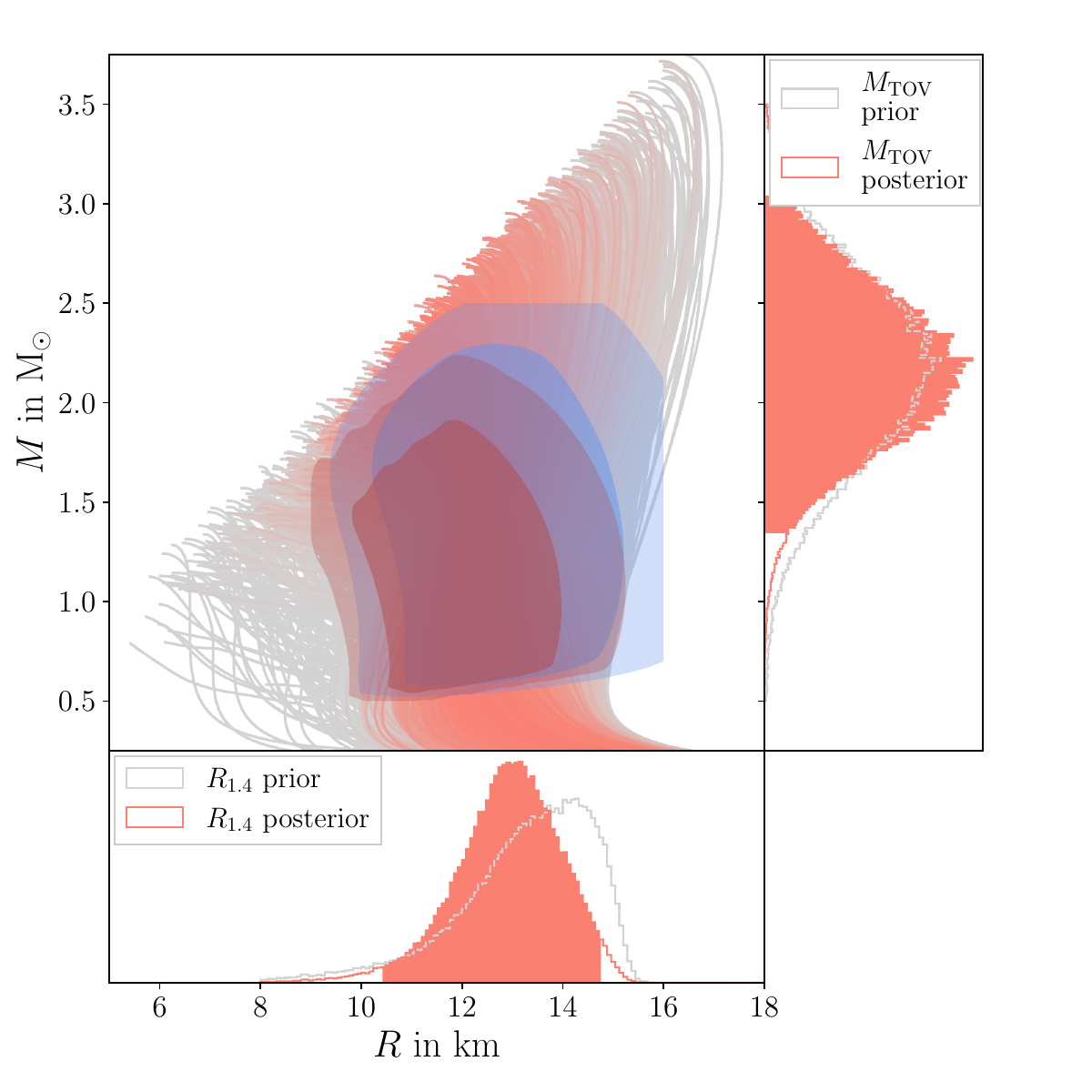}
    \includegraphics[width = \linewidth]{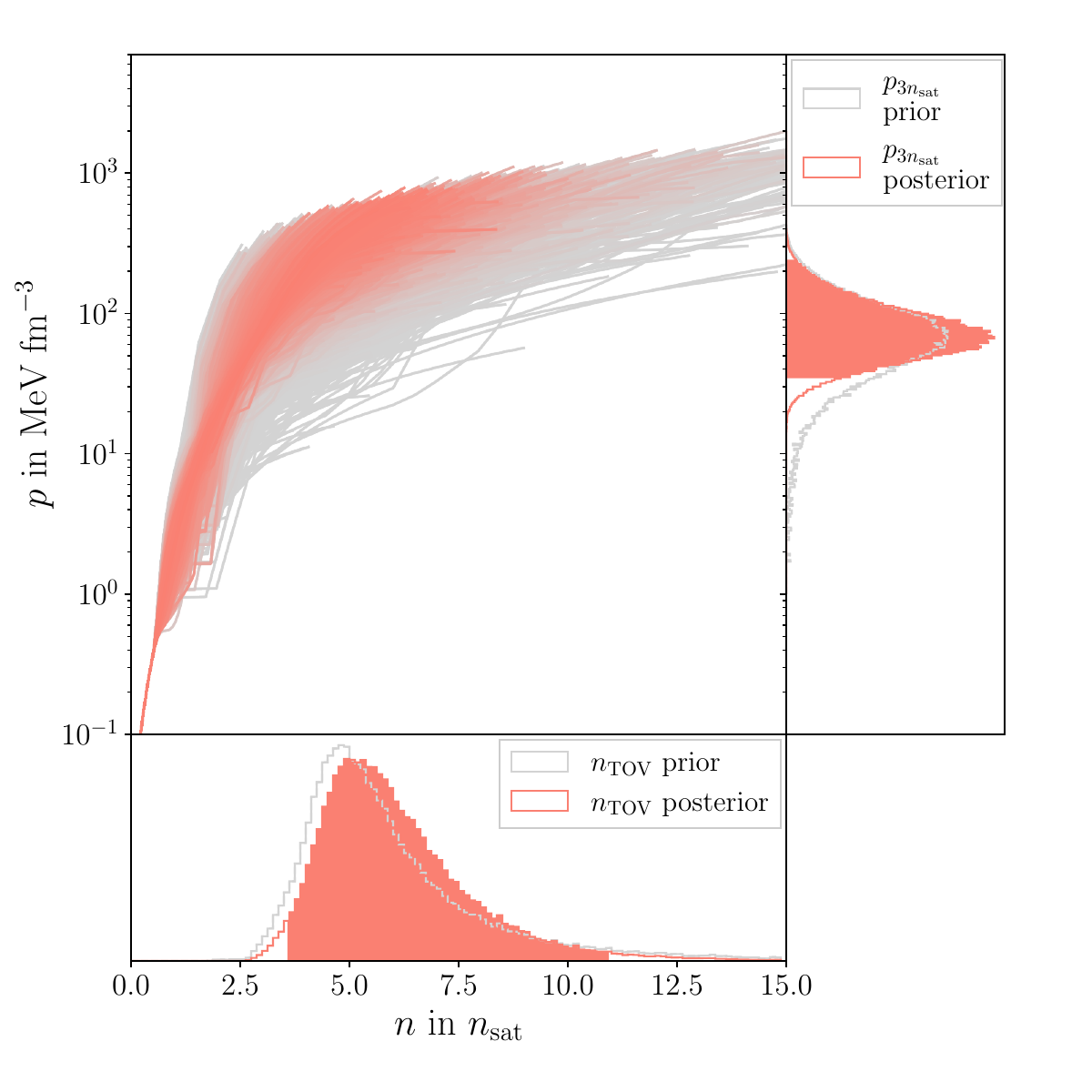}
    \caption{EOS inference based on mass-radius measurements of two NSs in qLMXBs in the globular clusters 47 Tucanae and $\omega$ Centauri. 
    Arrangement and color coding as in Fig.~\ref{fig:chiEFT}. 
    In the middle plot of the top panel, the contours show the 68\% and 95\% credible posterior regions for the $M$-$R$ measurements of the qLMXB in $\omega$ Centauri (blue) and of the qLMXB X5 in 47 Tucanae (red).}
    \label{fig:qLMXBs}
\end{figure}

\begin{figure}[ht!]
    \centering
    \includegraphics[width = \linewidth]{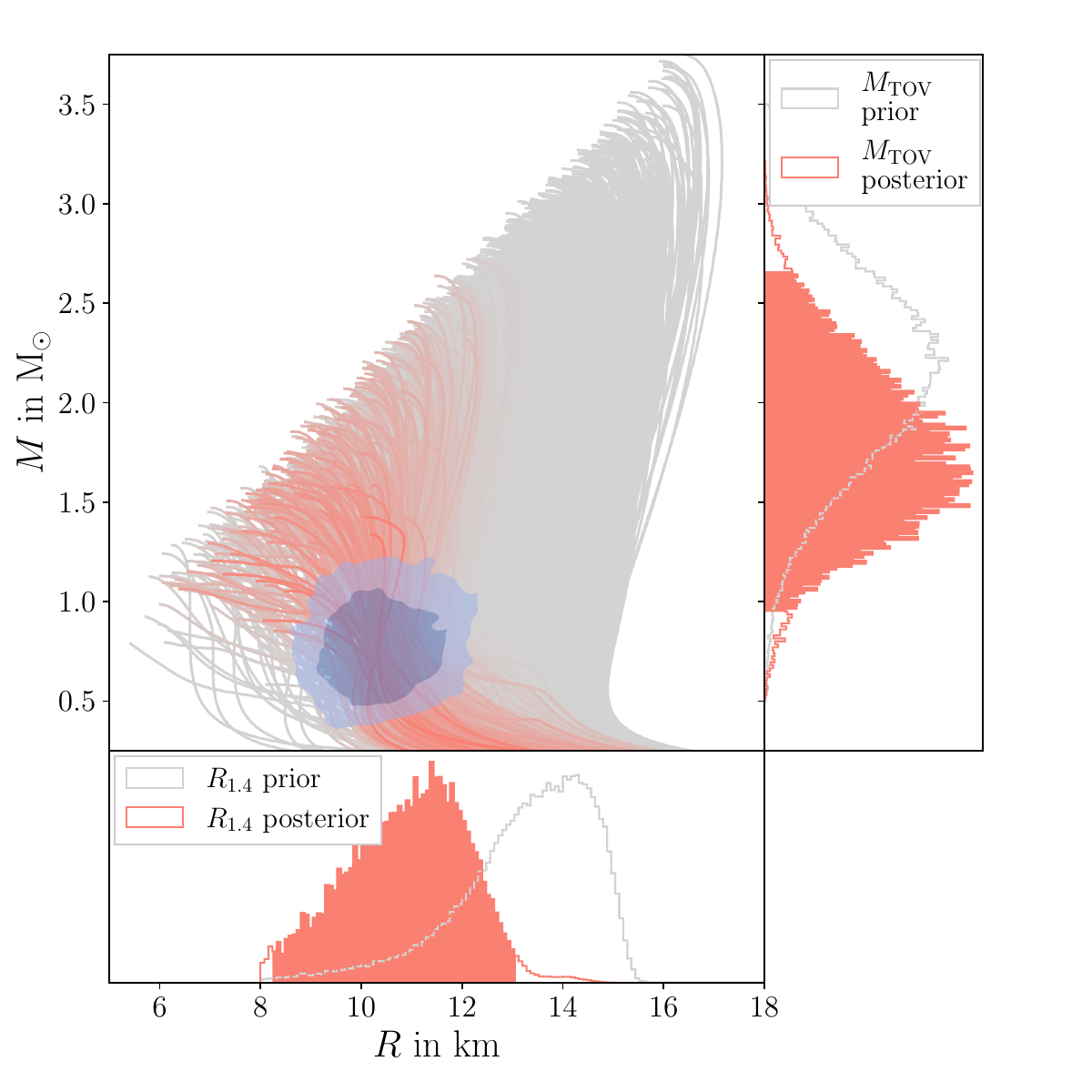}
    \includegraphics[width = \linewidth]{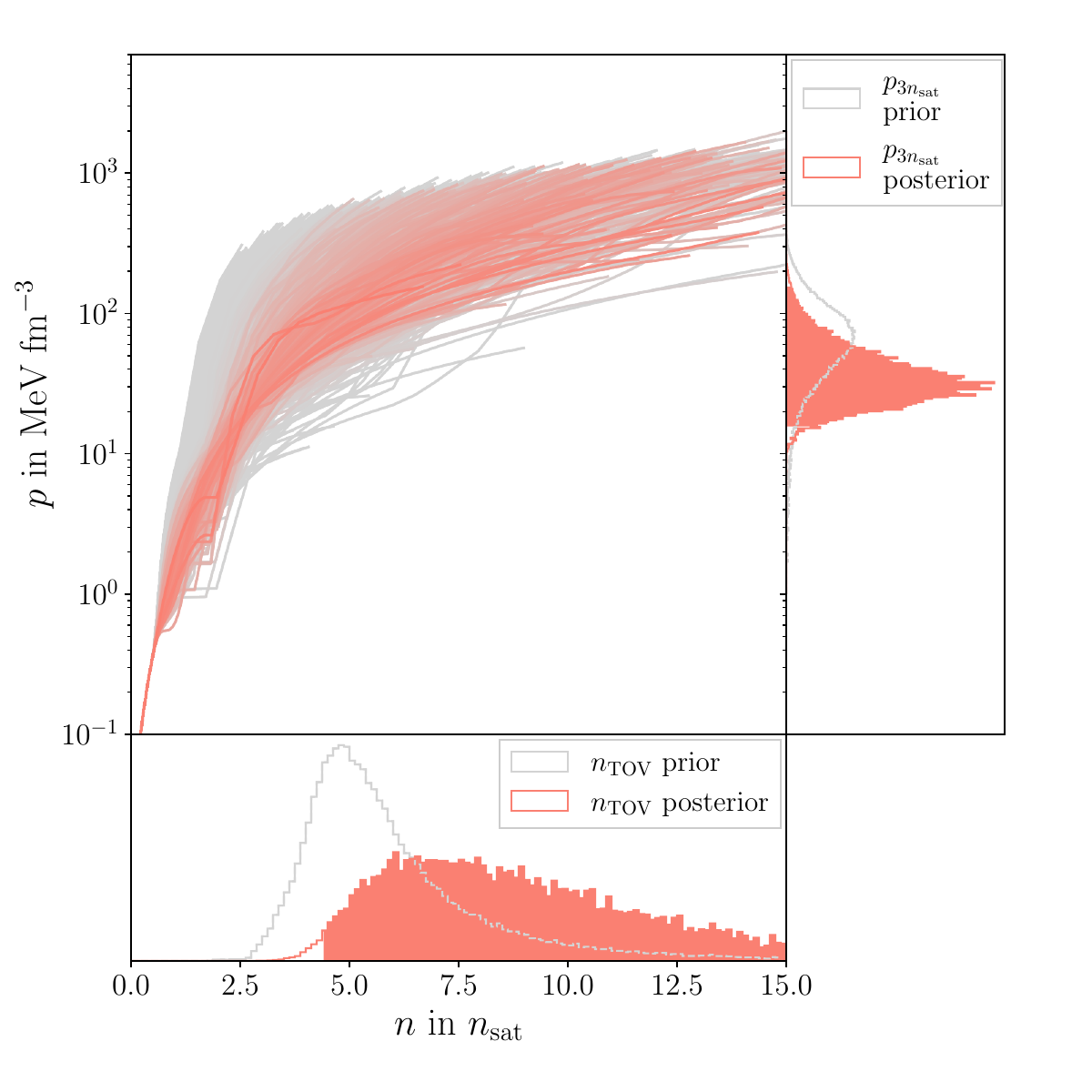}
    \caption{EOS inference based on the mass-radius measurement of HESS J1731-347. 
    Arrangement and color coding as in Fig.~\ref{fig:chiEFT}. 
    In the middle plot of the top panel, the blue contours show the 68\% and 95\% credible posterior regions for the $M$-$R$ measurement.}
    \label{fig:HESS}
\end{figure}

Even for NSs lacking any signs of pulsed emission, the radius and mass can also be deduced from its X-ray spectrum, in particular from the thermal component. 
As is the case with any ordinary star, the effective temperature, together with the absolute luminosity, allows one to determine the emitting surface area and hence the radius. 
For NSs, gravitational light bending needs to be taken into account, but by simultaneously measuring the gravitational redshift the mass $M$ and radius $R$ can be recovered~\citep{Brown_1998, Oezel_2016}. 
In practice, however, several caveats complicate this endeavor~\citep{Bhattacharyya_2010}.
For one, uncertain distance estimates to the sources make the measurement of the absolute luminosity difficult, especially in combination with interstellar extinction of high-energy photons. 
Additionally, some quiescent NSs can display nonthermal contributions in their spectra. Their nature is unclear, although it might be linked to residual accretion in binary systems~\citep{Heinke_2003, Chakrabarty_2014}.
Moreover, models for the surface emission usually need to assume uniform emission from the entire NS surface and suffer from uncertainties of the atmospheric composition. 
Several reports of thermal X-ray spectral measurements for NSs exist, e.g., Refs.~\citep{Rutledge_2002, Pavlov_2009, Bogdanov_2016, Oezel_2016B, Potekhin_2020}, but systematic uncertainties often impede inference of NS properties. 
Here, we focus on two instances of NS masses and radii reported from spectral analyses of thermal X-rays, namely, for the quiescent X-ray binaries from Ref.~\citep{Steiner_2018} and the compact object in \mbox{HESS J1731-347} from Ref.~\citep{Doroshenko_2022}. 

\textit{\citet{Steiner_2018}.} In low-mass X-ray binaries (LMXBs), a stellar or substellar companion with a mass below \qty{2}{\msun} orbits a stellar black hole or NS, so the companion is often lighter than the compact object itself. 
The majority of observationally known LMXBs are associated with NSs~\citep{Liu_2007, Avakyan_2023}, and we naturally restrict our discussion to LMXBs with NSs. 
The NS will in some instances accrete matter from either the companion or ambient gas. 
The accretion emission of the NS may change over time and can vary widely for different LMXBs, from (near) quiescence to X-ray emission at the order of the Eddington luminosity~\citep{Bhattacharyya_2010}. 
Quiescent NSs have been analyzed in certain types of LMXBs, accordingly named quiescent low-mass X-ray binaries (qLMXBs). 
Their accretion activity occasionally ceases for a time span of months to years due to instabilities in the accretion disk~\citep{King_2001}, before a new accretion outburst takes place. 
During this period, the luminosity is at a low level. 
The NS will mainly emit thermal radiation from its accretion heated surface. 
These qLMXBs usually appear with beneficial properties, including low magnetic fields and common occurrence in star clusters. 
The latter eases distance measurements, making them suitable targets for thermal X-ray spectral analysis~\citep{Bhattacharyya_2010}. 

Reference~\citep{Steiner_2018} reported measurements for the masses and radii of eight NSs hosted by qLMXBs in globular clusters, for which previous observations with the Chandra and/or XMM-Newton facilities had reported spectral data. 
Using the \textsc{xspec} framework~\citep{Arnaud_1996}, the authors performed Bayesian inference on the spectra. In particular, a predictive atmosphere model of either hydrogen or helium was used, taking distance uncertainties and possible hot spots into account.
The uncertainty in the atmosphere composition is one driving factor for systematics. 
In qLMXBs, previously accreted matter from the companion determines the NS atmosphere's ingredients. 
If that companion is devoid of hydrogen (e.g.,~a white dwarf), the NS atmosphere might comprise heavier elements. 
For their eight sources, Ref.~\citep{Steiner_2018} reported mass and radius values for both hydrogen and helium atmosphere models, respectively. 
For the NS X5 in Tucanae 47 and the NS in $\omega$ Centauri, only values with hydrogen atmospheres were reported. 
For the latter, hydrogen was reliably detected~\citep{Haggard_2004}, while the former has a long binary orbital period indicating a hydrogen-rich donor. 
Here we focus on these two cases, because they avoid any ambiguity regarding the atmosphere composition.

We show the corresponding $M$-$R$ contours in Fig.~\ref{fig:qLMXBs}, together with the posterior likelihood on the EOS according to
\begin{align}
    \mathcal{L}(\text{EOS}|\text{qLMXBs}) = \mathcal{L}(\text{EOS}|\omega \text{ Cen}) \times \mathcal{L}(\text{EOS}|\text{X5})\,.
\end{align}
The EOS likelihood from a single mass-radius measurement is again given by Eq.~(\ref{eq:l2}). 
The inference shows that information from these two qLMXBs mainly impact $R_{1.4}$ and reject the most extreme candidates in our EOS set, in particular, the very soft ones. 
As the $M$-$R$ results show comparatively wide statistical uncertainties in the NS mass, the main constraining power arises from restraining the radii to $R\gtrsim\qty{10}{km}$.

Apart from the atmospheric composition, additional systematic errors such as absorption variability and the robustness of hot-spot corrections remain. 
In particular, we point out that Ref.~\citep{Steiner_2018} excluded the NS X5 in Tucanae 47 from their baseline analysis due to its emission variability caused by its high inclination.
We still discuss it here because it is a frequently observed source that avoids any uncertainty stemming from the atmospheric composition. 
Other mass-radius measurements from qLMXBs are, for instance, given in Refs.~\citep{Catuneanu_2013, Oezel_2016B, Shaw_2018, Etivaux_2019, Al-Mamun:2020vzu}. 
The analysis of Ref.~\citep{Etivaux_2019} is similar to the one in Ref.~\citep{Steiner_2018}, but directly incorporates an EOS model to obtain the $M$-$R$ posterior and is therefore not practical for our study here. 

\textit{\citet{Doroshenko_2022}.} 
For the central compact object in the supernova remnant \mbox{HESS J1731-347}, the authors obtained X-ray spectra from XMM-Newton and the Suzaku telescope and robust parallax estimates through Gaia parallax measurements of the optical stellar counterpart~\citep{Doroshenko_2016}. 
The compact object in HESS J1731-347 is relatively bright and shows only small pulsations, making it a suitable observational target. 
For analysis in the \textsc{xspec} framework~\citep{Arnaud_1996}, Ref.~\citep{Doroshenko_2022} used a uniform temperature carbon atmosphere model including interstellar extinction and dust scattering. 
Their Bayesian data analysis led to an unusually small estimate for the central compact object's mass and radius. 
The $M$-$R$ posterior based on the samples provided in Ref.~\citep{Doroshenko_2022_samp} is shown in Fig.~\ref{fig:HESS} together with the posterior distribution on our EOS set.

The inferred results from \mbox{HESS J1731-347} clearly favor softer EOS and push the posterior to lower values for $M_{\text{TOV}}$, $p_{3\nsat}$, and $R_{1.4}$, requiring significantly higher TOV densities. 
It has been noted that this is in tension with many available nuclear models~\citep{Brodie_2023}, although it is still possible to reconcile the low-mass and radius values with very soft EOS models for NSs~\citep{Sagun_2023}. 
Recently, the authors of Ref.~\citep{Alford_2023} have pointed out that the resulting small mass and radius in Ref.~\citep{Doroshenko_2022} rely heavily on the assumption of a uniformly emitting carbon atmosphere and the observational distance being sampled from the Gaia distance estimate around~\qty{2.5}{kpc}. 
In fact, a previous analysis of \mbox{HESS J1731-347} that fitted carbon atmospheres at a distance above \qty{2.5}{kpc} led to significantly higher values for $M$ and $R$ and wider uncertainties (see Fig. 5 in Ref.~\citep{Klochkov_2015}).
Likewise, the supplementary analysis in Ref.~\citep{Doroshenko_2022} shows that a two-temperature model of carbon and hydrogen atmospheres fit the spectra of HESS J1731-347 similarly well when assuming a \qty{1.4}{\msun} NS (see Table 1 in Ref.~\citep{Doroshenko_2022}). 
It thus remains necessary to keep in mind the significance of systematic uncertainties governing the inference of the \mbox{HESS J1731-347} parameters based on thermal X-ray profile modeling. In Sec.~\ref{sec:outlier}, we discuss how the remaining constraints relate to this measurement.

\subsection{Thermonuclear accretion bursts in low-mass X-ray binaries}
\label{subsec:Burst}
\begin{figure}
    \centering
    \includegraphics[width = \linewidth]{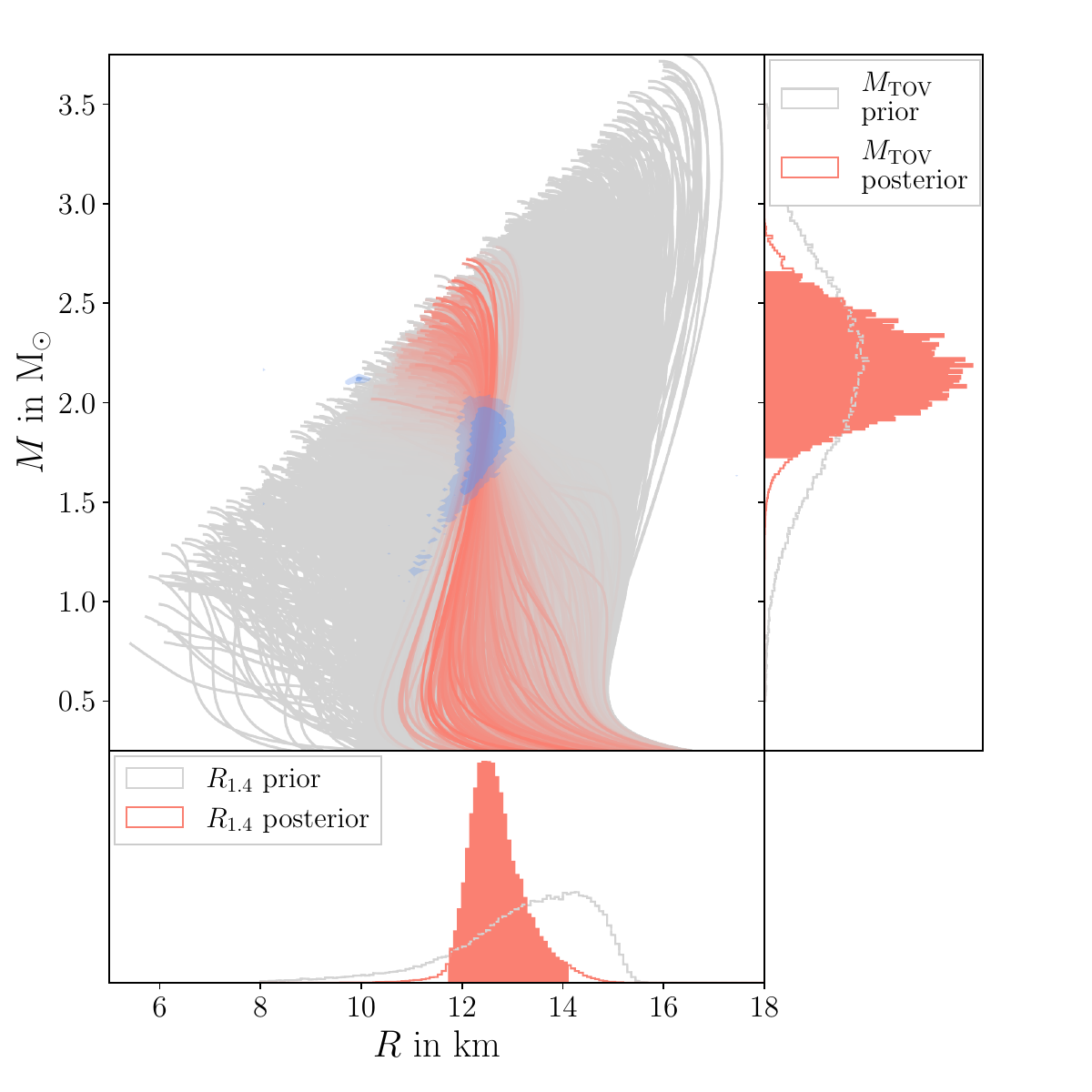}
    \includegraphics[width = \linewidth]{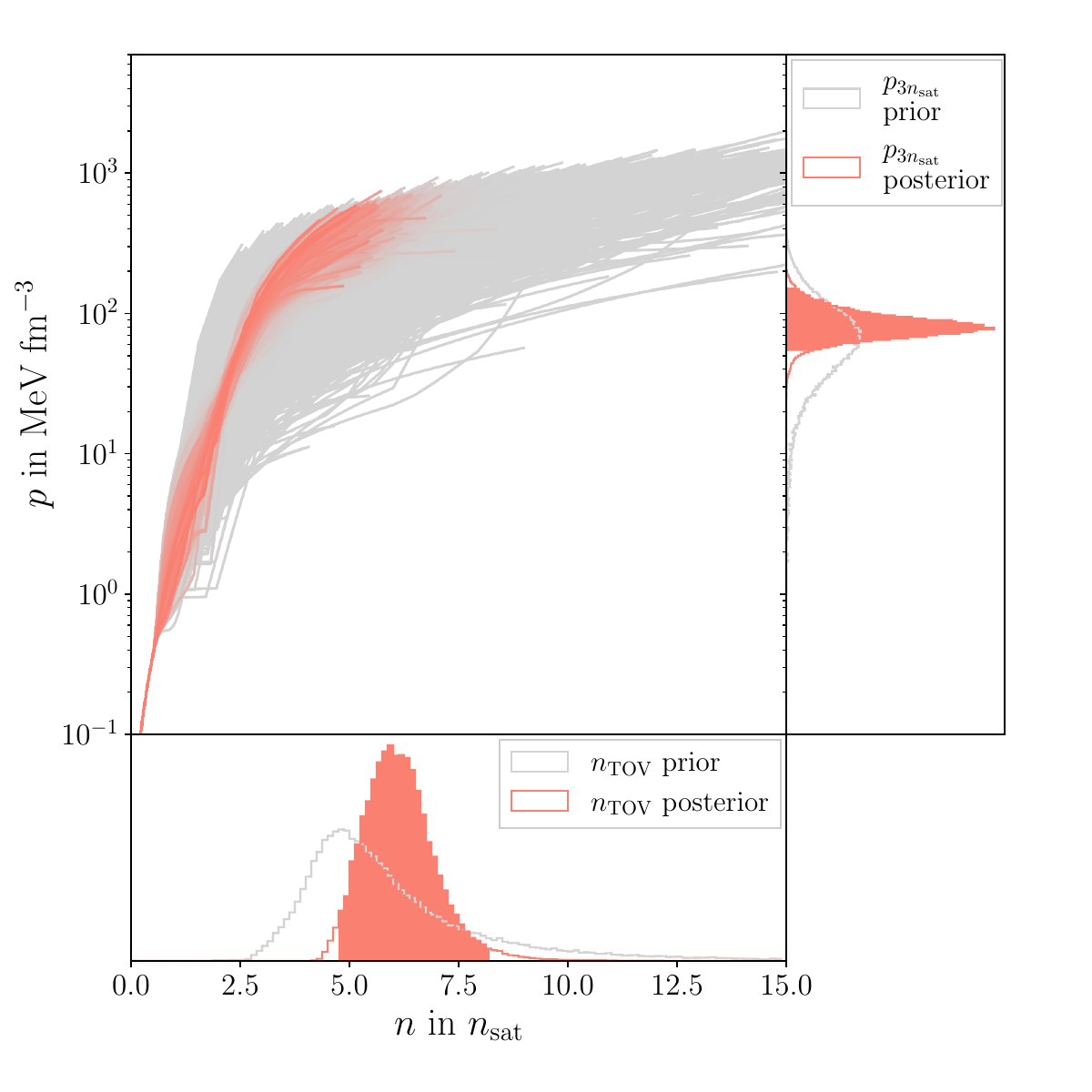}
    \caption{EOS inference based on the mass-radius measurement of \mbox{4U 1702-429}. Arrangement as in Fig.~\ref{fig:Shapiro}. In the top figure, blue contours show 68\% and 95\% credible posterior regions from the $M$-$R$ measurement.}
    \label{fig:Burst}
\end{figure}

\begin{figure}
    \centering
    \includegraphics[width = \linewidth]{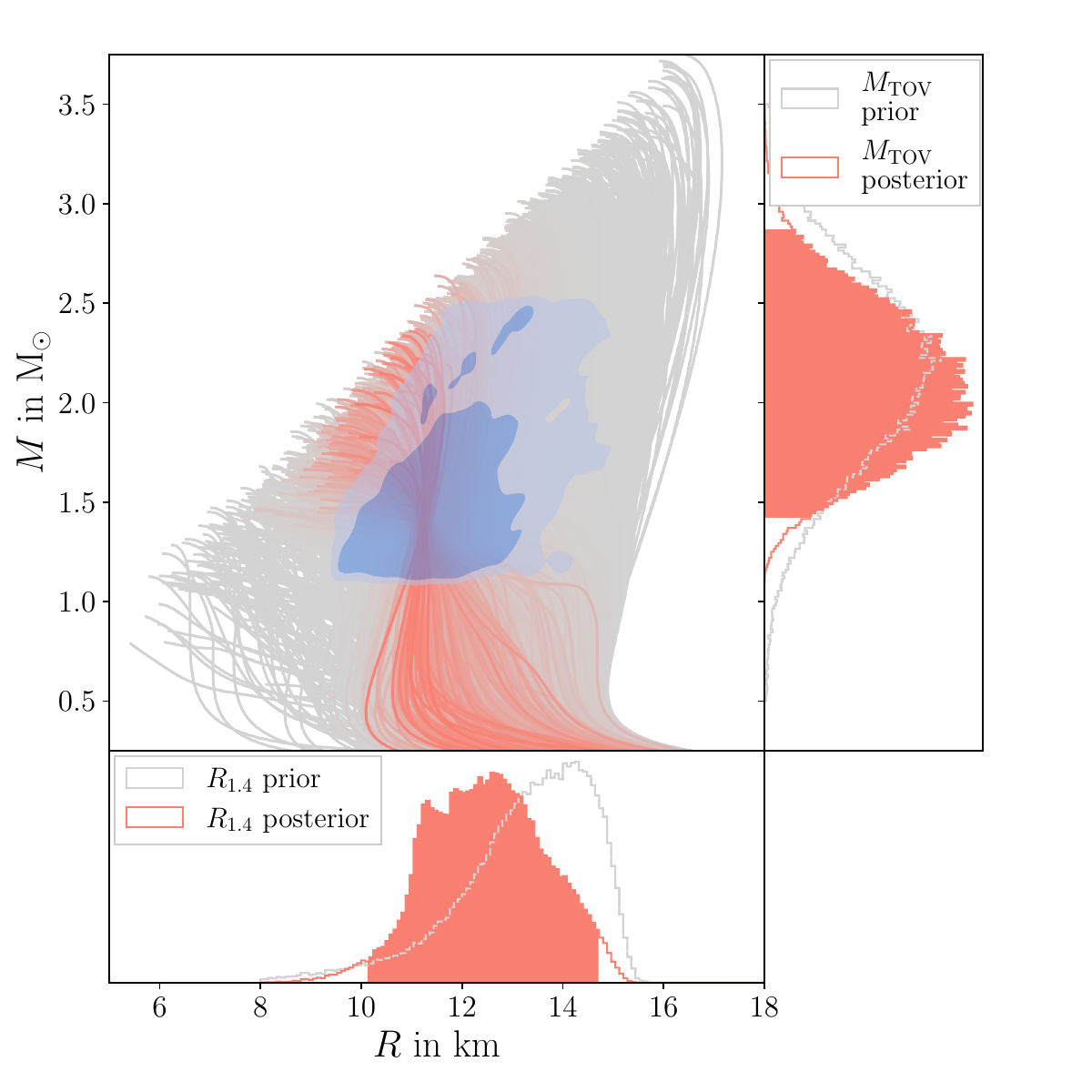}
    \includegraphics[width = \linewidth]{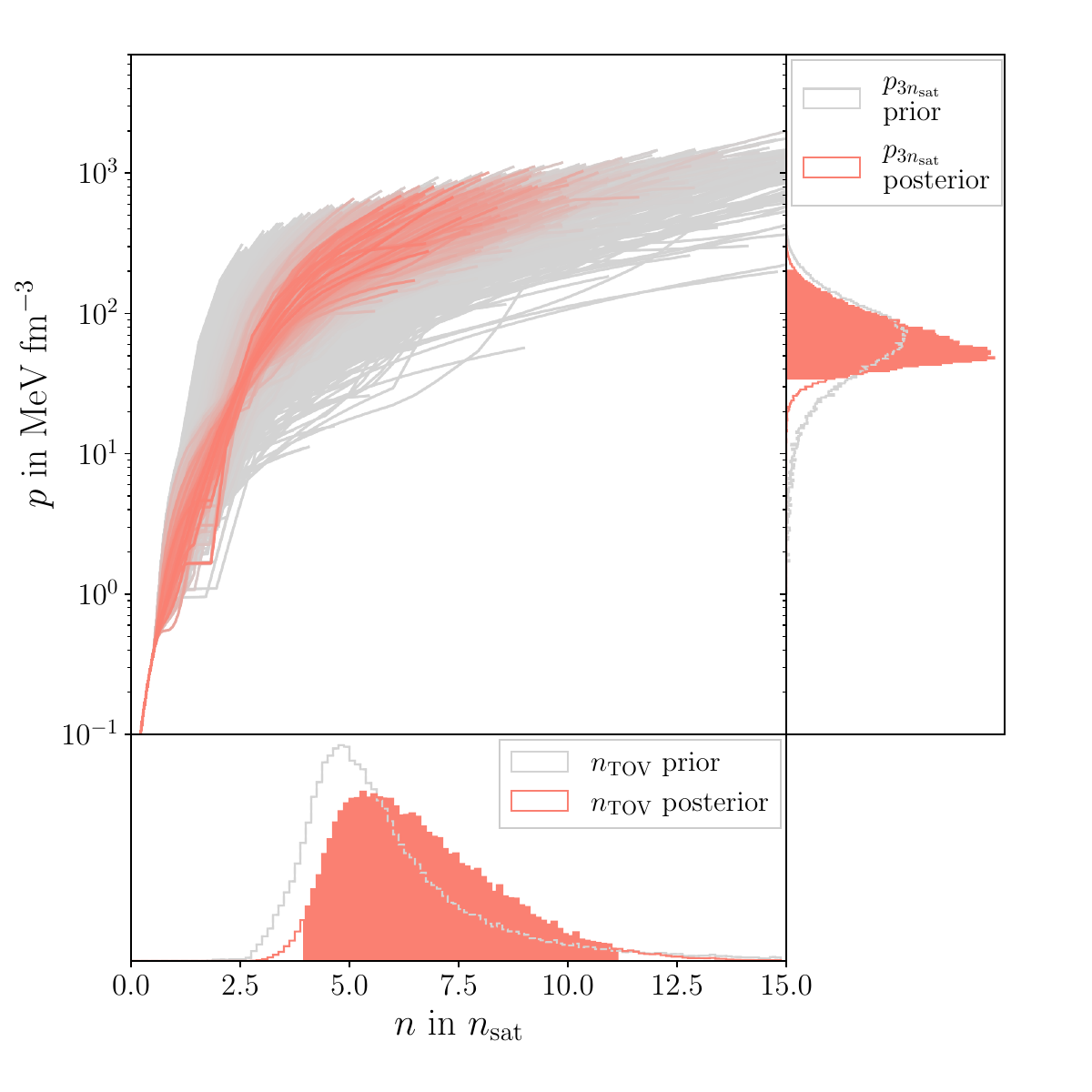}
    \caption{EOS inference based on the mass-radius measurement of \mbox{SAX J1808.4-3658}. Arrangement as in Fig.~\ref{fig:Shapiro}. In the top figure, blue contours show 68\% and 95\% credible posterior regions from the $M$-$R$ measurement.}
    \label{fig:SAX}
\end{figure}

If NSs are situated in low-mass X-ray binaries with sufficiently small orbital separation, the companion will overfill its Roche lobe, forming an accretion disk around the NS.
The magnitude of the accretion activity as well as its variations over time depend intricately on the binary's properties~\citep{Bhattacharyya_2010}. 
In certain cases, accretion causes a particular type of radiation outburst. 
These thermonuclear X-ray bursts, also called type-I X-ray bursts, occur when accreted material piles up on the NS surface until compression and pressure launch a runaway nuclear fusion reaction~\citep{Joss_1977, Narayan_2003, Cavecchi_2020}.
Because these bursts originate directly from the surface, they carry information about NS parameters such as temperature, spin, and its mass and radius. 
Furthermore, they are fairly bright as the luminosity typically increases by about a factor of 10 over a time span of seconds~\citep{Bhattacharyya_2010}, yielding high signal-to-noise ratios (SNRs). 
When an LMXB is observed over a longer period, repeated bursts can be combined for a joint analysis to obtain the mass and radius of the NS~\citep{Bhattacharyya_2005}. 
Bursting LMXBs have been used extensively for the inference of NS radii and masses, for instance in Refs.~\citep{Fujimoto_1986, Ebisuzaki_1987, Damen_1990, Haberl_1995}, with continued efforts over the past decade~\citep{Guever_2010, Steiner_2013, Oezel_2016B, Suleimanov_2020, Al-Mamun:2020vzu, Kim_2021}. 
Here, we focus on two modern investigations into thermonuclear X-ray bursters, namely the popular study of Ref.~\citep{Naettilae_2017} and the recent introduction of a Bayesian framework for X-ray bursters in Ref.~\citep{Goodwin_2019}.

\textit{\citet{Naettilae_2017}.} The authors reanalyzed five distinct X-ray bursts from the LMXB \mbox{4U 1702-429} that had been observed previously by the \textit{Rossi} X-ray explorer~\citep{Naettilae_2016}. 
Instead of simple thermal spectral fits, the burst spectra were fitted with a proper atmosphere model. 
This model was an adapted version of the stellar atmosphere code \textsc{atlas}~\citep{Kurucz_2014} that determines the emitted flux from the NS surface. 
Hierarchical Bayesian inference provided the NS parameters, where the most successful spectral model (model $\mathbb{D}$ in Ref.~\citep{Naettilae_2017}) samples additionally over the atmosphere metallicity and a systematic uncertainty parameter.
For the result, the authors recovered a comparatively narrow $M$-$R$ posterior. 
We show the contours together with the implications for our EOS set in Fig.~\ref{fig:Burst}. The likelihoods for our EOS candidates are again calculated through Eq.~(\ref{eq:l2}). 

Very soft and very stiff EOS candidates are rejected by this inference and the narrow uncertainty on the radius measurement restricts the canonical radius $R_{1.4}$ to ${12.62}^{+1.49}_{-0.87}$\,km. 
For the microscopic EOS, this translates to a relatively tight pressure constraint between \qtyrange{2}{3}{\nsat}. 
In total, only few EOSs with a high posterior likelihood remain, making this measurement one of the most restricting constraints available. 

The narrow radius uncertainties are partly attributed to the desirable properties of the source, e.g., its low accretion rate. 
However, potential biases in the analysis remain, even though they are partially accounted for with a systematic uncertainty parameter. 
They are related to the atmospheric composition and whether the typical assumption that the full surface of the NS contributes to the emission is justified. 
Likewise, the model assumes that the accretion environment in which the burst takes place remains unaffected by the sudden release of energy, which is not always appropriate~\citep{Worpel_2015, Kajava_2017, Naettilae_2022}.
While the $M$-$R$ posterior of Ref.~\citep{Naettilae_2017} coincides largely with expectations about the NS mass-radius relation from other sources such as NICER or GW data (see Sec.~\ref{sec:Multimessenger}), potential limitations should be kept in mind. 

\textit{\citet{Goodwin_2019}.} The authors introduced a new way to recover burst properties with a modified version of the semianalytical \textsc{settle} model~\citep{Cumming_2000, Galloway_2006}. 
Applying this method to bursts from \mbox{SAX J1808.4-3658} observed with the \textit{Rossi} X-ray explorer, they inferred the NS mass and radius through MCMC sampling. 
The $M$-$R$ prior was based on the final result of the baseline model in Ref.~\citep{Steiner_2018}, which investigated eight NSs in quiescent LMXBs (see Sec.~\ref{sec:Thermal}). 
We show the $M$-$R$ posterior and the posterior likelihood of our EOS set in Fig.~\ref{fig:SAX}. 
The statistical uncertainties on the mass and radius are much wider than in the study of Ref.~\citep{Naettilae_2017} above. 
The authors ascribed this to some possible degeneracies in their model's parameters. 
Similar to the analysis of qLMXBs in Ref.~\citep{Steiner_2018}, the mass uncertainty is very large, so the radius limit of $R<\qty{15}{km}$ delivers the main constraining power of this measurement.
They shift our posterior estimate for the canonical radius $R_{1.4}$ to lower values and rule out very stiff EOSs.
Likewise, any EOS with $\mtov<\qty{1.15}{\msun}$ is rejected, although this is due to the mass prior bound being set at that level. 

Systematics in the analysis are driven by uncertainties in the accretion disk geometry~\citep{Galloway_2006, Johnston_2018}. 
Assessing the model performance is further complicated by the fact that the model predicts six consecutive outbursts over the span of one week, of which only three were actually observed.
The other predicted outbursts lie outside the observation periods of the \textit{Rossi} X-ray explorer.
In any case, we expect systematic errors to be less dominant compared to Ref.~\citep{Naettilae_2017}, because of the larger statistical errors.

\section{Detections of binary neutron star mergers}
\label{sec:Multimessenger}

\begin{figure}
    \centering
    \includegraphics[width = \linewidth]{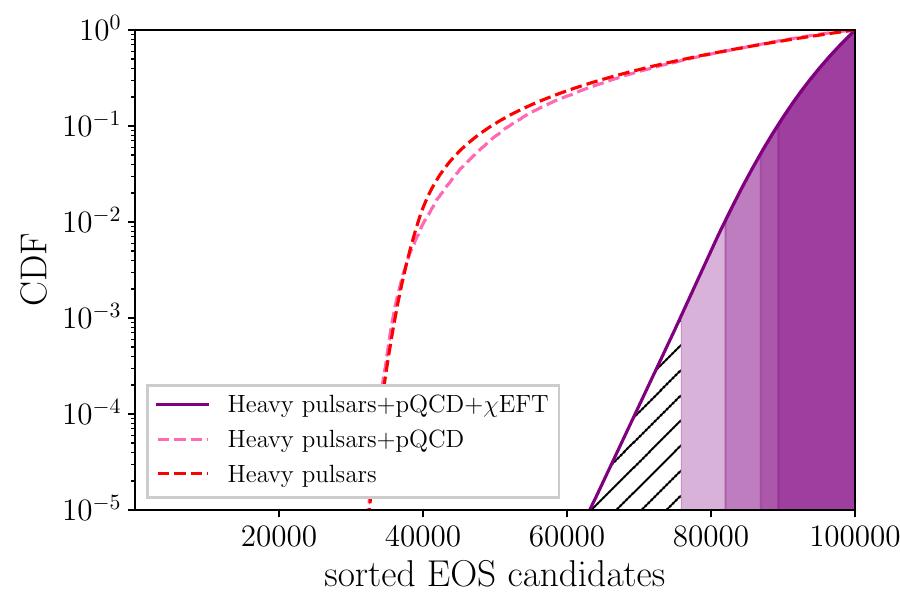}
    \caption{Cumulative distribution function (CDF) of the EOS after combining posterior likelihoods from the radio timing measurements of the three heavy pulsars, \chiEFT, and pQCD. 
    The EOS candidates are labeled from 1 to 100\,000 and sorted here by their posterior probability. 
    The filled areas mark the 0.001, 0.01, 0.05 and 0.1 quantiles, respectively, whereas the hatched area signifies those EOSs that are discarded for the inference of BNS signals in this section.}
    \label{fig:EOS_select}
\end{figure}

Besides observations from isolated NSs in mechanical equilibrium as discussed in the previous section, BNS coalescences have proven valuable for assessing the dense matter EOS. 
Here, we use the Bayesian multimessenger-analysis framework \textsc{NMMA}~\citep{NMMA_2023} to perform parameter estimations from observational data of the gravitational-wave event GW170817~\citep{LIGO_GW170817} and its electromagnetic counterparts AT2017gfo and GRB170817A~\citep{LIGO_GW170817_EM}, as well as the gravitational-wave event GW190425~\citep{LIGO_GW190425}, and the long gamma-ray burst (GRB) GRB211211A~\citep{Rastinejad_2022}. 
As a particular feature of \textsc{NMMA}, we can directly sample over the EOS as a parameter and therefore immediately obtain an EOS posterior distribution. 
Since the cost of Bayesian parameter estimation increases with the size of the parameter space, we restrict ourselves to a subset of the previously considered EOSs. 
We discard any EOS that lacks support from those constraints of Secs.~\ref{sec:Nuclear} and~\ref{sec:Astro} we deem most reliable.
These are the mass measurements of heavy pulsars through radio timing methods and the theoretical calculations from \chiEFT\ and pQCD. 

While this selection may be regarded as somewhat subjective, the reasoning here is that the techniques for the two theoretical inputs are well established and we impose their constraints in a conservative manner.
Likewise, the masses of \mbox{PSR J0740+6620}, \mbox{J1614-2230}, and \mbox{J0348+0432} obtained via radio timing techniques are the only pulsar observations that do not rely on intricate modeling of X-ray emission from NS surfaces and thereby constitute particularly reliable astrophysical observations. 
Combining the posterior likelihoods from these five independent constraints, we select all EOSs above the 0.001 quantile for our BNS inferences (corresponding roughly to a 3$\sigma$ credibility level), leading to a reduced number of 24\,288 remaining EOS candidates.
Figure~\ref{fig:EOS_select} shows the cumulative distribution function for the joint EOS probability combining these five constraints.
The analyses for this section show that only GW170817 and its electromagnetic counterparts provide meaningful restrictions on the EOS, mainly preferring EOS with canonical radii below \qty{13}{km}. 
If we also impose the postmerger constraints from the collapse of GW170817's remnant, we additionally obtain $\mtov \lesssim \qty{2.5}{\msun}$.

\subsection{Gravitational-wave signal GW170817}
\label{sec:GW170817}

\begin{figure}
    \centering
    \includegraphics[width = \linewidth]{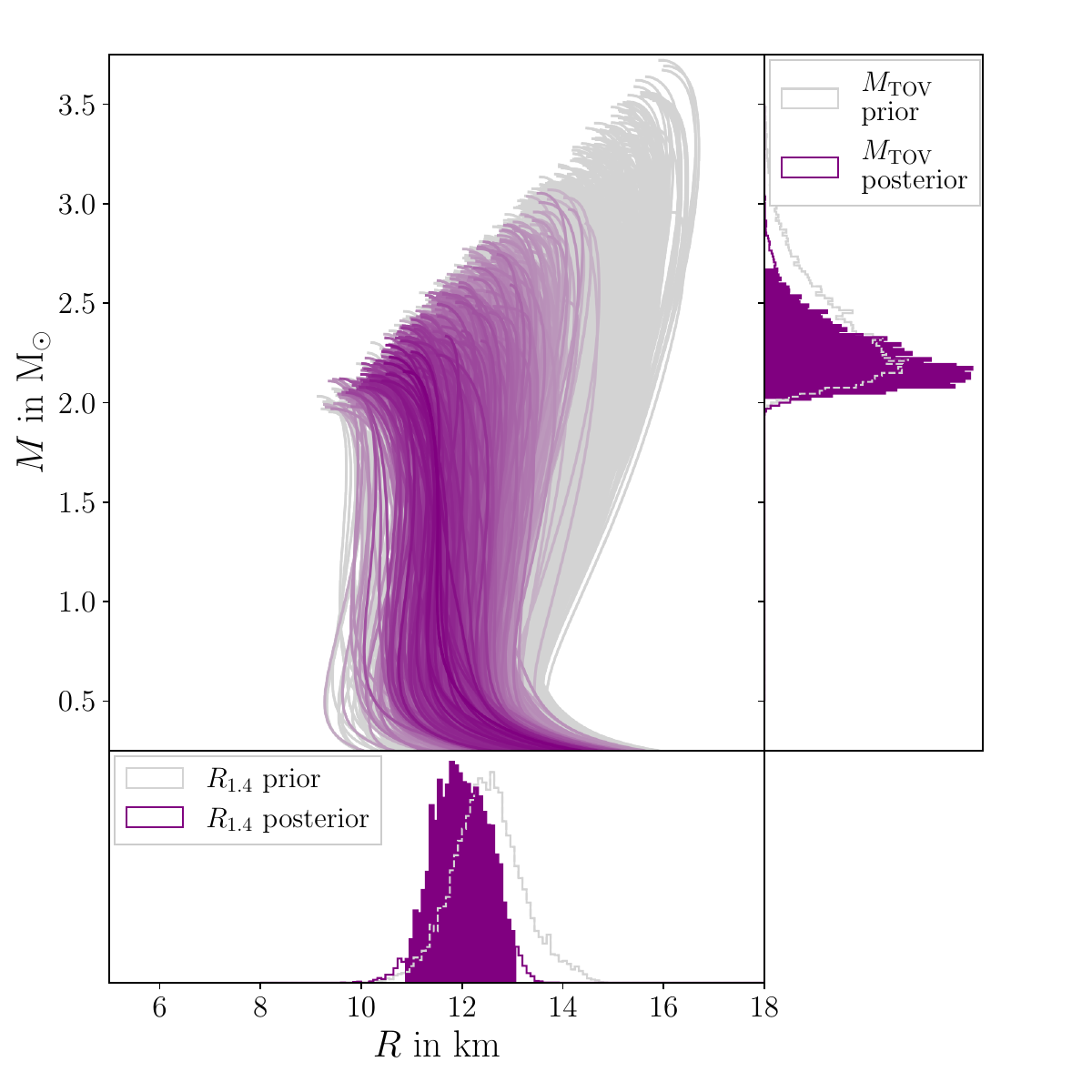}
    \includegraphics[width = \linewidth]{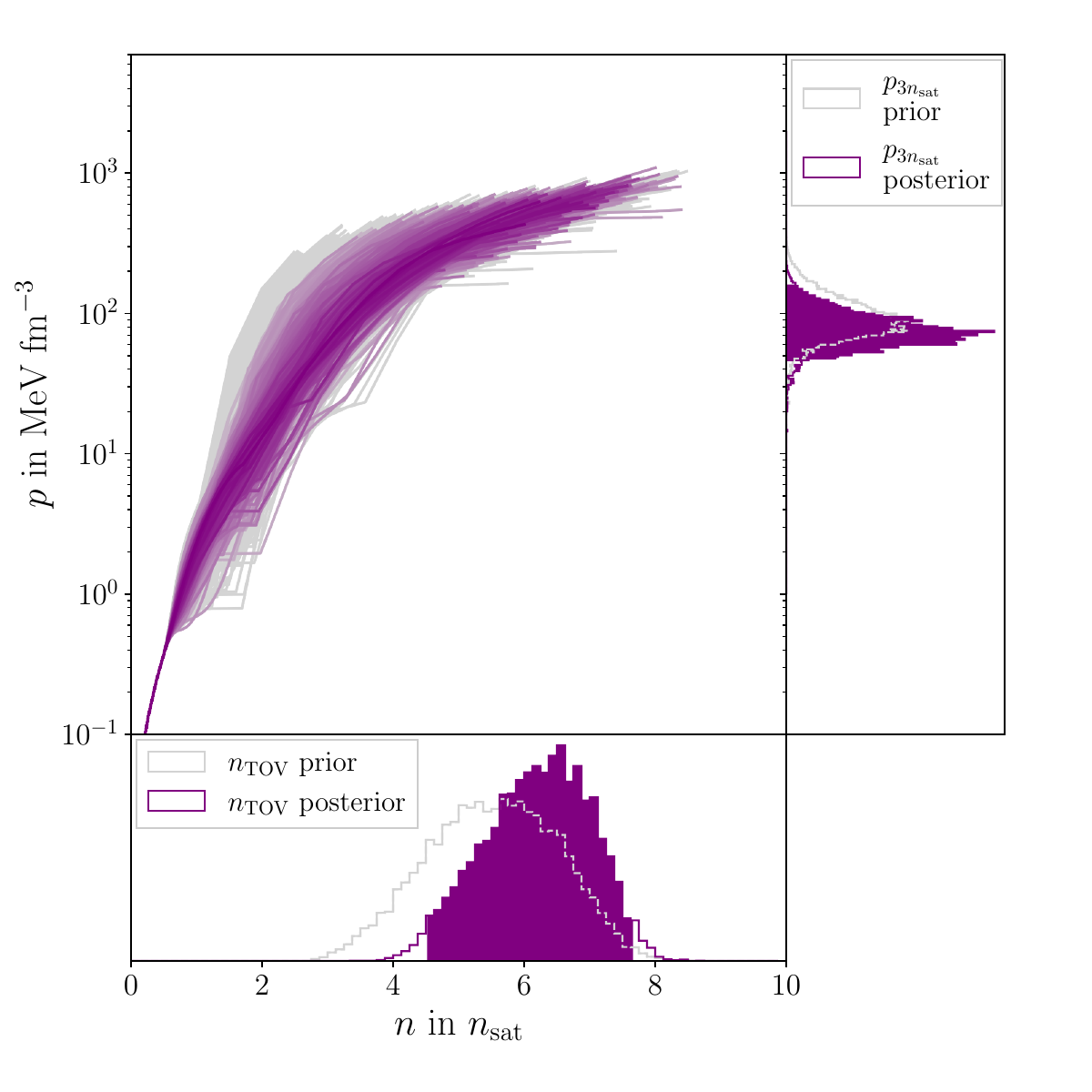}
    \caption{EOS inference based on the measurement of GW170817. Arrangement and color coding as in Fig.~\ref{fig:chiEFT}. The EOS shown here are the reduced prior set described at the beginning of the section.}
    \label{fig:GW170817}
\end{figure}

\begin{figure}
    \centering
    \includegraphics[width = \linewidth]{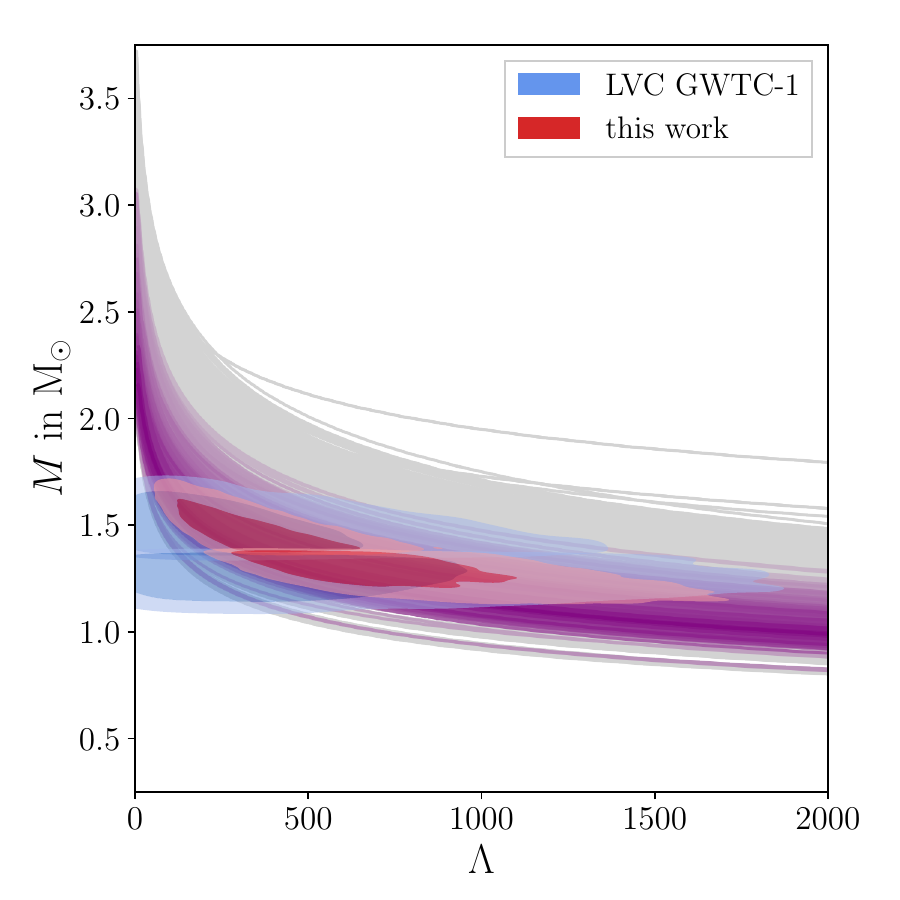}
    \caption{$M$-$\Lambda$ curves for our EOSs, color coded according to their posterior likelihood based on our inference of GW170817. 
    The contours show 68\% and 95\% $\Lambda$-$M$ posterior credibility regions for the primary and secondary NSs in GW170817. 
    Blue contours represent the posterior from LVC GWTC-1~\citep{LIGO_GWTC_1}, red contours the posterior of our own inference.}
    \label{fig:tidal_GW170817}
\end{figure}

The detection of GW170817 through the LIGO and Virgo collaboration (LVC)~\citep{LIGO_advanced_detector_2015, Acernese_2015} was the first GW observation of a BNS merger~\citep{LIGO_GW170817}.
Since NSs have finite size, they are susceptible to tidal deformation if placed in an inhomogeneous gravitational field. 
Specifically, if positioned in a binary system, the mutual gravitational attraction deforms both components. 
This in turn alters their quadrupole moment which is imprinted on the emitted gravitational waves. 
During the inspiral, tidal perturbations on the wave signal are measurable as a phase shift, which is primarily determined by the tidal deformability parameter $\Lambda$. 
The value for this parameter is determined by the EOS~\citep{Hinderer_2008, Vines_2011}. 

We reanalyze the GW170817 signal using the Bayesian multimessenger framework \textsc{NMMA}. 
Usually, GWs from a circular BNS system are analyzed through a waveform model with 17 parameters, two of which are the tidal deformabilities. 
This is the typical approach of, e.g., Refs.~\citep{LIGO_GW170817, LIGO_GW170817_properties, LIGO_GWTC_1}.
As mentioned above, \textsc{NMMA} samples directly over the EOS which is shared by both NSs, hence reducing the parameter space to 16 dimensions. 
A detailed list of all parameters and priors is given in Appendix \ref{app:GW_setup} in Table~\ref{tab:GW_prior}.
For brevity, we denote a sampling point in this parameter space as $\vec{\theta}$ and the corresponding waveform as $h(\vec{\theta})$.

For our analysis, we employ the \mbox{$\mathtt{IMRPhenomXP\_NRTidalv3}$} waveform model, which combines the \mbox{$\mathtt{IMRPhenomXP}$} model~\citep{Pratten_2021} with the newly developed \mbox{$\mathtt{NRTidalv3}$} model for the tidal effects~\citep{Dietrich_2019, Abac_2023}. \mbox{$\mathtt{IMRPhenomXP}$} describes the (2,2) mode~\footnote{Considering the small mass ratio in the BNS system, focusing on the (2,2)-mode seems justified as more than 99.5\% of the GW energy are released through the (2,2)-mode~\citep{Dietrich_2017}.} of a precessing circular binary of point masses based on a phenomenological ansatz.
Its advantage over previous models results from a refined description of the inspiral and calibration to a larger set of merger simulations in numerical relativity. 
For the description of BNS systems, \mbox{$\mathtt{NRTidalv3}$} adds tidal phase contributions to the \mbox{$\mathtt{IMRPhenomXP}$} waveform. 
In contrast to its predecessors, the model uses a larger set of numerical-relativity simulations covering systems with high mass ratio and a wide range of EOSs.
Its description also takes dynamical tides into account, where the tidal deformability is not adiabatic but a function of the GW frequency.

If the detectors measure a signal as data $d(f)$ as strain, we can express the likelihood for a given sampling point $\vec{\theta}$ as
\begin{align}
\begin{split}
   \log \mathcal{L}(\vec{\theta}|d) =& -2 \int_{f_{\rm min}}^{f_{\rm max}} \frac{\lvert d(f) - h(f,\vec{\theta})\rvert^2 }{S(f)} \ df+ {\rm const},
   \label{eq:GW_likelihood}
\end{split}
\end{align}
where $h(f, \vec{\theta})$ is the waveform at frequency $f$. 
Further, we assume stationary Gaussian noise with power spectral density $S(f)$ within the detector.
The strain data are taken from the first LVC GW transient catalog (GWTC-1)~\citep{LIGO_GWTC_1, GWOSC_2021}.
The Bayesian evidence and subsequently the posterior $P(\vec{\theta}|d)$ are then obtained by exploring the parameter space with the nested sampling algorithm as implemented in \textsc{dynesty}~\citep{Dynesty_2018} using 4096 live points.
Nested sampling is an established sampling algorithm and has been shown to be robust for the inference of BNS parameters~\citep{Veitch_2010, Ashton_2019, Buchner_2023}, also in comparison with MCMC based sampling methods~\citep{LIGO_GW170817_properties, Ashton_2021}.

To obtain the marginalized posterior on the EOS, one needs to account for selection effects. 
The GW detectors have an SNR threshold up to which they will detect an event.
Therefore, the data is inherently biased toward parameter values that are easier to detect, e.g., higher masses and smaller distances. 
Therefore, one needs to divide the marginalized posterior by the detection probability $P_{\text{det}}(\text{EOS})$ \citep{Mandel_2019, Kunert_2022} 
\begin{align}
    P_{\text{det}}(\text{EOS}) = \int P(\vec{\theta_I}|\text{EOS}) P_\text{det}(\vec{\theta_I})\ d\vec{\theta_I}\,.
    \label{eq:detection_probability}
\end{align}
Here, we assume that the distribution of intrinsic BNS parameters $P(\vec{\theta_I}|\text{EOS})$ is largely determined by the TOV mass. 
To evaluate Eq.~\eqref{eq:detection_probability}, we sample masses uniformly between 0.5 and $\mtov$ and determine the detection probability for these random events using the neural network classifier from Ref.~\citep{Gerosa_2020}. 
Thus, $P_{\text{det}}(\text{EOS})$ effectively becomes a function of $\mtov$ and we reweigh the EOS posterior samples from our nested sampling run with its inverse. 
Adjusting for the selection bias has negligible impact on the results for one event, but becomes important when combining constraints from multiple GW detections~\citep{Mandel_2019}.

The resulting distribution on the EOSs is shown in Fig.~\ref{fig:GW170817}. 
We note that our EOS sampling implicitly assumes that GW170817 indeed originated from a BNS. 
For this reason, and because the EOS prior is informed by nuclear theory and heavy pulsar measurements, our posterior on the component tidal deformabilities $\Lambda_{1,2}$ is significantly narrower compared to other analyses that sample $\Lambda$ uniformly over a certain range~\cite{LIGO_GW170817, LIGO_GW170817_properties, LIGO_GWTC_1}. 
Figure~\ref{fig:tidal_GW170817} shows preferred $M$-$\Lambda$ relations for our EOSs and compares our posteriors for $\Lambda_{1,2}$ to the corresponding posteriors from the LVC GWTC-1~\citep{LIGO_GWTC_1} that used the \mbox{$\mathtt{IMRPhenomPv2\_NRTidal}$} waveform model. 
The constraint from GW170817 pushes the posterior toward slightly softer EOSs and smaller radii, as high tidal deformabilities are disfavored by the data. 
This is consistent with previous studies~\citep{LIGO_GW170817_EOS, Raithel_2019A, Dietrich_2020, Capano_2020}. 

Most waveform models, including the $\mathtt{IMRPhenomXP\_NRTidalv3}$ employed here, rely in some way on the post-Newtonian (PN) approximation, with the very late inspiral and merger phase being described by fits to numerical-relativity simulations. 
The finite truncation of the PN expansion as well as the fit to a discrete set numerical-relativity data naturally introduces systematic biases that may impede parameter estimation, in particular for the tidal deformability~\citep{Dudi_2018, Samajdar_2018, Owen_2023}. 
Systematic effects become noticeable for detections with high SNRs ($\gtrsim 80$)~\citep{Gamba_2021} or when combining the results of multiple ($\gtrsim 30$) detections~\citep{Kunert_2022}. 
When determining tidal deformabilities from GW170817 with an SNR of about 33, however, the discrepancies are small compared to the relatively large statistical errors. 
Yet, for future observations with the LIGO-Virgo-KAGRA network operating at design sensitivity or with third-generation detectors, systematic uncertainties need to be accounted for \citep{Read_2023}.
Systematic effects would also arise if the assumptions about the realized physical setting are wrong, for instance if gravitational waves need to be described in modified theories of gravity~\citep{Saffer_2021, Brown_2022} or if dark matter is present in the NS interior~\citep{Das_2022, Giangrandi_2023, Jockel_2024, Karkevandi_2022}.

\subsection{Kilonova AT2017gfo and short gamma-ray burst GRB170817A}
\label{sec:KN+GRB}

\begin{figure}
    \centering
    \includegraphics[width = \linewidth]{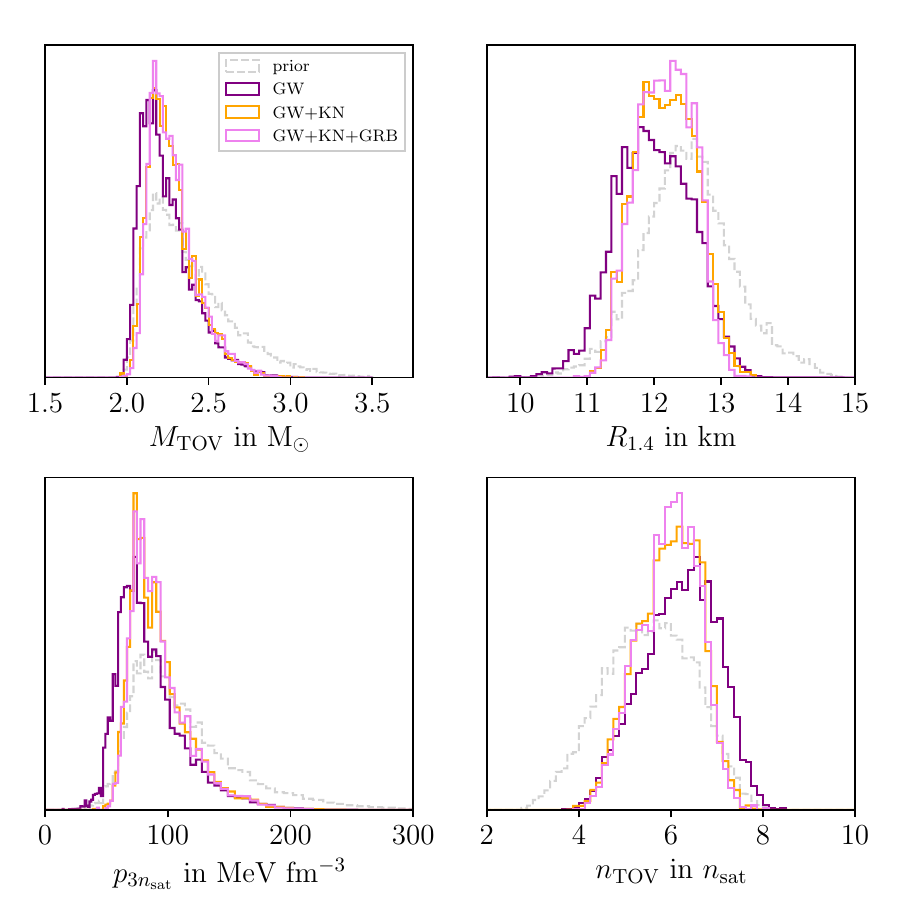}
    \caption{Posterior distributions of $R_{1.4}$, $\mtov$, $p_{3\nsat}$, and $n_{\text{TOV}}$ for inferences using only GW data (purple line), GW and KN data (orange line), and GW, KN, and GRB afterglow data (magenta line). Here, Bu2023 is the adapted KN model.}
    \label{fig:GW170817+EM}
\end{figure}

GW170817 was accompanied by different electromagnetic signals, namely the KN AT2017gfo~\citep{LIGO_AT2017gfo, Andreoni_2017, Coulter_2017, Lipunov_2017, Shappee_2017, Tanvir_2017, Utsumi_2017} and the short GRB170817A~\citep{GRB170817A, Goldstein_2017, Savchenko_2017} as well as its afterglow~\citep{Hallinan_2017, Alexander_2018, Margutti_2018, Ghirlanda_2019, Troja_2017, Avanzo_2018}. 
These electromagnetic counterparts allowed for the identification of the galaxy NGC 4993 as the signal's origin~\citep{Coulter_2017} and to place limits on the observation angle~\citep{Mooley_2018, Ghirlanda_2019, Mooley_2022}. 
The GRB likely originated from the launch of a relativistic jet~\citep{Ghirlanda_2019} and was observed $\sim \qty{1.7}{\second}$ after merger. 
The KN was fueled by pseudo-black-body emission from ejected material heated by the radioactive decay of heavy neutron-rich nuclei created in $r$-processes~\citep{Cowperthwaite_2017, Santos_2017}.
It was first detected $\sim \qty{11}{\hour}$ after the GW observation and continuously observed over the course of three weeks~\citep{Villar_2017}. The GRB afterglow was observed in the X-ray and radio after more than a week \citep{Troja_2017, Hallinan_2017} with continued observation over months \citep{Mooley_2018, Ghirlanda_2019}.

To use these electromagnetic signatures for inference of the EOS, one requires a model that links physical system parameters, such as the ejecta mass and velocity, to the emitted light curves. 
Several different models for the KN emission are available in the literature, e.g. from Refs.~\citep{Kasen_2017, Dietrich_2020, Anand_2021, Wollaeger_2021,Anand_2023}. 
For the present work, we employ the state-of-the-art KN model from Ref.~\citep{Anand_2023} (\mbox{Bu2023}) and an older version from Ref.~\citep{Dietrich_2020} (\mbox{Bu2019}). 
Both models are built with \textsc{possis}, a three-dimensional Monte Carlo radiative transfer code~\citep{Bulla_2019, Bulla_2023} in which the ejecta material is evolved through homologous expansion and the emitted photon packages are calculated from the temperature and opacity distributions. 
The \mbox{Bu2023} model uses five intrinsic parameters, namely the masses and velocities of the dynamical and wind ejecta as well as the dynamical ejecta's average electron fraction. 
On the other hand, the \mbox{Bu2019} model uses only dynamical and wind ejecta masses as well as the opening angle of the lanthanide-rich component.
The priors for these model parameters are listed in Table \ref{tab:EM_prior}. 
Compared to its predecessor, \mbox{Bu2023} profits from improved prescriptions of heating rates, thermalization efficiencies, and opacities in \textsc{POSSIS}; for further details we refer to Refs.~\citep{Bulla_2023, Anand_2023}. 
Since the computation time for one \textsc{POSSIS} light curve is on the order of hours and thus too large for the many likelihood evaluations required during sampling, the light curves for an arbitrary point in the parameter space are interpolated by a feedforward neural network over a fixed grid of \textsc{POSSIS} simulations~\citep{Almualla_2021, NMMA_2023}. 

Complementary, we model the light curve of the observed GRB afterglow with the package \textsc{afterglowpy}~\citep{Ryan_2020}. 
This model assumes a structured jet in the single shell approximation that is forward shocked with the ambient constant-density interstellar medium. 
It takes seven intrinsic parameters, listed together with their prior ranges in Table~\ref{tab:EM_prior}, to semianalytically determine the afterglow light curves at variable observation wavelengths. 
In this manner, we are able to predict the anticipated light curve given a certain set of model parameters $\vec{\theta}$ by combining the contribution from the GRB afterglow and KN model. 
Together with the observation angle and luminosity distance as the two observational parameters, we can directly deduce the expected AB~magnitudes $m^{f}(t_j, \vec{\theta} )$ for a wavelength filter $f$ at time $t_j$. Assuming a Gaussian error on the real magnitude measurements $m^f(t_j,d)$ with a statistical error $\sigma^{f}_{\text{stat}}(t_j)$ we set up the likelihood of the data at a given sample point $\vec{\theta}$ as
\begin{align}
    \log \mathcal{L}(\vec{\theta}|d) = -\frac{1}{2} \sum_{f,j} \frac{(m^{f}(t_j, \vec{\theta}) - m^f(t_j,d))^2}{(\sigma_{\text{syst}})^2+(\sigma^{f}_{\text{stat}}(t_j))^2} + {\rm const}\,.
    \label{eq:EM_likelihood}
\end{align}
We introduce the auxiliary systematic uncertainty $\sigma_{\text{syst}}$ to account for the systematic errors in the KN, and GRB afterglow models and set it conservatively to 1 mag following previous works~\citep{Heinzel_2021, NMMA_2023}.

To perform a full multimessenger analysis of the light-curve and GW data, we use \textsc{NMMA} to sample over the joint parameter space. 
The strain GW data are taken again from Ref.~\citep{GWOSC_2021} and the light-curve data are compiled from Refs.~\citep{Coughlin_2018, Troja_2019}. 
The joint parameter space includes the 16 parameters of the GW inference, 5 (respectively 3) parameters for the KN model, and 7 additional parameters for the GRB afterglow. 
To make full use of the multimessenger information in the data, we link the GW parameters to the parameters of the electromagnetic model. 
Specifically, we relate the dynamical ejecta masses $M_{\text{ej,dyn}}$ for the KN model to the GW parameters $M_1$, $M_2$ (the NS masses), and to the EOS via the following relation~\citep{Krueger_2020}:
\begin{align}
    \begin{split}
    \frac{M_{\text{ej,dyn}}}{10^{-3} \msun} =&M_1 \left[\frac{a}{C_1} + b\ \left( \frac{M_2}{M_1} \right)^n + c\ C_1\right] \\
    &+ (1 \leftrightarrow 2) + \alpha\,,
    \end{split}
\label{eq:dyn_eject}
\end{align}
with the compactness
\begin{align}
    C_j =& \frac{M_j}{R_j(M_j, \text{EOS})},  \qquad j \in \{1,2\}\,.
\end{align}
Here, the numerical coefficients $a=-9.3$, $b=114.2$, $c=-337.6$, and $n=1.5$ are fitted from numerical-relativity simulations, whereas $\alpha$ is drawn from a Gaussian distribution with mean 0 and standard deviation \qty{0.004}{\msun} as a fiducial parameter describing the error on the relation \citep{Krueger_2020}. 
Likewise, the wind ejecta $M_{\text{ej,wind}}$ for the KN model and the isotropic equivalent energy $E_0$ for the GRB afterglow model can be linked to the disk mass $M_{\text{disk}}$ that forms around the remnant after the merger:
\begin{align}
    M_{\text{ej,wind}} &= \zeta\ M_{\text{disk}}, \label{eq:wind_eject}\\
    E_0 &= \epsilon\ (1-\zeta)\ M_{\text{disk}}.\label{eq:iso_ener}
\end{align}
Here, we introduced $\zeta$, the fraction of the disk that gets unbound as wind, and $\epsilon$, the share of the remaining disk that is converted into jet energy. 
The former is assigned a uniform prior from 0 to 1, the latter is sampled log-uniformly from $10^{-7}$ up to 0.5. 
The disk mass itself can be determined from the total binary mass, the mass ratio $q$, and the EOS through the phenomenological relations~\citep{Dietrich_2020, Coughlin_2019, Agathos_2020}:
\begin{widetext}
\begin{align}
\log_{10}\left( \frac{M_{\text{disk}}}{\msun} \right)&= a \left[ 1+ b\,\tanh\left(\frac{c - (M_1+M_2)/M_{\text{threshold}}}{d} \right) \right]\,,
\end{align}
\end{widetext}
where the prompt collapse threshold mass is given by~\citep{Bauswein_2013}
\begin{align}
M_{\text{threshold}} &= \left( -3.6 \frac{\mtov(\text{EOS})}{R_{1.6}(\text{EOS})} + 2.38 \right) \mtov(\text{EOS})\,,
\end{align}
with
\begin{align}
\begin{split}
&a = -1.725 -2.337\xi,\quad b = -0.564 -0.437\xi, \\ 
&c = 0.953, \quad d = 0.057, \quad \xi = \frac{1}{2} \tanh\left(\beta  (q - q_{\text{trans}})\right), \\ 
&\beta = 5.879, \qquad q_{\text{trans}} = 0.886\,.
\end{split}
\end{align}
The luminosity distance and inclination angle are naturally shared by the GW and electromagnetic models. 

For sampling, we combine both likelihoods for the GW data and electromagnetic light curve by simply adding the log-likelihoods:
\begin{align}
    \log \mathcal{L}(\vec{\theta}|d) = \log \mathcal{L}_{\text{GW}}(\vec{\theta}|d) + \log \mathcal{L}_{\text{EM}}(\vec{\theta}|d).
\end{align}
Thus, the likelihoods $\mathcal{L}_{\text{GW}}(\vec{\theta}|d)$ taken from Eq.~(\ref{eq:GW_likelihood}) and the electromagnetic likelihood given through Eq.~(\ref{eq:EM_likelihood}) are taken as independent, but some of the parameters are linked on the prior level. 
To avoid prohibitively large computation times, we restrict the number of live points to 1024.

The two KN models perform differently, as evident by variations in the posterior estimates for certain parameters.
Meanwhile the differences with regards to the EOS are fairly mild. 
Statistically, the \mbox{Bu2019} model is preferred with a Bayes factor $\ln \mathcal{B}^{\text{Bu2019}}_{\text{Bu2023}}$ of 12.73 for the inference with GW data and KN, and 6.12 when the GRB afterglow is added. 
However, the luminosity distance and inclination are not well estimated in the GW+KN+GRB inference with \mbox{Bu2019}.
For the present work, we therefore quote our results with respect to the \mbox{Bu2023} model unless stated otherwise. 
We discuss the performance of the two models in more detail in Appendix \ref{app:KN_model_comp}.

In Fig.~\ref{fig:GW170817+EM} we compare the resulting posterior distributions on EOS quantities, when we perform a parameter inference with the data from GW170817 and KN AT2017gfo, and the joint inference combining GW170817, AT2017gfo, and GRB afterglow.
Adding the electromagnetic signals leads to the rejection of the softest EOS still contained in the posterior of the GW-only inference.
Thus, we are able to place narrower limits on $R_{1.4}$ and $n_{\text{TOV}}$. 
Compared to previous works, we find slightly larger statistical uncertainties in the NS radii, e.g.; Ref.~\citep{Pang_2023} quotes $R_{1.4} = 11.86^{+0.41}_{-0.53}~(11.98^{+0.35}_{-0.40})$\,km for their GW+KN (GW+KN+GRB) inference, whereas we find $R_{1.4} = 12.19^{+0.73}_{-0.73}~(12.19^{+0.71}_{-0.63})$\,km, all values quoted here at 90\% credibility. 
We attribute this difference to our larger prior EOS set, since we implemented the constraint from \chiEFT\ more conservatively.

In our analysis of the electromagnetic counterparts to GW170817, systematic uncertainties arise from many possible sources.
They are partially accounted for during sampling. 
For example, we include ancillary fit-error parameters to acknowledge biases that arise from the phenomenological relations used in Eqs.~(\ref{eq:dyn_eject}) and (\ref{eq:wind_eject}). 
Likewise, uncertain modeling assumptions, such as heavy-nuclei opacities~\citep{Tanaka_2020}, nuclear heating rates~\citep{Zhu_2022}, local thermalization~\citep{Barnes_2016, Kasen_2019}, or idealized geometries in the KN models~\citep{Heinzel_2021}, as well as, for example, the lack of reverse shocks and self-Compton emission in the GRB afterglow model, may affect the results. 
Though they are somewhat addressed by setting $\sigma_{\text{syst}}$ to 1 mag in Eq.~(\ref{eq:EM_likelihood})~\citep{Heinzel_2021}, the differences in the posteriors of the two KN models, as discussed in Appendix~\ref{app:KN_model_comp}, still exemplify current uncertainties in KN inference arising from a lack of knowledge about the relevant microphysical processes.

\subsection{Gravitational-wave signal GW190425}
\label{sec:GW190425}

\begin{figure}
   \centering
   \includegraphics[width = \linewidth]{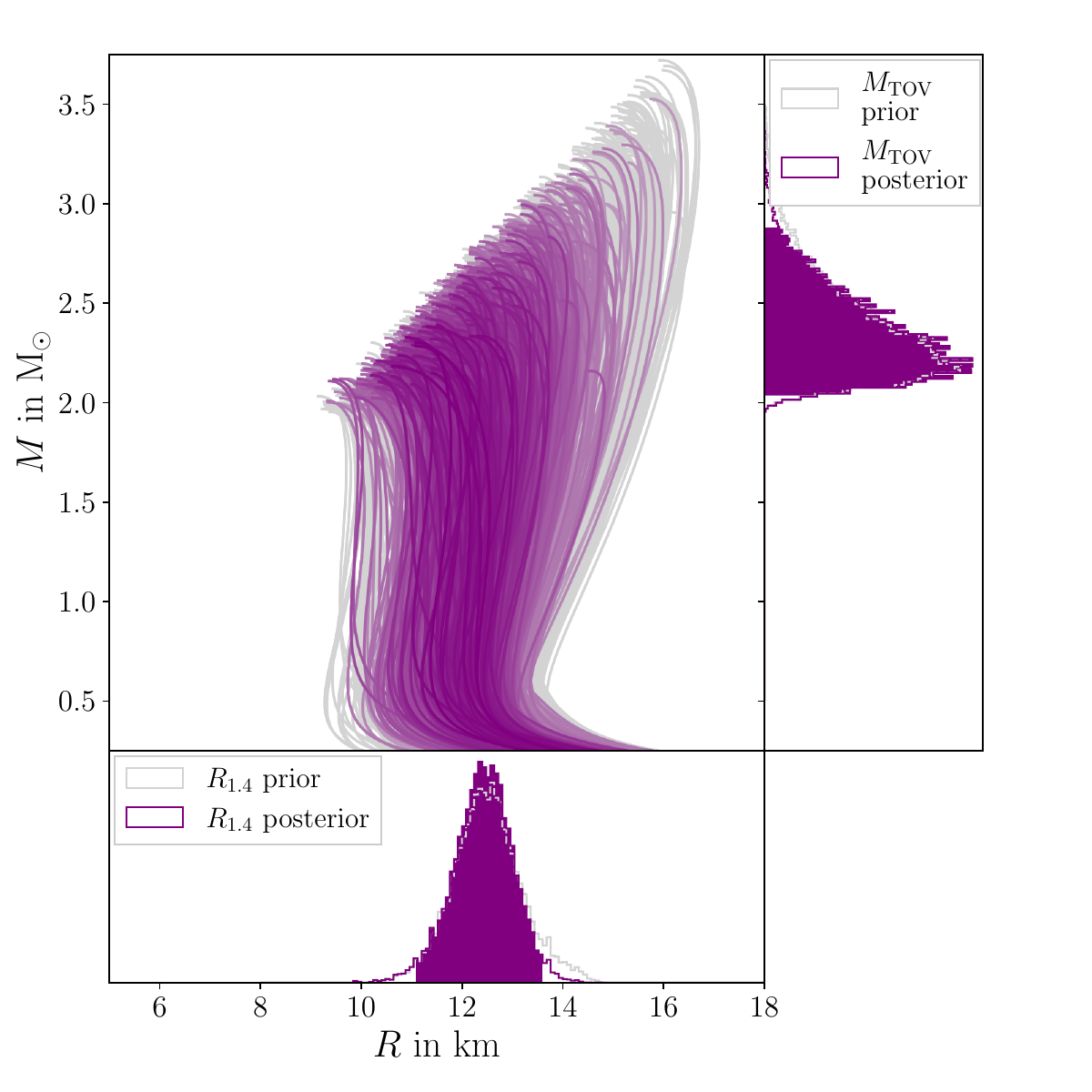}
   \includegraphics[width = \linewidth]{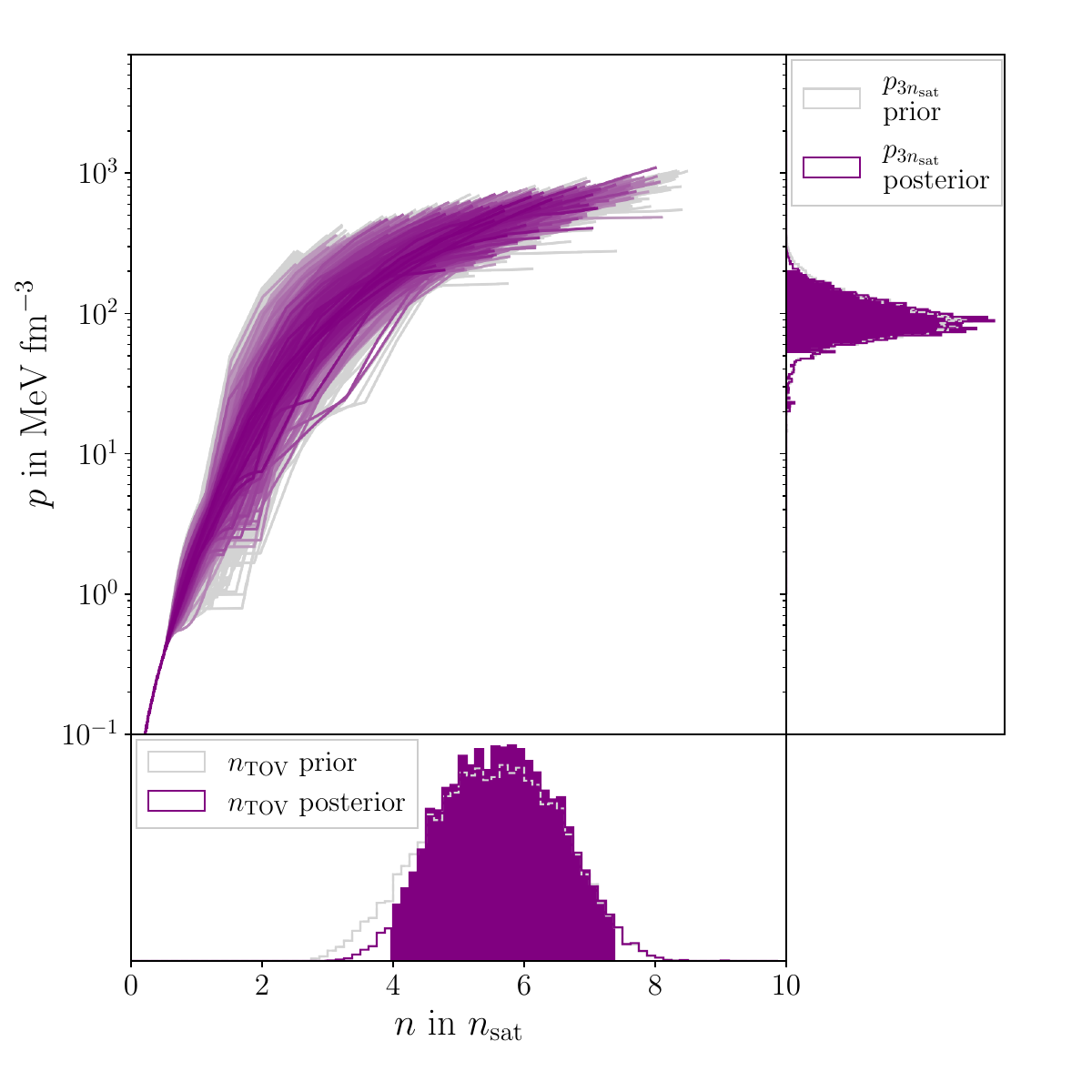}
   \caption{EOS inference based on GW190425. Arrangement and color coding as in Fig.~\ref{fig:GW170817}.}
   \label{fig:GW190425}
\end{figure}

\begin{figure}
    \centering
    \includegraphics[width = \linewidth]{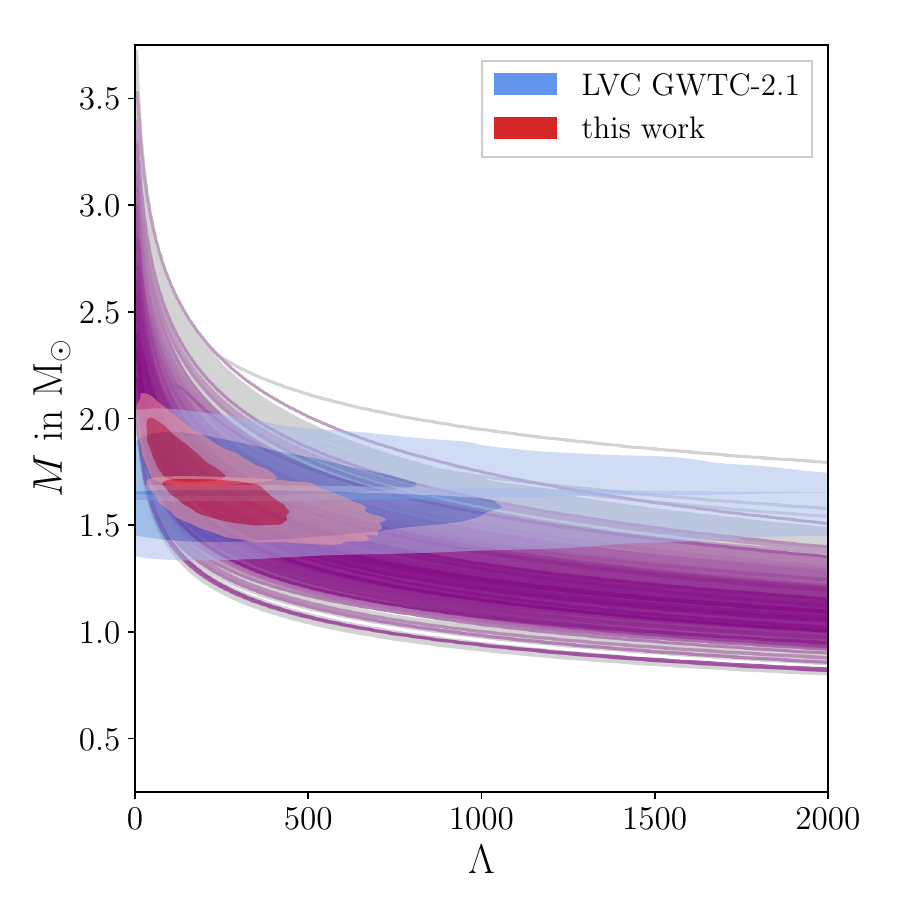}
    \caption{$M$-$\Lambda$ curves of our selected EOS set, color coded according to inference based on the measurement of GW190425. The red and blue credible regions are as in Fig. \ref{fig:tidal_GW170817}, with the blue contours from LVC GWTC-2.1 posterior~\citep{LIGO_GWTC_2.1}.}
    \label{fig:tidal_GW190425}
\end{figure}

GW190425 is the second convincing candidate for a GW signal from a BNS merger~\citep{LIGO_GW190425} and was observed through the LIGO detectors \citep{LIGO_advanced_detector_2015}. 
However, GW190425 was not accompanied by a firm detection of an electromagnetic signal~\citep{Coughlin_2019A, Lundquist_2019, Hosseinzadeh_2019} and had an overall weaker SNR of 12.4, compared to 33 for GW170817.
As in Sec.~\ref{sec:GW170817} for GW170817, we perform parameter inference of GW190425 using \textsc{NMMA} with the same \mbox{$\mathtt{IMRPhenomXP\_NRTidalv3}$} waveform model, sample again directly over the EOS candidates and reweigh the result to account for selection effects. 
The priors are listed in Table~\ref{tab:GW_prior} and the strain data were taken from~Ref.~\citep{GWOSC_2021}. 
We note that the primary component in GW190425 is significantly heavier than in GW170817, and thus the sampling of $M_1$ may occasionally reach above the TOV mass of the sampled EOS. 
In this case, \textsc{NMMA} sets $\Lambda_1$ to 0 and hence assumes an NSBH. 
However, the EOS candidates we sample over all fulfill $\mtov\gtrsim\qty{2}{\msun}$ and the source frame mass of the primary component is $M_1<\qty{1.94}{\msun}$ at 95\% credibility. 
Hence, no NSBH sample point is present in our posterior files.

We find no further constraints on the EOS set, as our analysis simply recovers the input prior, as can be seen in Fig.~\ref{fig:GW190425}. 
The outcome for the tidal deformabilities is shown in Fig.~\ref{fig:tidal_GW190425} and compared to GWTC-2.1 data of Ref.~\citep{LIGO_GWTC_2.1}. 
Like for GW170817, our posterior estimates on $\Lambda_{1,2}$ are tighter compared to the analysis of the corresponding LVC GWTC catalog estimates~\citep{LIGO_GWTC_2.1}, because we sample over an EOS set. 
While it appears as if indeed some stiffer EOS candidates are outside of the credible region of the tidal constraint, this impression is misleading since we sample our prior uniformly in mass ratio and chirp mass and use the weighted EOS set to determine the tidal deformabilities. 
In that way, the mass measurement of the stars in GW190425 affects the inference of $\Lambda_{1,2}$, but no real information about the EOS is recovered~\citep{Kastaun_2019}.
In fact, Fig.~\ref{fig:GW190425} confirms that the posterior distribution on $R_{1.4}$ and other EOS-derived quantities does not change significantly compared to the prior. 
A very weak tendency toward softer EOSs with smaller radii seems noticeable, but this effect may also arise from undersampling the EOS space in regions with low prior likelihood.

Overall, the larger NS masses and an only moderate SNR make GW190425 unsuitable for rigorous constraints on tidal deformabilities~\citep{Kastaun_2019, LIGO_GW190425, Raaijmakers_2021}. 
For similar reasons, we do not perform inferences for the black hole-neutron star merger candidates \mbox{GW200105\_162426} and \mbox{GW200115\_042309}~\citep{LIGO_BHNS}.

\subsection{Gamma-ray burst GRB211211A}

GRB211211A is one of the closest GRBs observed so far. 
It was first detected on 11 December 2021 at 13:09:59 (UTC) by the Burst Alert Telescope of the Swift Observatory. 
Subsequent observations in the optical and near-infrared were reported, e.g., in Refs.~\citep{Rastinejad_2022, Troja_2022}. 
Its particularly long duration of \qty{51.37(80)}{\second}~\citep{Rastinejad_2022} suggests that it could have originated from the collapse of a massive star, as evidence exists that long GRBs are linked to supernovae~\citep{Woosley_2006}.
However, multiple studies have shown that the observed electromagnetic emission in the optical and near-infrared is best described when invoking a KN associated with a compact binary merger~\citep{Rastinejad_2022, Troja_2022, Yang_2022, Mei_2022, Kunert_2024}. 
The authors of Ref.~\citep{Kunert_2024} performed a large set of multiwavelength analyses for GRB211211A assuming four different scenarios, namely, a BNS merger, an NSBH merger, a core-collapse supernova, and an $r$-process-enriched core-collapse supernova. 
Their model comparison revealed that GRB211211A is best fit when jointly inferring the data with the GRB afterglow model of Refs.~\citep{van_Eerten_2010, Ryan_2020} and KN contribution modeled as in Ref.~\citep{Kasen_2017}. Therefore, the likeliest explanation for this GRB is a BNS system as progenitor.

Assuming for the sake of argument that GRB211211A originated from a BNS merger, we explore what constraints it can provide on the EOS. 
The idea is to use the ejecta posterior from Ref.~\citep{Kunert_2024}  with the highest evidence (cf. BNS-GRB-M$^{\rm{Kasen}}_{\rm{top}}$ in Tables 1 and~2 in Ref.~\citep{Kunert_2024}) as likelihood to sample over the EOS and an agnostic prior for the BNS parameters.
To that end, we employ the relations of Eqs.~(\ref{eq:dyn_eject}) and (\ref{eq:wind_eject}) to link the EOS candidates and ejecta masses. 
The result of that calculation is then compared to the marginalized posterior $P(M_{\text{ej,dyn}}, M_{\text{ej,wind}} |d_{\text{EM}})$ obtained from the inference of GRB211211A light-curve data $d_{\text{EM}}$~\citep{Kunert_2024}.
In practice, we sample uniformly over the chirp mass $\mathcal{M}$ in the range of \qtyrange{0.7}{3.7}{\msun}, and mass ratio $q$ on the interval \qtyrange{0.125}{1}{}, and additionally sample over the uncertainty parameters $\zeta$ and $\alpha$ in the phenomenological relations; hence our parameter space consists of $\mathcal{M}$, $q$, $\zeta$, $\alpha$.
Then, the full likelihood reads
\begin{align}
\begin{split}
    \mathcal{L}(\mathcal{M},q,\zeta,\alpha,\text{EOS} |d_{\text{EM}}) = P(M_{\text{ej,dyn}}, M_{\text{ej,wind}} |d_{\text{EM}})\,.
\end{split}
\end{align}
The light-curve posterior on the right-hand side is evaluated numerically through the kernel density estimation from discrete samples of Ref.~\citep{Kunert_2024}.
This approach is implemented in \textsc{NMMA}, using \textsc{pymultinest} for nested sampling~\citep{Buchner_2014}. 
Although our analysis finds a preference for a chirp mass of $\mathcal{M} =1.39^{+0.59}_{-0.67}\,\msun$ and a mass ratio of $q = 0.82^{+0.17}_{-0.15}$, corresponding to $M_1=1.76^{+0.79}_{-0.86}\,\msun$ and $M_2=1.46^{+0.52}_{-0.70}\,\msun$, we find no posterior constraint on the EOS, as the impact of the EOS on the light curve is mitigated by the necessary, but unconstrained $\zeta$ parameter during sampling.
Given the additional systematic uncertainty from the KN and GRB afterglow model (see Sec.~\ref{sec:KN+GRB}) and the unknown nature of the source, GRB211211A thus provides no valuable information on the EOS.
We expect similar findings for future GRB detections without accompanying GW signal.

\subsection{Postmerger constraints from GW170817}
\label{subsec:postmerger}
\begin{figure}
    \centering
    \includegraphics[width = \linewidth]{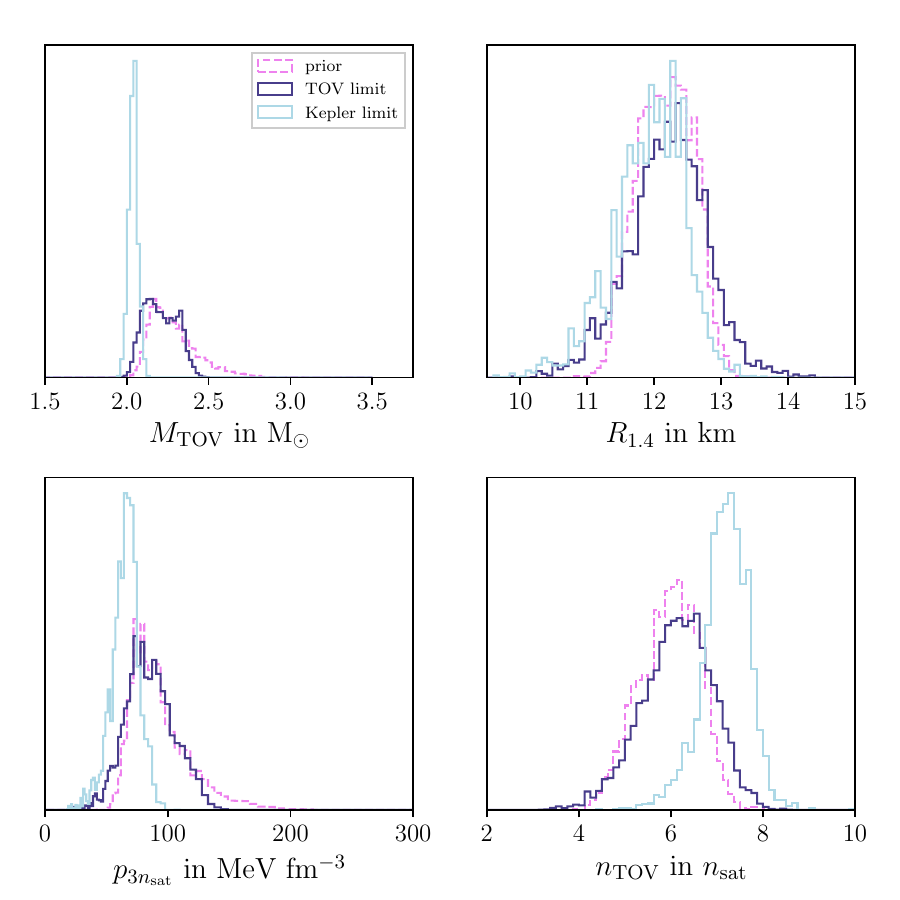}
    \caption{Posterior distributions of $R_{1.4}$, $\mtov$, $p_{3\nsat}$, and $n_{\text{TOV}}$ based on the postmerger constraint from the collapse of GW170817's remnant. The dashed magenta line represents the posterior from the joint GW+KN+GRB inference of GW170817 that served as prior for imposing the postmerger collapse criterion. The light blue lines display the posterior when choosing $M_{\text{coll},b} = M_{\text{TOV},b}$, the dark blue line for $M_{\text{coll},b} = M_{\text{Kep},b}$.}
    \label{fig:postmerger}
\end{figure}

The postmerger fate of the BNS remnant in GW170817 provides information on the EOS, complementary to that obtained from the tidal deformabilities during inspiral. 
Various arguments, such as the lack of spin-down luminosity~\citep{Pooley_2018} and the launch of a relativistic jet for the GRB~\citep{Ghirlanda_2019, Putten_2023}, support the hypothesis that a black hole was created in the aftermath of GW170817. 
At the same time, based on the presence of an electron-rich, blue component in KN AT2017gfo, there is general agreement that the remnant did not collapse immediately to a black hole, but instead formed a spinning (supramassive or hypermassive) NS for a brief intermediate period~\citep{Margalit_2017, Bauswein_2017, Cowperthwaite_2017, Gill_2019}.
Numerical-relativity simulations support this picture~\citep{Shibata_2019, Nedora_2021}. 
The remnant slowed down until angular momentum became insufficient to sustain the star, triggering the collapse to a black hole. 
The exact spin-down timescale remains a matter of debate, as there are two possibilities (excluding prompt collapse). GW170817 could have formed a hypermassive NS which was sustained through differential rotation and collapsed on the order of milliseconds. 
Alternatively, it might have created a supramassive, rigidly spinning star that collapsed on a longer timescale. 
The moderate strength of the blue component as well as the released structured jet favor the first scenario~\citep{Margalit_2017, Gill_2019, Murguia_2021}. 

In any case, to eventually produce a black hole, the remnant mass needs to exceed the threshold mass $M_{\text{coll}}$. 
The exact value for this threshold depends on the rotational state of the remnant and the EOS, but will generally scale with the TOV mass; i.e., the higher the TOV mass, the less likely it would appear that for a given remnant mass a black hole was created. 
Hence, by rejecting any EOS that predicts a threshold mass lower than the actual remnant mass, the black-hole formation hypothesis can place upper limits on the TOV mass. 
Such limits have been proposed, for example, in Refs.~\citep{Margalit_2017, Rezolla_2018, Ai_2020, Annala_2022, Ai_2023}. 
In Ref.~\citep{Bauswein_2017}, the authors use the total mass of GW170817 as a lower limit to the threshold mass for prompt collapse to infer a lower limit on the radii.

To perform Bayesian inference on our EOS set with this constraint, we combine the approaches of Refs.~\citep{Annala_2022, Ai_2023}. We use our joint parameter inference of GW170817, AT2017gfo, and the GRB170817A afterglow to determine the remnant mass and compare this value to the threshold $M_{\text{coll}}$.
During the merger, gravitational mass is not conserved, as some share is radiated away in the form of gravitational waves.
Instead, the mass balance for the remnant mass $M_{\text{rem}}$ has to be set up in terms of baryonic masses, which we subscript with $b$. The mass balance reads
\begin{align}
    M_{\text{rem},b} = M_{1,b} + M_{2,b} - M_{\text{ej,dyn}, b} - M_{\text{disk}, b}\,.
    \label{eq:remnant_mass}
\end{align}
To determine the posterior likelihood of one of our EOSs, we sample the relevant parameters $M_1$, $M_2$, $M_{\text{ej,dyn}}$, $M_{\text{disk}}$, and EOS from the joint GW+KN+GRB posterior of Sec.~\ref{sec:KN+GRB}. 
The EOS of the current sample point is used to convert the gravitational masses to baryonic masses. 
Ejecta and disk mass are already expressed in terms of the baryonic mass. 
The likelihood for such a sample point is then
\begin{align}
\begin{split}
    \mathcal{L}(M_1, M_2, M_{\text{ej,dyn}}, M_{\text{disk}}, \text{EOS}|\text{BH collapse}) \\ = \begin{cases}
        1 \qquad \text{if $M_{\text{rem},b} > M_{\text{coll},b}$}\\
        0 \qquad \text{else,}
    \end{cases}
\end{split}
\label{eq:collapse}
\end{align}
where the baryonic remnant mass on the right-hand side is calculated with Eq.~(\ref{eq:remnant_mass}) and the threshold for collapse $M_{\text{coll},b}$ is determined from the EOS. The sampling was implemented with \textsc{pymultinest}~\citep{Buchner_2014}.

There are two natural choices for the threshold mass, namely the Kepler mass limit $M_{\text{Kep}}$, i.e., the maximum mass that can be supported by an EOS when the star rotates rigidly, and more conservatively the TOV mass $\mtov$. 
Choosing the TOV mass as threshold is more conservative because it is a valid cutoff even if GW170817's remnant was a only supramassive NS, while selecting the Kepler mass limit implies that the collapse was precipitated by a hypermassive star. 
On the other hand, the multimessenger evidence of GW170817 hints toward a short-lived hypermassive NS as remnant~\citep{Margalit_2017, Gill_2019, Murguia_2021}, and thus, both choices are justified.
Obtaining the baryonic TOV mass from the EOS is straightforward, but determining the Kepler mass limit requires the construction of rotation sequences of relativistic stars. 

This is achieved with the \textsc{rns} code~\citep{Stergioulas_1995, Nozawa_1998}, which we use to determine the baryonic Kepler mass limit $M_{\text{Kep}, b}$ for each of our EOS candidates. 
However, the calculation with \textsc{rns} will carry a numerical error that depends on the stiffness and presence of phase transitions in the specific EOS. 
Therefore, for each EOS, we ran \textsc{rns} twice, one time with standard and the other time with high resolution. The difference between the two values is used as uniform error on the result from the high resolution run. 
Accordingly, the likelihood in Eq.~\eqref{eq:collapse} is then given by the cumulative distribution over that uniform range for $M_{\text{Kep}, b}$. 
The error is typically on the order of \qtyrange{0.01}{0.03}{\msun}, but for some difficult EOS goes up to \qty{0.08}{\msun}. 
For about 700 of our EOS candidates, \textsc{rns} fails to converge at the higher resolution. For these cases we use a simple fit relation derived from the successful calculations,
\begin{align}
    M_{\text{Kep}, b} = 1.737\ \mtov -0.698\,
    \label{eq:Kepler_fit}
\end{align}
with a Gaussian error of \qty{0.06}{\msun}.

In Fig.~\ref{fig:postmerger}, we compare the impact of the postmerger constraint on our EOS candidate set when setting the threshold to the TOV mass or the Kepler limit. 
For the former we find $\mtov = {2.20}^{+0.19}_{-0.16}$\,$\msun$, whereas the latter yields $\mtov = {2.04}^{+0.06}_{-0.06}$\,$\msun$.
In both cases, the posterior is shifted to softer EOS. 
Especially when $M_{\text{coll}}$ is set to the Kepler limit, the posterior largely resides in regions with relatively low prior likelihood, since the criterion in Eq.~\eqref{eq:collapse} requires EOSs with a TOV mass of around \qty{2}{\msun} and these are not considered very likely in the GW+KN+GRB posterior. 
This causes the posterior on $R_{1.4}$ and other quantities to widen again compared to Sec.~\ref{sec:KN+GRB}. 
The resampling also rejects any disk mass above \qty{0.1}{\msun} and $M_{\text{ej,dyn}}$ superseding \qty{0.01}{\msun}, as these values would imply a small remnant mass that could only collapse if the TOV mass was significantly below \qty{2}{\msun}.

Although there are multiple indications that point toward the collapse of a hypermassive remnant in GW170817, there is only a cautious consensus and remaining alternative scenarios would invalidate the deduced strong upper limit on $\mtov$~\citep{LIGO_2020}. 
Therefore, we adopt the EOS posterior obtained by setting $M_{\text{coll}}$ to the TOV limit as the default when combining different constraints in the sections below. 
If the assumption that the GW170817 remnant collapsed to a black hole is accurate, the derived limits can be seen as relatively robust. 
Systematic uncertainty mainly carries over from the GW and light-curve data models, as discussed in Secs.~\ref{sec:GW170817} and~\ref{sec:KN+GRB}. In particular, the disk mass $M_{\text{disk}}$ is relatively unconstrained. 
When relying on \textsc{rns} to determine $M_{\text{Kep}}$, systematic errors could be introduced by the numerical errors or by the fit in Eq.~\eqref{eq:Kepler_fit}. 
In Appendix~\ref{app:Kepler_limit} we thus show how the results drawn from this particular constraint remain the same if the Kepler mass limits are determined through quasiuniversal relations.
We point out that the method for GW170817's postmerger fate presented here combines all data available from the GW170817 event, since the postmerger constraint is implemented on top of the GW+KN+GRB posterior. 
In future work, one could also incorporate this directly into the inference of the raw data.

\section{Comparing and combining the results}
\label{sec:Comparing_and_combining}
\begin{table*}[t]
\renewcommand{\arraystretch}{1.1}
\caption{Summary of the individual constraints. 
We describe the measured quantities, mutual dependencies, and  the credible intervals for EOS  quantities at the 95\% level. 
The Kullback-Leibler divergence of each posterior is shown in brackets directly below each credible interval.\\}
\label{tab:summary}
\begin{tabular}{>{\centering\arraybackslash} p {2.5 cm} >{\centering\arraybackslash} p {2.5 cm} >{\centering\arraybackslash} p {2.5 cm} >{\centering\arraybackslash} p { 1.5 cm} >{\centering\arraybackslash} p {1.5 cm} >{\centering\arraybackslash} p {1.5 cm} >{\centering\arraybackslash} p {1.5 cm} >{\centering\arraybackslash} p {1.5 cm} }
\toprule
\toprule
Constraint & Measurement & Dependencies & $R_{1.4}$\ \ \ \ (km) & $\mtov$ ($\msun$) & $p_{3\nsat}$ (MeV\,fm$^{-3}$) & $n_{\text{TOV}}$ ($\nsat$) & $p_{\text{TOV}}$ (MeV\,fm$^{-3}$) \\
\midrule
Prior & \dots & \dots & $13.49^{+ 1.61}_{-3.83}$ & $2.17^{+ 0.97}_{-1.06}$ & $90^{+ 175}_{-66}$ & $5.44^{+ 7.95}_{-2.10}$ & $425^{+ 752}_{-360}$ \\
\midrule 
$\chi$EFT & Theoretical & \dots & $12.11^{+1.96}_{-3.39}$ & $2.05^{+1.07}_{-1.16}$ & $69^{+186}_{-53}$ & $6.51^{+10.70}_{-3.11}$ & $519^{+983}_{-455}$ \\
  & &  & (0.561)  & (0.022)  & (0.093)  & (0.13)  & (0.093)  \\
\rule{0pt}{4ex}
pQCD & Theoretical & \dots & $13.47^{+1.62}_{-3.78}$ & $2.14^{+0.92}_{-1.00}$ & $88^{+154}_{-62}$ & $5.46^{+6.67}_{-2.07}$ & $403^{+679}_{-341}$ \\
  & &  & (0.0)  & (0.006)  & (0.011)  & (0.006)  & (0.007)  \\
\rule{0pt}{4ex}
CREX & $E_{\text{sym}}$, $L_{\text{sym}}$ & \dots & $12.33^{+1.74}_{-3.59}$ & $2.09^{+1.01}_{-1.16}$ & $76^{+175}_{-61}$ & $6.19^{+10.77}_{-2.74}$ & $497^{+931}_{-430}$ \\
  & &  & (0.454)  & (0.009)  & (0.059)  & (0.081)  & (0.057)  \\
\rule{0pt}{4ex}
PREX-II & $E_{\text{sym}}$, $L_{\text{sym}}$ & \dots & $13.44^{+1.18}_{-3.43}$ & $2.17^{+0.96}_{-1.07}$ & $90^{+173}_{-62}$ & $5.50^{+7.12}_{-2.12}$ & $432^{+714}_{-366}$ \\
  & &  & (0.141)  & (0.001)  & (0.002)  & (0.003)  & (0.002)  \\
\rule{0pt}{4ex} 
$^{208}$Pb dipole & $E_{\text{sym}}$, $L_{\text{sym}}$ & \dots & $13.02^{+1.37}_{-3.62}$ & $2.13^{+0.98}_{-1.11}$ & $84^{+174}_{-63}$ & $5.76^{+8.62}_{-2.34}$ & $457^{+809}_{-392}$ \\
  & &  & (0.192)  & (0.002)  & (0.008)  & (0.017)  & (0.01)  \\
\rule{0pt}{4ex}
HIC & $e_{\text{sym}}(n)$ & \dots & $13.34^{+1.69}_{-3.53}$ & $2.16^{+0.96}_{-1.06}$ & $88^{+172}_{-62}$ & $5.52^{+7.54}_{-2.14}$ & $431^{+731}_{-367}$ \\
  & &  & (0.017)  & (0.002)  & (0.003)  & (0.002)  & (0.001)  \\
\midrule 
Heavy pulsars & $M$ & \dots & $13.70^{+1.41}_{-2.17}$ & $2.35^{+0.73}_{-0.29}$ & $111^{+140}_{-49}$ & $5.15^{+1.89}_{-1.66}$ & $435^{+330}_{-259}$ \\
 & &  & (0.069)  & (0.466)  & (0.279)  & (0.217)  & (0.128)  \\
\rule{0pt}{4ex}
Black Widow & $M$ & \dots & $13.87^{+1.30}_{-2.03}$ & $2.52^{+0.73}_{-0.45}$ & $136^{+156}_{-62}$ & $4.73^{+1.82}_{-1.50}$ & $423^{+274}_{-244}$ \\
J0952-0607 & &  & (0.116)  & (0.529)  & (0.433)  & (0.315)  & (0.174)  \\
\rule{0pt}{4ex}
NICER & $M$, $R$ & \dots & $13.17^{+1.65}_{-2.24}$ & $2.16^{+0.83}_{-0.71}$ & $89^{+143}_{-46}$ & $5.62^{+4.43}_{-1.91}$ & $434^{+480}_{-324}$ \\
J0030+0451 & &  & (0.102)  & (0.051)  & (0.054)  & (0.054)  & (0.028)  \\
\rule{0pt}{4ex}
NICER & $M$, $R$ & Radio timing & $13.39^{+1.57}_{-1.72}$ & $2.34^{+0.65}_{-0.32}$ & $107^{+125}_{-40}$ & $5.34^{+1.63}_{-1.61}$ & $460^{+279}_{-262}$ \\
J0740+6620 & &J0740+6620 & (0.111)  & (0.459)  & (0.319)  & (0.277)  & (0.182)  \\
\rule{0pt}{4ex}
qLMXBs & $M$, $R$ & \dots & $12.97^{+1.77}_{-2.54}$ & $2.17^{+0.86}_{-0.82}$ & $88^{+154}_{-52}$ & $5.75^{+5.18}_{-2.13}$ & $461^{+540}_{-354}$ \\
  & &  & (0.135)  & (0.025)  & (0.021)  & (0.041)  & (0.031)  \\
\rule{0pt}{4ex}
HESS & $M$, $R$ & \dots & $10.99^{+2.04}_{-2.73}$ & $1.74^{+0.91}_{-0.79}$ & $42^{+113}_{-24}$ & $8.59^{+10.13}_{-3.99}$ & $654^{+1048}_{-558}$ \\
J1731-347 & &  & (1.276)  & (0.34)  & (0.621)  & (0.659)  & (0.369)  \\
\rule{0pt}{4ex}
X-ray burster & $M$, $R$ & \dots & $12.62^{+1.49}_{-0.87}$ & $2.17^{+0.49}_{-0.44}$ & $85^{+68}_{-30}$ & $6.09^{+2.11}_{-1.31}$ & $512^{+291}_{-305}$ \\
4U 1702-429 & &  & (0.709)  & (0.346)  & (0.434)  & (0.543)  & (0.224)  \\
\rule{0pt}{4ex}
X-ray burster & $M$, $R$ & qLMXBs & $12.48^{+2.21}_{-2.34}$ & $2.09^{+0.78}_{-0.66}$ & $75^{+130}_{-40}$ & $6.33^{+4.80}_{-2.34}$ & $507^{+549}_{-356}$ \\
J1808.8-3658 & &  & (0.281)  & (0.083)  & (0.074)  & (0.153)  & (0.102)  \\
\midrule 
GW170817 & $M$, $\Lambda$ & $\chi$EFT, pQCD, & $11.98^{+1.07}_{-1.09}$ & $2.21^{+0.45}_{-0.18}$ & $80^{+81}_{-32}$ & $6.25^{+1.39}_{-1.70}$ & $562^{+293}_{-307}$ \\
  & &Heavy pulsars & (0.312)  & (0.183)  & (0.26)  & (0.258)  & (0.126)  \\
\rule{0pt}{4ex}
GW170817+KN & $M$, $\Lambda$, $M_{\text{ej}}$ & $\chi$EFT, pQCD, & $12.18^{+0.84}_{-0.80}$ & $2.25^{+0.42}_{-0.19}$ & $88^{+73}_{-27}$ & $6.01^{+1.16}_{-1.44}$ & $528^{+257}_{-275}$ \\
  & &Heavy pulsars, GW170817 & (0.256)  & (0.151)  & (0.188)  & (0.208)  & (0.066)  \\
\rule{0pt}{4ex}
GW170817 & $M$, $\Lambda$, $M_{\text{ej}}$, $E_0$ & $\chi$EFT, pQCD, & $12.18^{+0.73}_{-0.78}$ & $2.26^{+0.40}_{-0.17}$ & $88^{+73}_{-26}$ & $6.01^{+1.09}_{-1.38}$ & $535^{+238}_{-272}$ \\
+KN+GRB & &Heavy pulsars, GW170817 & (0.304)  & (0.168)  & (0.201)  & (0.236)  & (0.091)  \\
\rule{0pt}{4ex}
GW190425 & $M$, $\Lambda$ & $\chi$EFT, pQCD, & $12.44^{+1.14}_{-1.35}$ & $2.31^{+0.56}_{-0.26}$ & $100^{+104}_{-46}$ & $5.64^{+1.72}_{-1.64}$ & $483^{+318}_{-266}$ \\
  & & Heavy pulsars & (0.029)  & (0.018)  & (0.016)  & (0.026)  & (0.005)  \\
\rule{0pt}{4ex}
GRB211211A & $M_{\text{ej}}$, $E_0$ & $\chi$EFT, pQCD, & $12.53^{+1.52}_{-1.33}$ & $2.35^{+0.69}_{-0.29}$ & $106^{+129}_{-50}$ & $5.45^{+1.80}_{-1.89}$ & $466^{+321}_{-258}$ \\
  & &Heavy pulsars & (0.002)  & (0.007)  & (0.005)  & (0.004)  & (0.002)  \\
\rule{0pt}{4ex}
GW170817 & $M_{\text{TOV}}$ & GW170817 & $12.22^{+1.12}_{-1.26}$ & $2.20^{+0.19}_{-0.16}$ & $84^{+42}_{-35}$ & $6.17^{+1.36}_{-1.69}$ & $520^{+314}_{-330}$ \\
postmerger & &+KN+GRB & (0.525)  & (0.235)  & (0.156)  & (0.12)  & (0.133)  \\
\rule{0pt}{4ex} GW170817 & $M_{\text{TOV}}$, $M_{\text{Kep}}$ & GW170817 & $11.98^{+0.86}_{-1.31}$ & $2.04^{+0.06}_{-0.06}$ & $66^{+20}_{-28}$ & $7.22^{+0.88}_{-1.28}$ & $658^{+261}_{-347}$ \\
postmerger Kepler limit & &+KN+GRB & (0.74)  & (3.377)  & (1.258)  & (2.208)  & (0.621)  \\
\bottomrule 
\end{tabular}
\end{table*}
\begin{table}[t]
\renewcommand{\arraystretch}{1.5}
\caption{Summary of the three combination sets. 
We provide the credible intervals for EOS quantities at 95\% credibility.\\}
\label{tab:set_final}
\begin{tabular}{>{\centering\arraybackslash} p {1.7 cm}  >{\raggedright\arraybackslash} p { 2 cm} >{\raggedright\arraybackslash} p {2 cm} >{\raggedright\arraybackslash} p {2 cm} }
\toprule
\toprule
Set & A& B &C \\
\midrule
 \multirow{13}{8.5pt}{\rotatebox{90}{Description}}  & \chiEFT & Set A& Set B\\
   & pQCD & HIC & PREX-II\\
   & Heavy pulsars & Black Widow \mbox{J0952-0607}  & CREX\\
   & NICER \mbox{J0030+0451} & qLMXBs& $^{208}$Pb dipole\\
   & NICER \mbox{J0740+6620} & GW170817+ KN+GRB & Burster \mbox{4U 1702-429}\\ 
   & GW170817 & & Burster \mbox{J1808.8-3658}\\
   &  & &  HESS \mbox{J1731-347} \\
   &  & &  GW190425\\
   &  & &  GRB211211A\\
   &  & &  GW170817 postmerger\\
\midrule
$R_{1.4}$\ \ \ \ \ (km) & $12.26^{+0.80}_{-0.91}$ & $12.41^{+0.57}_{-0.78}$ & $12.17^{+0.46}_{-0.51}$  \\
$\mtov$ ($\msun$) & $2.25^{+0.42}_{-0.22}$ & $2.33^{+0.35}_{-0.22}$ & $2.30^{+0.07}_{-0.20}$  \\
$p_{3\nsat}$ (MeV\,fm$^{-3}$) & $90^{+71}_{-31}$ & $99^{+65}_{-29}$ &  $93^{+33}_{-20}$  \\
$n_{\text{TOV}}$\ \ \ \ ($\nsat$) & $5.92^{+1.34}_{-1.38}$ & $5.67^{+1.09}_{-1.09}$ & $5.73^{+0.95}_{-0.83}$  \\

\bottomrule
\end{tabular}
\end{table}
\begin{figure*}[t!]
    \centering
    \includegraphics[width = \linewidth]{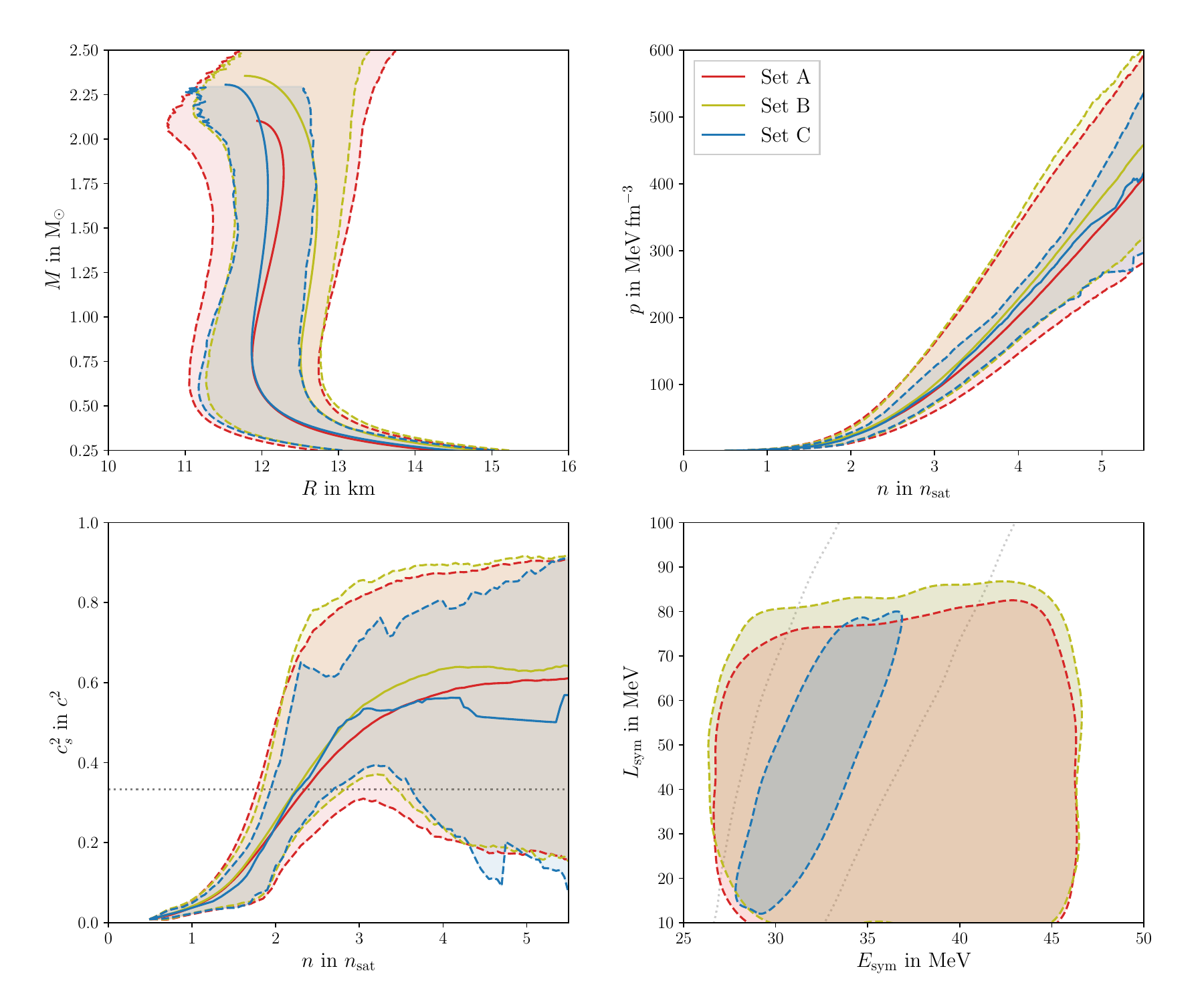}
    \caption{Final posterior estimate for the $M$-$R$ and $p$-$n$ relationship, speed of sound, and symmetry-energy parameters from the three different combinations of constraints A (red), B (yellow), C (blue). 
    The solid lines in the top left-hand panel show the $M$-$R$ curves with highest posterior likelihood, the dashed lines mark the 95\% credibility intervals in radius at a given mass. 
    Similarly, the dashed lines in the top right-hand panel indicate the 95\% credible intervals for the pressure as a function of number density, the solid line there refers to the median.
    The bottom left-hand panel shows the same quantities for the speed of sound as a function of number density, with the medians drawn again as solid lines.
    For the calculation of these posterior properties, we only include EOS~samples below their TOV density, i.e., we consider $P(p|n, n<n_{\text{TOV}})$. 
    The gray dotted line in the bottom left panel indicates the conformal limit $c_s^2= c^2/3$. 
    The bottom right-hand panel shows the 95\% credible regions for the nuclear symmetry parameters $E_{\text{sym}}$ and $L_{\text{sym}}$. 
    The grey dotted line indicates the 95\% credible region for these parameters from the combined results of PREX-II and CREX.}
    \label{fig:set_plot}
\end{figure*}

\begin{figure*}
    \centering
    \includegraphics[width = \linewidth]{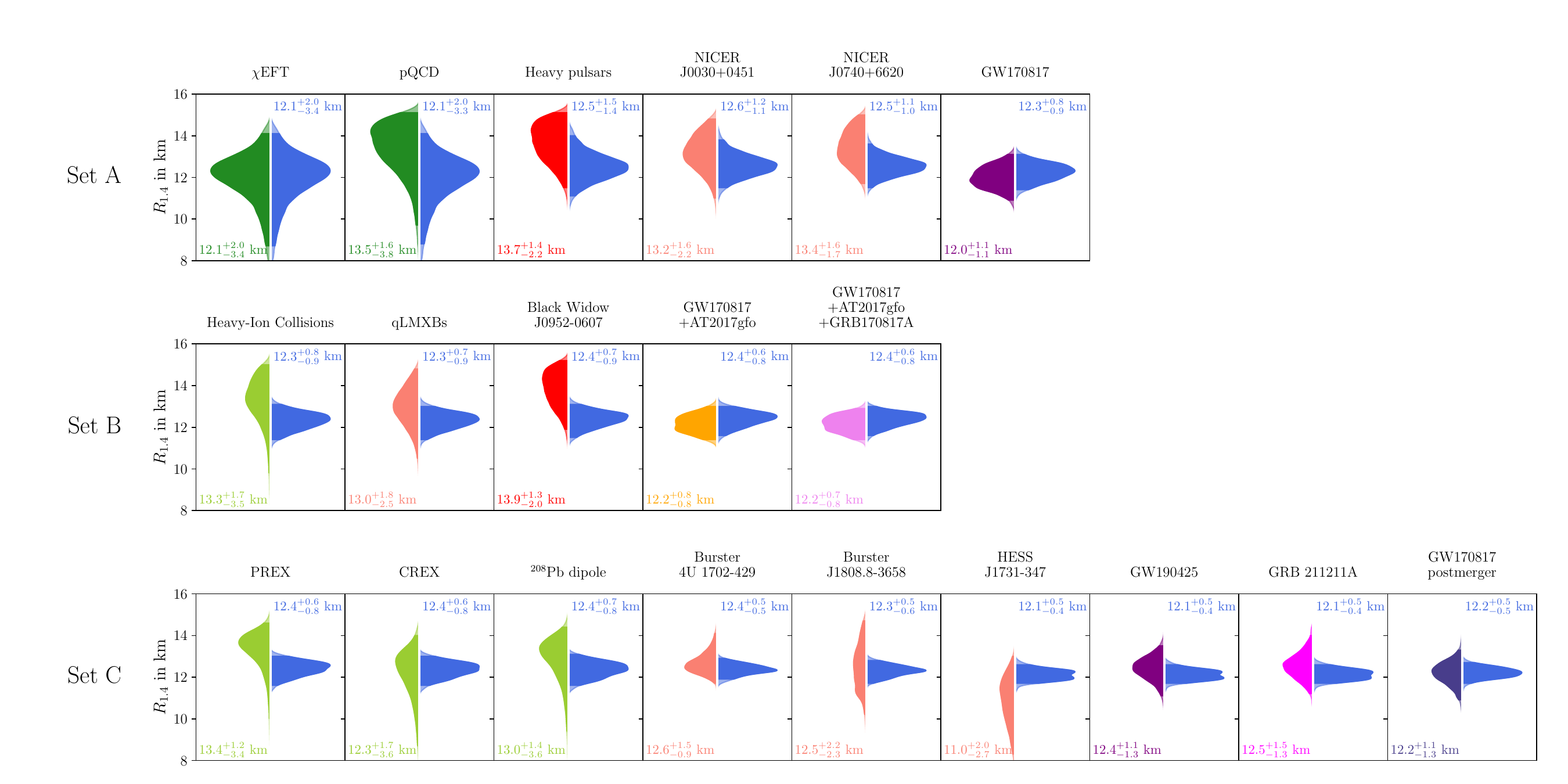}
    \caption{Distribution of $R_{1.4}$ for different constraints. 
    The left-hand wings of the violin plots show the posterior distribution of $R_{1.4}$ from the constraint individually. 
    The right-hand wings show the posterior distribution when multiple constraints are successively combined. 
    Constraints are combined sequentially from top to bottom and left to right, with dependencies taken into account as discussed in the text. 
    Symmetric 95\% credible intervals are indicated by the solid areas under the curves and stated in the bottom left for the constraint individually and upper right-hand corner for the combination.}
    \label{fig:R14_comparison}
\end{figure*}

\begin{figure*}
    \centering
    \includegraphics[width = \linewidth]{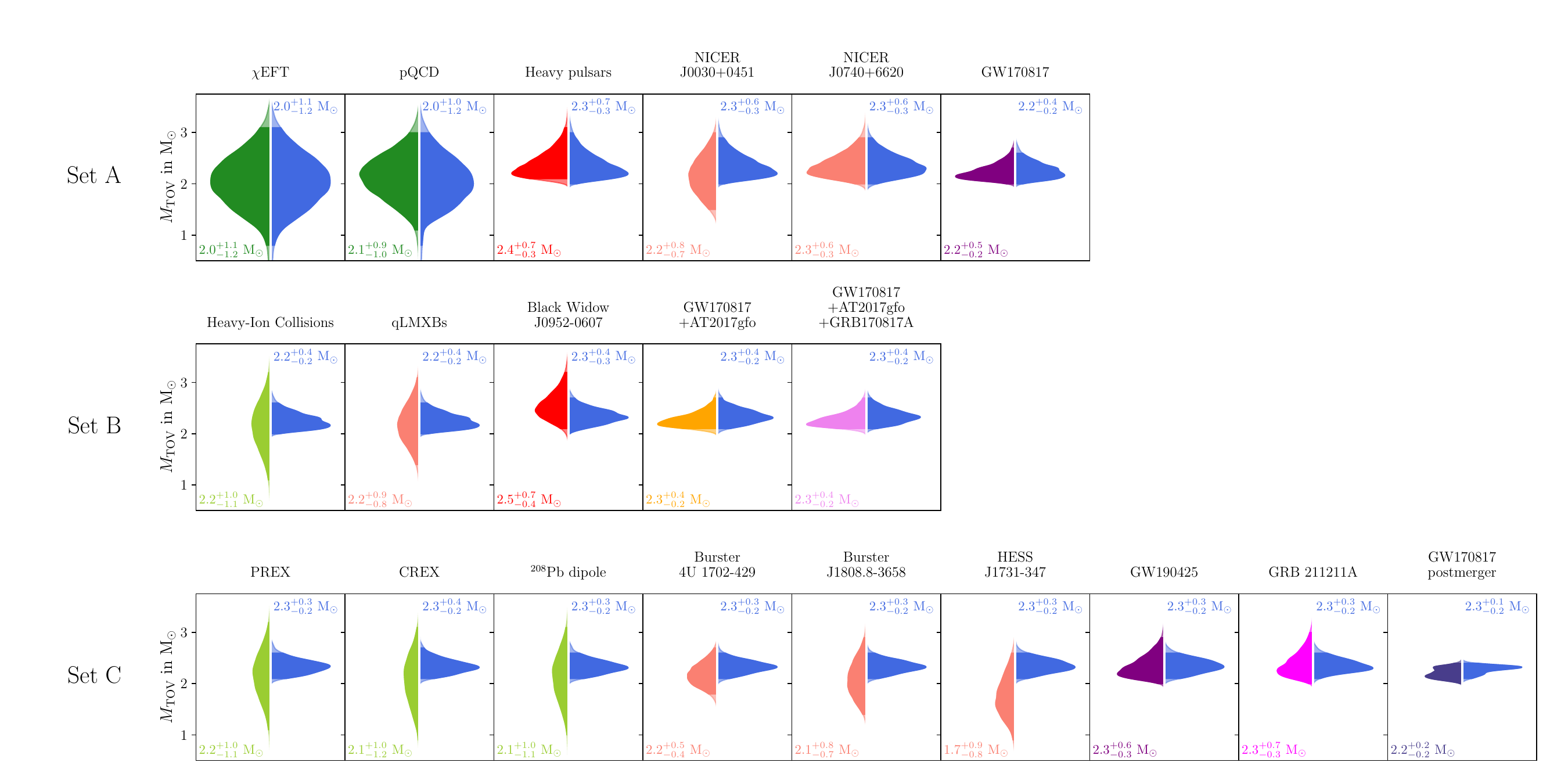}
    \caption{Distribution of $\mtov$ for different constraints. 
    Arrangement as in Fig.~\ref{fig:R14_comparison}.}
    \label{fig:MTOV_comparison}
\end{figure*}

\begin{figure}
    \centering
    \includegraphics[width = \linewidth]{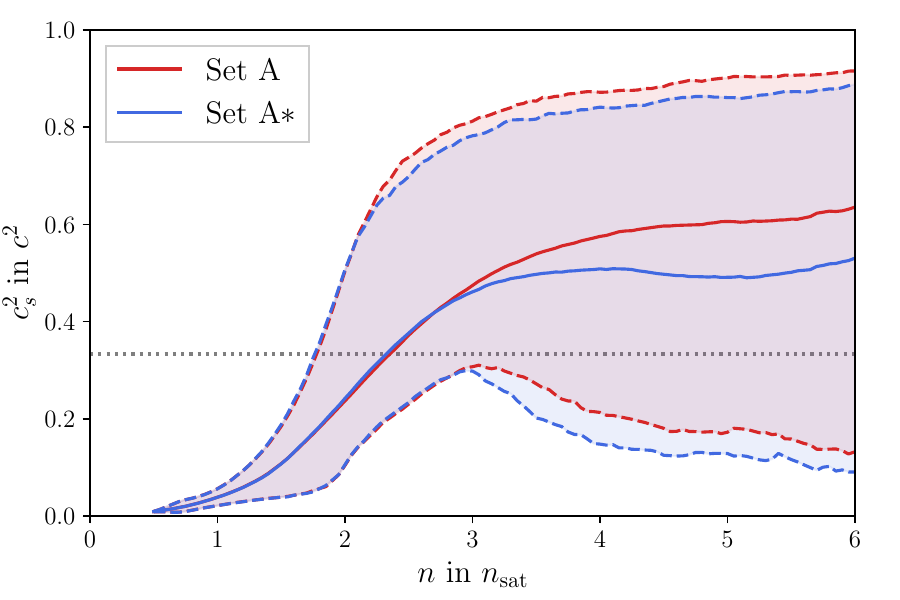}
    \caption{Credibility intervals on the speed of sound when the conservative pQCD matching condition (Fig.~\ref{fig:pQCD}) is replaced with pQCD$*$ (Fig.~\ref{fig:pQCD_GPR}). The red lines are as in the middle panel of Fig.~\ref{fig:set_plot}. The blue lines show the resulting  posterior median and 95\% credibility interval when pQCD is replaced with pQCD$*$ in set A, denoted as set A$*$.}
    \label{fig:cs2_reweighted}
\end{figure}

In the preceding sections, we studied many different constraints on the EOS.
Table~\ref{tab:summary} summarizes the credible ranges for certain EOS quantities given the individual constraints. 
It also quantifies which constraints have a particular impact on specific properties of the EOS through the Kullback-Leibler divergence (KLD)~\citep{Kullback_1951}:
\begin{align}
    \text{KLD}(d,x) = \int dx\ P(x|d) \ln\left(\frac{P(x|d)}{P(x)}\right)\,.
\end{align}
This metric compares the posterior $P(x|d)$ of a quantity $x$ to the prior $P(x)$. 
A value of zero means the prior and posterior are equal. 
The KLD values are shown as bracketed values in Table~\ref{tab:summary} for each inference.
We point out that for the BNS inferences of Sec.~\ref{sec:Multimessenger} the prior was informed by the inferences of \chiEFT, pQCD, and heavy pulsar mass measurements, while the prior for the postmerger constraint is given by the posterior from the joint inference of GW170817, the KN, and the GRB. 
Our determination of the KLD accounts for that.

From the KLD values it is apparent that currently no type of source is able to restrict all properties of the EOS on its own~\footnote{It is worth pointing out that the Kullback-Leiber divergence not only rewards narrow posteriors but also shifts in the expected range of the quantity, so for instance the posteriors from the X-ray burster \mbox{SAX J1808.4-3658} have generally a higher divergence value compared to the inference with \mbox{NICER J0030+0451}, but they also display larger uncertainties.}.
The nuclear constraints primarily affect $R_{1.4}$ and less so $\mtov$ or any other listed quantity. 
The constraints from BNS inferences also primarily affect the radii and intermediate part of the EOS. 
Constraints that restrict $\mtov$ are also the only ones that have considerable impact on $n_{\text{TOV}}$ and $p_{\text{TOV}}$, but are insensitive to the shape of the EOS at lower densities, as expected. 
In general, the relative uncertainties on the microphysical EOS description are significantly higher, e.g., when estimating $p_{3\nsat}$ or $p_{\text{TOV}}$, than the uncertainties of macroscopic properties like NS radii.

This underscores that several constraints need to be combined to obtain an EOS posterior that is restricted over the full density range.
We describe our method of combining our constraint collection in Sec.~\ref{subsec:howtocombine}. 
In Sec.~\ref{subsec:combined_constraints} we group the constraints into three different sets and look at the resulting limits on the EOS. 
As an application of the inferred total knowledge, we afterward compare and discuss the impact of the pQCD matching prescription when combined with the other constraints and apply our joint EOS constraints to determine whether the HESS~\mbox{J1731-347} compact object is consistent with them.

\subsection{How to combine different constraints}
\label{subsec:howtocombine}

For the purpose of combining different inputs, we emphasize that not all constraints studied in the previous sections are independent of each other. 
Some constraints, namely \chiEFT, pQCD, and the heavy pulsars, have been used as prior input for the BNS inferences of Sec.~\ref{sec:Multimessenger}. 
Additionally, other constraints also have been used as prior for results of previous works we rely upon.
Table~\ref{tab:summary} lists the dependencies between the different inputs.

Further, the NS for our constraints may originate from the same underlying mass population, and hence, the priors for their masses should be linked. 
When combining multiple measurements, one would ideally address this through hierarchical sampling, where the masses of those NSs thought to originate from the same population are drawn from some population model with free hyperparameters.
In practice though, the neutron stars we consider might be part of different subpopulations; e.g., the population of merging BNSs is potentially different from the population of galactic double pulsars~\citep{LIGO_GWTC_3, Landry_2021, Farrow_2019}, which, in turn, does not need to coincide with the qLMXB population~\citep{Heinke_2003, Riccio_2022}. 
It has been demonstrated that population effects in GW analysis from BNS mergers only become significant after $\gtrsim 10$ events~\citep{Wysocki_2020, Golomb_2022}.
Our constraint collection encompasses two BNS events and similarly low numbers of NS in other population categories; therefore, we do not expect any impact on the final result by neglecting hierarchical population effects. 
Moreover, since the form of the different population models is not known, the assumptions of a uniform NS mass range up to the TOV mass seems appropriate and is a common choice~\citep{Biswas_2022, Kedia_2024}. 
This population model is implicitly assumed, e.g., in Eq.~\eqref{eq:l1}.

To now actually combine the individual constraints, we take the immediate dependencies into account by only multiplying independent likelihoods and omitting the dependent ones. 
For instance, if we include the PSR \mbox{J0740+6620} NICER constraint, we will not simultaneously factor in the radio timing measurement of PSR \mbox{J0740+6620}, because both concern the same NS and the latter was used as prior for the analysis of the NICER data.
Likewise, the BNS inferences of Sec.~\ref{sec:Multimessenger} are not combined again with the constraints from \chiEFT, pQCD, and radio timing measurements, as these constituted their prior. 
When we combine two constraints that share a common prior, we reweigh the combination accordingly~\footnote{We note that this is effective, because we are just interested in the marginalized posterior for only one parameter, namely the EOS. 
If we were interested in the full posterior of GW170817 with the dependency on e.g. \mbox{PSR J0740+6620} removed, we would need to take the evidence ratio for the two different priors into account.}. 
For example, the GW170817 and GW190425 posteriors both carry information from \chiEFT, pQCD, and the heavy pulsars in their prior; hence we combine them as follows
%\begin{widetext}
\begin{align}
\begin{split}
     &P(\text{EOS}|\{\text{GW170817, GW190425}\})  \propto \\
     & \quad \frac{P(\text{EOS}|\text{GW170817})\ P(\text{EOS}|\text{GW190425})}{P(\text{EOS}|\{\text{\chiEFT, pQCD, Heavy pulsars}\})}.
    %P(\text{EOS}|\{\text{GW170817, NICER J0740+6620}\}) \propto    P(\text{EOS}|\text{NICER J0740+6620}) \frac{P(\text{EOS}|\text{GW170817})}{P(\text{EOS}|\text{Radio timing J0740+6620})}\,.
\label{eq:reweighting}
\end{split}
\end{align}
We proceed analogously when combining a BNS inference with the NICER result for \mbox{PSR J0740+6620} and simply divide by the radio timing measurement of that pulsar one more time.
Because of the finite sample size in the BNS inferences, undersampling effects will come into play when too many of these constraints are added together. 
However, in our collection at most three independent BNS inferences can be combined (GW190425, GRB211211A, and one instance of the GW170817 analyses) and we verified that the undersampling effects cause only very minor deviations, since both GW190425 and GRB211211A have no constraining effect on the EOS.

\subsection{Results from combined constraints}
\label{subsec:combined_constraints}
We are now able to combine the various constraints from our EOS inference, taking the dependencies into account as just described. 
We will do so by defining three different sets.
Set A includes those constraints used most often in the literature, namely \chiEFT, pQCD, the three heavy pulsar mass measurements, GW170817, and the NICER observations. 
Note that the first three are already implicitly included in our inference of GW170817. 
Set B complements set A with the heavy-ion collision data, the qLMXBs, the black widow pulsar \mbox{PSR J0952-0607} and adds the KN and GRB afterglow to the inference of GW170817. 
Set C additionally includes the symmetry energy constraints from PREX-II, CREX, and the $^{208}$Pb dipole measurements, as well as the X-ray bursters from Sec.~\ref{subsec:Burst}, HESS \mbox{J1731-347}, GW190425, GRB211211A, and the postmerger constraint from GW170817. 
Hence, set C combines every information discussed so far, except pQCD$*$ and the GW170817 postmerger constraint with the Kepler limit. 
Sets A, B, and C are formed in order of increasing model dependency, with set A including those constraints we deem reliable and set C containing also data points where the systematic uncertainties are potentially large.
Because this assessment is subjective and other combinations are permissible, we created an online interface~\citep{Rose_2024} where we provide the opportunity to freely combine constraints and obtain the resulting posterior distributions of the EOS parameters. 
We summarize the composition and results of sets A, B, and C in Table~\ref{tab:set_final}.

We show the resulting constraints on the NS masses and radii, pressure and speed of sound, and the symmetry energy in Fig.~\ref{fig:set_plot}. 
The radii of NSs can be determined within an uncertainty of $\sim$ \qtyrange{0.5}{1}{km}, depending on the number of constraints employed. 
With higher NS mass, the uncertainty in the radius grows slightly.
In Figs.~\ref{fig:R14_comparison} and~\ref{fig:MTOV_comparison} we also show how the posterior estimates on $R_{1.4}$ and $\mtov$ evolve when the individual constraints are sequentially added. 
We find that the posterior median for the canonical radius settles around $\approx$\qty{12}{km} even when adding only a few constraints, relatively independent of the exact combination. 
In contrast to the narrow limits on NS radii, meaningful restrictions on the speed of sound and the empirical nuclear parameters are harder to obtain. 
While the inclusion of \chiEFT\ allows us to restrict $L_{\text{sym}}<100$\,MeV, narrower constraints, in particular on $E_{\text{sym}}$, can be achieved only when the results of PREX-II and CREX are directly included \citep{Gueven_2020}. 
Regarding the speed of sound, our inference implies that its value likely exceeds the conformal limit of $1/\sqrt{3}$\,$c$ around 3\,$\nsat$, but for higher number densities the posterior distribution reaches a flat plateau and the upper limits on $c_s^2$ remain weak and not particularly informative.

\subsection{Impact of the pQCD matching prescription}

In Sec.~\ref{sec:pQCD}, we presented two different methods of how the pQCD constraint can be applied to the EOS candidates.
One option is to use the step function in Eq.~\eqref{eq:pQCD_condition} for the matching condition and to marginalize over the uncertainties in the renormalization scale and other parameters. 
Significantly stronger results from pQCD can be obtained when instead marginalizing over the set of possible high-density extrapolations, as shown in Fig.~\ref{fig:pQCD_GPR}. This comes at the cost of additional model dependence. 
We denote the latter version as pQCD$*$. 

To determine how the conclusions would change with the stronger pQCD$*$ constraint, we reweigh the posteriors of sets A, B, and C in a similar manner as Eq.~\eqref{eq:reweighting}, to make them independent of the default pQCD.
We then factor in again the likelihoods of the EOSs under pQCD$*$. 

When including pQCD$*$ instead of the usual pQCD constraint in set A, we find a decrease in the estimated TOV mass from originally ${2.25}^{+0.42}_{-0.22}$\,$\msun$ to ${2.20}^{+0.40}_{-0.18}$\,$\msun$, while the tail in the posterior for large $p_{3\nsat}$ is cut from ${90}^{+71}_{-31}$\,MeV\,$\fmiq$ to ${92}^{+59}_{-31}$\,MeV$\,\fmiq$. 
The estimates for the canonical NS radius and TOV density do not change significantly. 
This behavior is similar when pQCD$*$ is introduced in sets B and C, where $\mtov$ reduces from ${2.33}^{+0.35}_{-0.22}\,\msun$ to ${2.30}^{+0.34}_{-0.21}\,\msun$ and from ${2.30}^{+0.07}_{-0.20}\,\msun$ to ${2.28}^{+0.10}_{-0.18}\,\msun$, respectively. 
Hence, combining pQCD$*$ with astrophysical constraints allows one to exclude some stiff EOS candidates that otherwise would remain likely.

Figure~\ref{fig:cs2_reweighted} also compares the evolution in the speed of sound when the usual pQCD constraint is replaced by pQCD$*$ in set A.
The initial rise of $c_s^2$ above the conformal limit is very similar to the rise of $c_s^2$ with the original pQCD constraint as shown in Fig.~\ref{fig:set_plot}; however, with pQCD$*$ we observe a slightly stronger upper limit for $c_s^2$ across all densities and a slight decrease of the median $c_s^2$ after \qty{4}{\nsat}, though it increases again after \qty{5}{\nsat}. 
Other studies generally find that the speed of sound decreases at higher densities after peaking above the conformal limit~\citep{Kojo_2020, Pang_2023}, especially when pQCD constraints are implemented~\citep{Altiparmak_2022, Somasundaram_2023, Gorda_2023a, Komoltsev_2023}. 
We do not observe any softening or stringent upper limits on the speed of sound within the usual sets A, B, and C, since our EOS candidates are constructed through a model-agnostic speed-of-sound extrapolation and were matched to the conservative pQCD constraint at $n_{\text{TOV}}$. 
When implementing pQCD$*$, we observe a slight softening, even though the constraint is applied at $n_{\text{TOV}}$.
This is in agreement with other studies, that generally find that the speed of sound decreases at higher densities after peaking above the conformal limit~\citep{Kojo_2020, Pang_2023}, especially when information from pQCD is included~\citep{Altiparmak_2022, Somasundaram_2023, Gorda_2023a, Komoltsev_2023, Fan_2023}. 
However, the median as well as the uncertainty range for the speed of sound increases again at $n>\qty{5}{\nsat}$. This is possibly due to the decrease in sample size for $c_s^2(n)$ at higher density, as we include only those EOS with $n<n_{\text{TOV}}$. Hence, above $\sim 5$\,$\nsat$, statistical limitations prevent reliable restrictions on the speed of sound.

\subsection{Outlier detection}
\label{sec:outlier}

\begin{table}[t]
\renewcommand{\arraystretch}{1.5}
\caption{Posterior predictions about the \mbox{HESS J1731-347} radius. The table lists different metrics explained in the text to compare the posterior prediction about the radius to the actual measurement. The measured radius of Ref.~\citep{Doroshenko_2022} is $R = {10.40}^{+1.62}_{-1.37}\,$km. Set C here excludes the original HESS measurement.\\}
\label{tab:HESS_outlier}
\begin{tabular}{>{\centering\arraybackslash} p {2.2 cm}  >{\centering\arraybackslash} p { 2 cm} >{\centering\arraybackslash} p {2 cm} >{\centering\arraybackslash} p {2 cm} }
\toprule
\toprule
Set & A& B &C \\
\midrule
R (km) & $12.07^{+0.75}_{-0.94}$& $12.18^{+0.65}_{-0.83}$& $12.15^{+0.55}_{-0.68}$\\
KWM (km) & 1.61& 1.73& 1.71\\
KLD & 2.231& 2.659& 2.825\\
$\mathcal{C}$ & 0.4& 0.55& 0.57\\
$\mathcal{P}$ & 6.98& 11.35&  16.42\\

\bottomrule
\end{tabular}

\end{table}

The mass-radius measurement of \mbox{HESS J1731-347} causes the overall highest deviation from the EOS prior, as it shifts the posterior to very soft EOSs. 
Given the other inputs and their Kullback-Leibler divergences, it appears that \mbox{HESS J1731-347} is the likeliest candidate for a statistical outlier in the constraint dataset. 

To examine this impression quantitatively, we employ posterior predictive checking. 
Specifically, we may compare the $M$-$R$ posterior of \mbox{HESS J1731-347} to the radius range we would expect for this object from some given set of other constraints $d = \{d_1, d_2, \dots\}$, e.g., from set A.
To determine the expected radius range, we resample the masses from the original posterior $P(M,R|\text{HESS})$ of Ref.~\citep{Doroshenko_2022}, but determine the radius samples from our EOS set weighted with the posterior likelihood from $P(\text{EOS}|d)$~\footnote{The EOS on its own does not allow a statement about the expected mass range of NSs, so we assume here for the sake of argument that the masses for \mbox{HESS J1731-347} were accurately determined and only compare the expected radius. 
The small correlation coefficient of 0.12 between mass and radius of in the actual measurement allows for this approach.}. 
We denote the mock posterior obtained this way by $P(M,R|\text{HESS}[d])$. 

To quantify the difference between the actually measured radius and the posterior from our mock prescription, we may resort again to the Kullback-Leibler divergence or alternatively to the Kantorovich-Wasserstein metric (KWM)~\citep{Kantorovich_1960}. 
For the different constraint sets A, B, and C, the comparison of these radius distributions is done in Table~\ref{tab:HESS_outlier}. 
Moreover, we can use the mock posterior to perform again an inference on the EOS set with Eq.~\eqref{eq:l2}. 
This allows us then to compare the posterior predictive value for the true measurement to the value that would be achieved if the measurement agreed perfectly with the expectation from the constraints $d$. 
The posterior predictive ratio $\mathcal{P}$,
\begin{align}
\begin{split}
    \mathcal{P} = &\frac{P(\text{HESS}[d]|d)}{P(\text{HESS}|d)}\\
    =&\frac{\sum_{\text{EOS}}\mathcal{L}(\text{EOS}|\text{HESS}[d]) P(\text{EOS}|d)}{\sum_{\text{EOS}} \mathcal{L}(\text{EOS}|\text{HESS}) P(\text{EOS}|d)}\,,\\
\end{split}
\end{align}
can be interpreted like a Bayes factor, in the sense that, assuming the constraints in $d$ for the EOS, the mock radius value of $P(M, R| \text{HESS}[d])$ is $\mathcal{P}$ times more plausible than the actual measurement.

We can also explicitly analyze the hypothesis of whether the compact object in \mbox{HESS J1731-347} should be described by the same EOS as the constraints in $d$. 
We call this hypothesis $H_0$, and its evidence is given by
\begin{align}
\begin{split}
    Z_{H_0} &= \sum_{\text{EOS}} P(\text{HESS}, d|\text{EOS}) P(\text{EOS}) \\
            &= \sum_{\text{EOS}} \mathcal{L}(\text{EOS}|\text{HESS}) \mathcal{L}(\text{EOS}|d)  P(\text{EOS}) \\
            &= Z_{d} \sum_{\text{EOS}} \mathcal{L}(\text{EOS}|\text{HESS}) P(\text{EOS}|d)\,, \\
            \label{eq:evidence_H0}
\end{split}
\end{align}
where $P(\text{EOS})$ is the prior on the EOS and $Z_{d}$ the evidence for the constraint set $d$. 
The alternative hypothesis $H_1$ is that \mbox{HESS J1731-347} is not necessarily described by the same EOS as the rest of the data, but by some other $M$-$R$ relationship in our candidate set. 
The evidence for $H_1$ is given by
\begin{align}
\begin{split}
    Z_{H_1} &= \sum_{\text{EOS}_j, \text{EOS}_k} P(\text{HESS}|\text{EOS}_j) P(d|\text{EOS}_k)\\
            & \qquad \qquad  \times P(\text{EOS}_j)  P(\text{EOS}_k)\\
            &= Z_{d}  \sum_{\text{EOS}_j} \mathcal{L}(\text{EOS}_j|\text{HESS}) P(\text{EOS}_j)\\
            & \qquad \qquad  \times  \sum_{\text{EOS}_k} P(\text{EOS}_k|d)\\
            &= Z_{d}  \sum_{\text{EOS}_j} \mathcal{L}(\text{EOS}_j|\text{HESS}) P(\text{EOS}_j)\\
            \label{eq:evidence_H1}
\end{split}
\end{align}
The coherence ratio is then defined as $\mathcal{C} = Z_{H_1}/Z_{H_0}$, where the factors $Z_d$ in Eqs.~(\ref{eq:evidence_H0} and (\ref{eq:evidence_H1}) cancel.

The values for $\mathcal{P}$ and $\mathcal{C}$ are provided in Table~\ref{tab:HESS_outlier}. 
Altogether, the different metrics listed there show a growing tension between the expected radius of \mbox{HESS J1731-347} and the actual measurement as we add more constraints to the EOS.
Yet, the posterior predictive ratios are not very large, especially for set A; hence we cannot immediately reject the HESS measurement as an outlier. 
Also for all of the constraint sets A, B, C, we have $\mathcal{C}<1$, so we never reject the null hypothesis $H_0$. 
Still, we find some substantial deviation of the HESS measurement from the posterior EOS expectation, especially for set C. 
We note that testing the hypothesis $H_0$ against $H_1$ is mainly a statement about the measured radius and does not comment on the low mass of the object. 
To properly address the nature of the HESS compact object, evolutionary aspects should be investigated and the systematic uncertainties mentioned in Sec.~\ref{sec:Thermal}, also in regards to the mass measurement, need to be kept in mind.

The postmerger constraints from GW170817 and the X-ray burster \mbox{4U 1702-429} also stand out with respect to the KLD of their posteriors. 
However, the narrow radius estimate from the latter falls in the range expected from the remaining constraints and hence it is not an outlier in the statistical sense. 
The postmerger constraint is the only one that places a strong upper limit on $\mtov$, and, therefore, it is difficult to assess its validity based on the available data, though it seems to be consistent with the trend from pQCD$*$.
Additionally, PREX-II and CREX could be classified as mutual outliers based on their different predictions for $L_{\rm sym}$. 
However, their impact on neutron-star properties is minor and, hence, statistical discrimination based on the remaining data seems difficult. 
We may resort such analyses to future work.

\section{Conclusions}

In the present work, we discussed a diverse collection of available constraints on the EOS of dense, neutron-rich matter. 
We applied these constraints one by one to our broad set of EOS candidates constructed with the metamodel approach and speed-of-sound extrapolation, while assessing their respective impact and commenting on remaining uncertainties. This allowed us to compare the impact different constraints have on distinct parts of the EOS.
In the end, we combined the constraints to obtain stringent limits on the canonical NS radius, maximum NS mass, and other EOS-derived quantities. In doing so, we introduced three sets of constraint combinations. Set A includes those constraints we consider less prone to large systematic errors, whereas in sets B and C we subsequently add every other constraint we discussed in the present work. 

In total, we find for set A $R_{1.4} = 12.26^{+0.80}_{-0.91}$\,km, as well as $\mtov = {2.25}^{+0.42}_{-0.22}$\,$\msun$. For sets B and C we have $R_{1.4} = 12.41^{+0.57}_{-0.78}$\,km, $\mtov= {2.33}^{+0.35}_{-0.22}$\,$\msun$ and $R_{1.4} = 12.17^{+0.46}_{-0.51}$\,km, $\mtov = {2.30}^{+0.07}_{-0.20}$\,$\msun$ respectively.
This is in good agreement with previous studies, e.g., Refs.~\citep{Capano_2020, Annala_2022, Dietrich_2020, Huth_2022, LIGO_GW170817_EOS, Raaijmakers_2021, Gueven_2023, Breschi_2024, Brandes_2023}.
An overview on the sets and their posterior results is given in Table~\ref{tab:set_final}. When we use the Kepler limit instead of the TOV mass for the postmerger GW170817 constraint in set C, we are able to place an even stronger limit on $\mtov$ of ${2.05}^{+0.07}_{-0.04}$\,$\msun$. 
Other custom combinations of EOS constraints can be obtained from our Web interface in Ref.~\citep{Rose_2024} created for this purpose.

As an application, we combined the new pQCD matching criterion pQCD$*$ with the astrophysical data, showing how this further excludes stiffer EOSs that are allowed by the remaining data, at the cost of additional assumptions about the extrapolation. We also investigated whether the $M$-$R$ measurement of the \mbox{HESS J1731-347} compact objects qualifies as an outlier, finding that the more constraints we apply to the EOS, the less likely the small radius of the actual measurement becomes compared to the expectation from the posterior. The evidence shows substantial deviation, but especially for the conservative set A we cannot immediately reject the HESS measurement as an outlier.

Explicit dependencies between the different data points are taken into account in our combination effort. 
At the same time, we ignore any population effects when combining inferences from multiple NSs. Though we do not expect this to impact the final results significantly based on the low number of events in each possible subpopulation (see Sec.~\ref{subsec:howtocombine}), we emphasize that incorporating population effects through hierarchical inference will become important when a large number of events is available in the future~\citep{Wysocki_2020}. Furthermore, the results are also subject to potential systematic biases in the inputs; thus care should be taken when quoting strict limits on EOS quantities.

The selection of constraints explored here is not complete, e.g., we have not considered limits on $\mtov$ from NS population studies~\citep{Alsing_2018, Shao_2020, Fan_2023},  or further $M$-$R$ measurements from X-ray observations~\citep{Oezel_2016B} or quasiperiodic oscillations~\citep{Sotani_2023}.
When supplementing the EOS model with a prescription for thermodynamic properties like heat capacity or neutrino emissivity, NS cooling curves also place limits on the EOS~\citep{Potekhin_2015, Wijnands_2017, Brown_2022, Marino_2024}.
Yet, our extensive set of constraints shows how combining different pieces of information in a fully Bayesian fashion can place narrow limits on NS masses and radii. 
Although it has been noted that simple hard cuts on the EOS candidates based on empirical limits deliver similar outcomes compared to a Bayesian likelihood function on the EOS space~\citep{Jiang_2023}, we believe performing Bayesian analysis directly over the space of possible EOSs has multiple advantages.
For one, including the full Bayesian uncertainty of a measurement will maintain consistency even if future measurements with improved uncertainties become available \citep{Miller_2020}. 
Moreover, it allows us to combine different types of measurements in a more flexible way and without the need to relate different measured and derived quantities through phenomenological relations.

We note that even given our large collection of constraints, microscopic quantities such as $p_{3\nsat}$ are still subject to large relative uncertainty. 
The relative uncertainty on the pressure at a given density grows strongly with density.
At \qty{5}{\nsat} the 95\% credibility range for the pressure already encompasses a factor of 2, even for set C.
One reason for this is that most of our constraints are astrophysical in nature and therefore affect only the microscopic EOS in an integrated form across the range of several $\nsat$. 
Likewise, our broad posterior estimates of the empirical nuclear parameters show that to effectively constrain the low-density regime of the EOS, current NS observations alone are not sufficient, and conversely, knowing the symmetry energy parameters alone has only mild impact on NS properties.

The uncertainty about the EOS will be reduced in the future, when new measurements with higher data precision become available. For instance, even the relatively well-constrained NS radii would only improve slowly if further NS observations with present statistical uncertainties were added. 
This emphasizes the need for high-precision observations, e.g., from next-generation GW detectors~\citep{Puecher_2023, Rose_2023A, Iacovelli_2023, Hild_2011, Reitze_2019, Evans_2021, Sathyaprakash_2012, Walker_2024, Punturo_2010, Pradhan_2023}, moment-of-inertia measurements from long-term pulsar timing observations~\citep{Greif_2020, Carreau_2019, Kramer_2009, Kramer_2021}, or new X-ray observatories such as Strobe-X~\citep{Ray_2019}. More stringent constraints on the symmetry energy and other nuclear parameters could be obtained from high-precision scattering experiments like MREX~\citep{Becker_2018} or a new generation of HIC experiments~\citep{Ivanov_2024, Tahir_2020, Russotto_2021, Durante_2019}.

\begin{acknowledgments}
We thank A. Steiner for their support with the data from the qLMXBs, A. Goodwin for providing the $M$-$R$ posterior of SAX 1808.4-3658, and J. Nätillä for providing the $M$-$R$ posterior of the X-ray burster 4U 1702-429. We thank S. Anand, M. Bulla, P. Landry, J. Margueron, C. Miller, J. Read, and A. Schwenk for helpful discussions and comments.

H.~K. and T.~D. acknowledge funding from the Daimler and Benz Foundation for the project “NUMANJI” and from the European Union (ERC, SMArt, 101076369). 
Views and opinions expressed are those of the authors only and do not necessarily reflect those of the European Union or the European Research Council. Neither the European Union nor the granting authority can be held responsible for them.
P.~T.~H.~P. is supported by the research program of the Netherlands Organization for Scientific Research (NWO).
M.~W.~C. and B.~F.~H. are supported by the National Science Foundation with Grants No. PHY-2308862 and No. PHY-2117997.
R.S. acknowledges support from the Nuclear Physics from Multi-Messenger Mergers (NP3M) Focused Research Hub which is funded by the National Science Foundation under Grant No. 21-16686, and by the Laboratory Directed Research and Development program of Los Alamos National Laboratory under Project No. 20220541ECR.
I.T. was supported by the U.S. Department of Energy, Office of Science, Office of Nuclear Physics, under Contract No.~DE-AC52-06NA25396, by the Laboratory Directed Research and Development program of Los Alamos National Laboratory under Projects No. 20220541ECR and No. 20230315ER, and by the U.S. Department of Energy, Office of Science, Office of Advanced Scientific Computing Research, Scientific Discovery through Advanced Computing (SciDAC) NUCLEI program.

This research used resources of the National Energy Research Scientific Computing Center (NERSC), a U.S. 
Department of Energy Office of Science User Facility located at Lawrence Berkeley National Laboratory, operated under Contract No. DE-AC02-05CH11231 using NERSC award ERCAP0023029. 
This research used resources on the national supercomputer HPE Apollo Hawk at the High Performance Computing (HPC) Center Stuttgart (HLRS) under the Grant No. GWanalysis/44189, on the GCS Supercomputer SuperMUC\_NG at the Leibniz Supercomputing Centre (LRZ) [project pn29ba], and on the HPC systems Lise/Emmy of the North German Supercomputing Alliance (HLRN) [project bbp00049].

This material is based upon work supported by NSF's LIGO Laboratory which is a major facility fully funded by the National Science Foundation.

\end{acknowledgments}

\appendix
\section{IMPACT OF THE \texorpdfstring{\chiEFT}{χEFT} LIKELIHOOD FUNCTION}
\label{app:chiEFT}
\begin{figure}
    \centering
    \includegraphics[width = \linewidth]{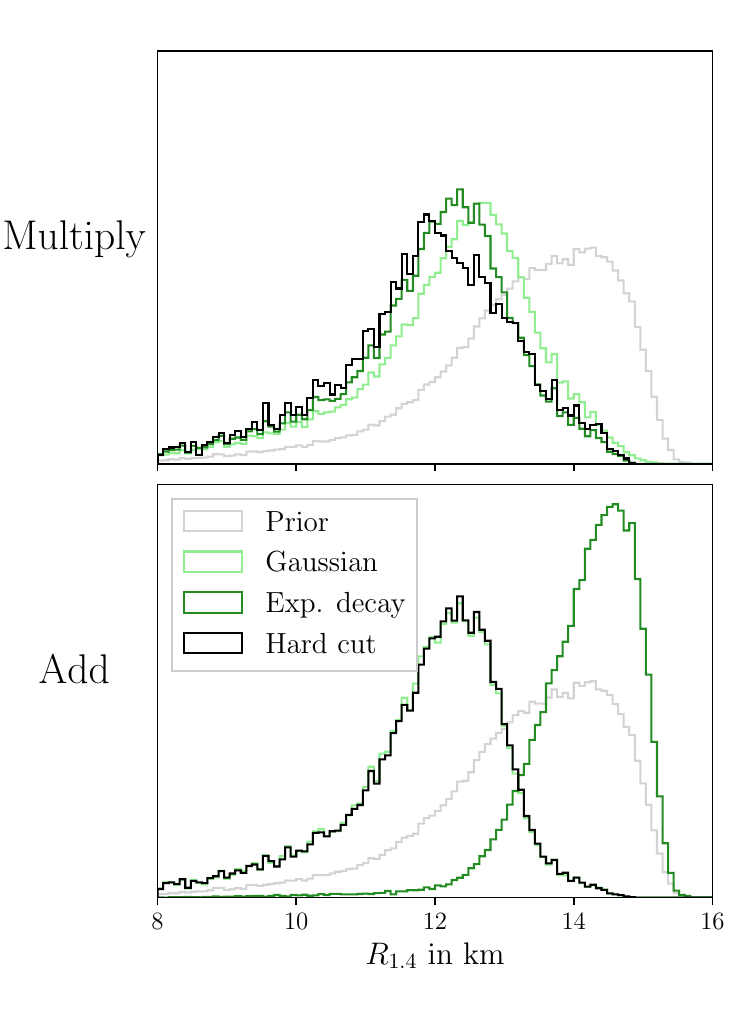}
    \caption{$R_{1.4}$ posterior from the \chiEFT\ constraint when different versions for the likelihood $\mathcal{L}(\text{EOS}|\chi\text{-EFT})$ are used. The top shows the posterior when the score values are multiplied along the EOS (Eq.~\ref{eq:chiEFT_likelihood}), whereas the bottom displays the same when the score values instead are added (Eq.~\ref{eq:chiEFT_likelihood_add}). The color coding in both panels refers to the form of the score function $f$. Exponential decay means $f$ is chosen as Eq.~(\ref{eq:score_function}) with $\beta =6$, Gaussian refers to Eq.~(\ref{eq:score_function_gaussian}), Hard cut to Eq.~(\ref{eq:score_function_hardcut}). The likelihood adopted in the main section is to multiply the score values with an exponential-decay-score function.}
    \label{fig:chiEFT_explore}
\end{figure}
\begin{figure}
    \centering
    \includegraphics[width = \linewidth]{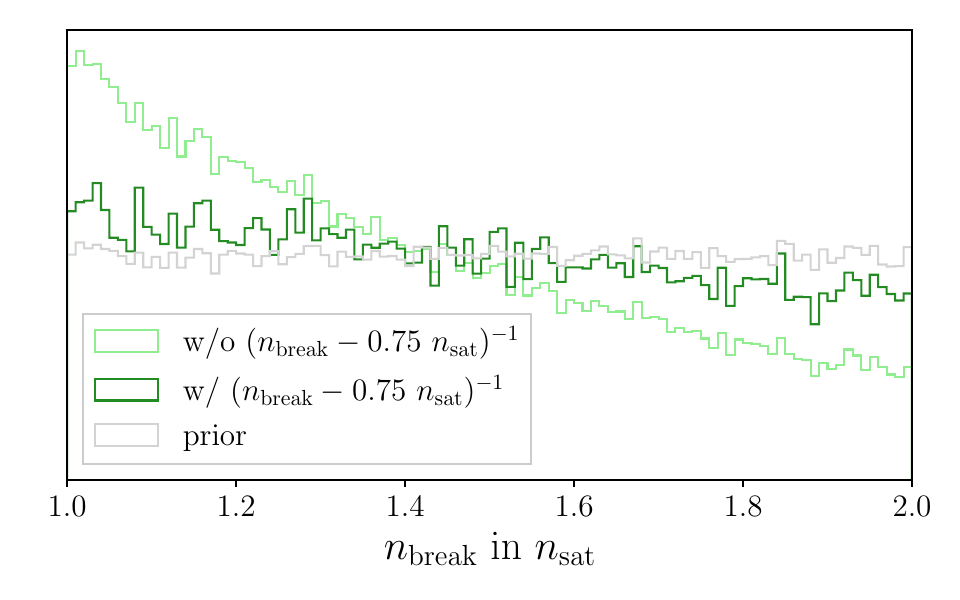}
    \caption{Posterior distribution of $n_{\text{break}}$ when different versions for the likelihood (Eq.~\ref{eq:chiEFT_likelihood}) are used. The skew in the posterior distribution decreases when the exponent in the likelihood contains an additional factor of $(n_{\text{break}} - 0.75~\nsat)^{-1}$. Here, $f$ was taken from Eq.~\eqref{eq:score_function}.}
    \label{fig:nbreak_hist}
\end{figure}
Since it remains difficult to interpret the pressure band proposed by \chiEFT\ calculations in a genuinely Bayesian fashion, the choice of the likelihood function Eq.~(\ref{eq:chiEFT_likelihood}) and subsequently the score function $f$ in Eq.~(\ref{eq:score_function}) is ambiguous, though it naturally impacts the conclusions drawn from this constraint.

The score function $f$ is often reduced to a hard cut, i.e.,
\begin{align}
    f_{\text{HC}}(p, n) = \begin{cases} 1 \qquad \text{if $p_{-} < p < p_{+}$,}\\
                            0 \qquad \text{else.}
    \label{eq:score_function_hardcut}
    \end{cases}
\end{align}
This is usually implemented implicitly during the construction of the EOS prior set \citep{Huth_2022, Pang_2023}. 
However, alternative approaches are equally justified. 
For instance, in Ref.~\citep{Drischler_2020}, Gaussian process regression for the speed of sound is used to assess the uncertainties of \chiEFT\ predictions, finding the posterior distributions for the pressure well approximated by a Gaussian distribution. Therefore, one may as well impose a Gaussian $f$
\begin{align}
\begin{split}
    f_{\text{G}}(p, n) &= \exp\left(-8 \left(\frac{p-\mu}{p_{+}-p_{-}}\right)^2 \right)\\
    \text{with}\quad \mu &= \frac{p_{+} - p_{-}}{2}.
\label{eq:score_function_gaussian}
\end{split}
\end{align}
In a similar manner, Ref.~\citep{Brandes_2023} construct a Gaussian likelihood function on the squared speed of sound.
In any case, a further question arises of whether the individual score values $f(p(n, \text{EOS}), n)$ along an EOS pressure curve should be added, 
\begin{align}
\begin{split}
    \mathcal{L}(\text{EOS}|\chi\text{EFT}) \propto
    \int_{0.75~\nsat}^{n_{\text{break}}} dn\ f(p(n, \text{EOS}), n)
\label{eq:chiEFT_likelihood_add}
\end{split}
\end{align}
or instead be multiplied as in Eq.~({\ref{eq:chiEFT_likelihood}). The former is more forgiving should an EOS deviate from the band only on a short density range, but in principle also allocates posterior likelihood to EOSs that deviate heavily from the band in some region and coincide perfectly with it on a short interval. 

The approach adopted in the main text followed two deliberations. 
First, if the \chiEFT\ constraint is applied, the prediction from this theory should match to the EOS across the entire nucleonic regime, i.e., from $0.75~\nsat$ to $n_{\text{break}}$, and awkward behavior where the constraint is only fulfilled on a subinterval should be excluded. 
Hence, the results in the main section use the likelihood that multiplies the score values. 
Second, small deviations from the band should be allowed, with the uncertainty arising in the theoretical prediction being systematic and not Gaussian.
Consequently, the score function adopted in the main section is neither the hard cut nor Gaussian distribution, but rather sets the $f(p,n)$ constant across the AFDMC band and then exhibits exponential decay beyond it. 

The exact choice of the likelihood prescription does not impact the inference too much, as long as the score function is sufficiently narrow. 
In Fig.~\ref{fig:chiEFT_explore}, we compare the \chiEFT-posterior distribution on $R_{1.4}$ when different likelihood prescriptions are used. 
When multiplying the score functions, the results are relatively independent on the score function's exact form, though the hard cut enables slightly stricter EOS constraints.
Adding the score values along the EOS with the slow-exponential-decay score function is too forgiving and distorts the posterior toward very stiff EOSs.
However, employing the hard cut or narrow Gaussian score functions yields posterior distributions that resemble expectations from other studies more closely~\citep{Hebeler_2013, Tews:2018kmu, Drischler_2021}.
It also matches the case where score values are multiplied well.

A challenge of our setup lies in the fact that the density regime in which we apply the \chiEFT\ constraint is not constant, but depends on the EOS through the individual breakdown density $n_{\text{break}}$. 
If the breakdown density is large, more score values are multiplied, which decreases the total likelihood and introduces a bias toward smaller $n_{\text{break}}$. To mitigate this effect, we introduced the factor $(n_{\text{break}} - 0.75~\nsat)^{-1}$ in Eq.~(\ref{eq:chiEFT_likelihood}).
Figure~\ref{fig:nbreak_hist} shows the posterior distribution of $n_{\text{break}}$ for the case where this factor in the exponent is absent, and when it is included. With our prescription as adopted in the main text, the bias toward smaller $n_{\text{break}}$ is significantly reduced. 
We point out that $f(p,n)$ is not normalized, i.e.,\begin{align}
    \int dp\ f(p,n) \neq 1.
\end{align}
This is because normalizing would foster the bias toward smaller breakdown densities, since the score values along the EOS would then decrease in absolute value the larger the uncertainty in the \chiEFT\ band becomes.

\section{PRIOR SETUPS FOR THE BINARY NEUTRON STAR INFERENCES}
\label{app:GW_setup}
\begin{table}[t!]
    \centering
    \caption{Parameters and priors for the GW inferences. The top two blocks show the intrinsic and observational parameters as used for all inferences involving GW170817. In the bottom block we present those priors that were modified for the analysis of GW190425. We denote uniform priors by $\mathcal{U}$, Cosine signifies uniform sampling of the cosine of that angle. Luminosity distances are sampled uniformly in comoving volume. The subscript $j$ indicates the two binary components, i.e., $j\in \{1,2\}$.}
    \label{tab:GW_prior}
    \begin{tabular}{clrlr}
    \toprule
    \toprule
       &   Parameter  &   \multicolumn{2}{c}{symbol} & \multicolumn{1}{c}{prior} \\
      \midrule
      \multirow{6}{8.5pt}{\rotatebox{90}{intrinsic}} & Chirp mass $(\rm M_\odot) $ &&$\mathcal{M}$ &  $\mathcal{U}(1.18,1.21)$\\
       & Mass ratio && $q$ & $\mathcal{U}(0.125,1)$\\
        & Spin magnitude && $a_{j}$ & $\mathcal{U}(0,0.05)$\\
        & Tilt && $\sin \varphi_j$ & $\mathcal{U}(-1,1)$\\
        & Misalignment (rad) && $\phi_{12},\ \phi_{JL}$ & $\mathcal{U}(0, 2\pi)$\\
        & Equation of State &&EOS & see Sec.~\ref{sec:Multimessenger} \vspace{0.5 cm}\\
       \multirow{6}{8.5pt}{\rotatebox{90}{observational}}& Luminosity distance (Mpc) &&$d_{L}$ & $\mathcal{U}_{\text{com. vol.}}(15,75)$ \\
       & Right ascension (rad) && $\alpha$ & 3.44616\\
       & Declination (rad)&& $\delta$ & -0.408084\\
       & Trigger time (GPS) && $t_c$ & 1187008882.43\\
       & Inclination && $\theta_{JN}$& Cosine$(0,\pi)$\\
      & Phase (rad) &&$\phi$ & $\mathcal{U}(0,2\pi)$\\
       & Polarization (rad) &&$\psi$ & $\mathcal{U}(0,2\pi)$ \vspace{0.25cm}\\
       \midrule \addlinespace[0.25cm]
       \multirow{6}{8.5pt}{\rotatebox{90}{GW190425}} & Chirp mass $(\rm M_\odot) $ &&$\mathcal{M}$ & $\mathcal{U}(1.485,1.49)$ \\
       & Mass ratio && $q$ & $\mathcal{U}(0.4,1)$ \\
       & Luminosity distance (Mpc) &&$d_{L}$ & $\mathcal{U}_{\text{com. vol.}}(10, 300)$\\
       & Right ascension (rad) && $\alpha$ & Cosine$(0,\pi)$\\
       & Declination (rad) && $\delta$ & $\mathcal{U}(0,2\pi)$\\
       & Trigger time (GPS) && $t_c$ & 1240215503.0\\
       \bottomrule
    \end{tabular}
\end{table}
\begin{table}[t!]
    \centering
    \caption{Parameters and priors for the light-curve inferences. The top block lists parameters for the \mbox{Bu2019} KN model, the middle one for \mbox{Bu2023}. Below, we list the parameters for the GRB-afterglow model and at the bottom one finds the fit coefficients linking EM and GW parameters. We denote uniform priors by $\mathcal{U}$, Gaussian priors by $\mathcal{N}$, and log-uniform priors by Log.}
    \label{tab:EM_prior}
    \begin{tabular}{clrlr}
    \toprule
    \toprule
       &  Parameter  &   \multicolumn{2}{c}{symbol} & \multicolumn{1}{c}{prior} \\
      \midrule
        \multirow{3}{8.5pt}{\rotatebox{90}{Bu2019}} & Dyn. ejecta mass $(\rm M_\odot)$ && $M_{\text{ej,dyn}}$& Eq.~(\ref{eq:dyn_eject})\\
        & Wind ejecta mass $(\rm M_\odot)$ && $M_{\text{ej,wind}}$ & Eq.~(\ref{eq:wind_eject})\\
        & Opening angle (deg) && $\Phi$ & $\mathcal{U}(15, 75)$\vspace{0.5cm}\\
        
       \multirow{6}{8.5pt}{\rotatebox{90}{Bu2023}}& Dynamical ejecta mass $(\rm M_\odot)$ && $M_{\text{ej,dyn}}$& Eq.~(\ref{eq:dyn_eject})\\
       & Dynamical ejecta velocity $(c)$&& $v_{\text{ej,dyn}}$ & $\mathcal{U}(0.12, 0.25)$\\
       & Wind ejecta mass $(\rm M_\odot)$ && $M_{\text{ej,wind}}$ & Eq.~(\ref{eq:wind_eject})\\
       & Wind ejecta velocity $(c)$&& $v_{\text{ej,wind}}$ & $\mathcal{U}(0.03, 0.15)$\\
       & Average electron fraction && $Y_{e, \text{ dyn}}$ & $\mathcal{U}(0.15, 0.3)$ \vspace{0.5cm}\\

       \multirow{8}{8.5pt}{\rotatebox{90}{GRB afterglow}}& Isotropic energy equivalent  (erg) && $E_0$& Eq.~(\ref{eq:iso_ener})\\
       & Half opening angle (rad)&& $\theta_{\text{core}}$ & $\mathcal{U}(0, \pi/2)$\\
       & Cutoff angle (rad)&& $\theta_{\text{wing}}$ & $\mathcal{U}(0, \pi/2)$\\
       & Interstellar medium density (cm$^{-3}$) && $n_0$ & Log$(10^{-8}, 0)$\\
       & Electron power law index && $p$ & $\mathcal{U}(2,3)$\\
       & Electron energy fraction  && $\epsilon_e$ & Log$(10^{-4}, 1)$\\
       & Magnetic energy fraction && $\epsilon_B$ & Log$(10^{-5}, 1)$\vspace{0.5cm}\\

       \multirow{3}{8.5pt}{\rotatebox{90}{Fit}} & Wind-disk fraction && $\zeta$ & $\mathcal{U}(0,1)$ \\
       & Dynamical ejecta fit error && $\alpha$ & $\mathcal{N}(0, 0.04)$ \\
       & Jet energy fraction && $\epsilon$ & Log$(10^{-7}, 0.5)$\\
       \bottomrule
    \end{tabular}
\end{table}
The 16 parameters for the GW waveform model are listed in Table~\ref{tab:GW_prior} together with their priors for the analyses of GW170817 and GW190425. For GW170817 we fixed the sky location to its electromagnetic counterpart, trigger times were fixed for both GW170817 and GW190425. Instead of sampling over the component masses $M_1$ and $M_2$, it is common practice in GW analysis to use the equivalent parameters of chirp mass $\mathcal{M}$ and mass ratio $q$, because the latter appear explicitly in the waveform models. However, to prevent unreasonably small component masses, we restrict the resulting $M_1$, $M_2$ that are sampled from the uniform prior in chirp mass and mass ratio to lie between \qty{1}{\msun} and \qty{3.3}{\msun}, which effectively yields a minimum mass ratio of 0.33. The cutoff at \qty{1}{\msun} is relatively sharp for the lower mass NS in GW170817, though the posterior appears unimpaired by that bound. We chose this cutoff, because $\mathtt{NRTidalv3}$ was not yet sufficiently tested for very low masses at that time, and to reduce computational cost especially in the expensive joint inference of the GW, KN and GRB afterglow.

Similarly, the parameters of the light-curve models are listed in Table~\ref{tab:EM_prior}. The ejecta masses and isotropic energy equivalent are obtained from the GW parameters, but are additionally constrained to specific ranges. This is to prevent sample points with unphysical values and, in the case of the KN models, to ensure that the ejecta masses are contained within the grid of \textsc{possis} simulations. 
Thus, we require $M_{\text{ej,dyn}}$ to lie in between $10^{-3}\ \msun$ and $10^{-1}\ \msun$ for the Bu2019 model, and between $10^{-3}\ \msun$ and $10^{-1.7}\ \msun$ for the \mbox{Bu2023} model. The wind ejecta are restricted to the range of $10^{-3}-10^{-0.5}\ \msun$ for Bu2019 and $10^{-2}-10^{-0.89}\ \msun$ for \mbox{Bu2023}. The isotropic energy equivalent has to fall in the range of $10^{48}$ to $10^{60}$ erg.

All priors and constraints were implemented using the \textsc{bilby} prior module~\citep{Ashton_2019}. 

\section{COMPARISON BETWEEN THE \mbox{Bu2019} AND \mbox{Bu2023} KILONOVA MODEL}
\label{app:KN_model_comp}
\begin{figure}[t!]
    \centering
    \includegraphics[width = \linewidth]{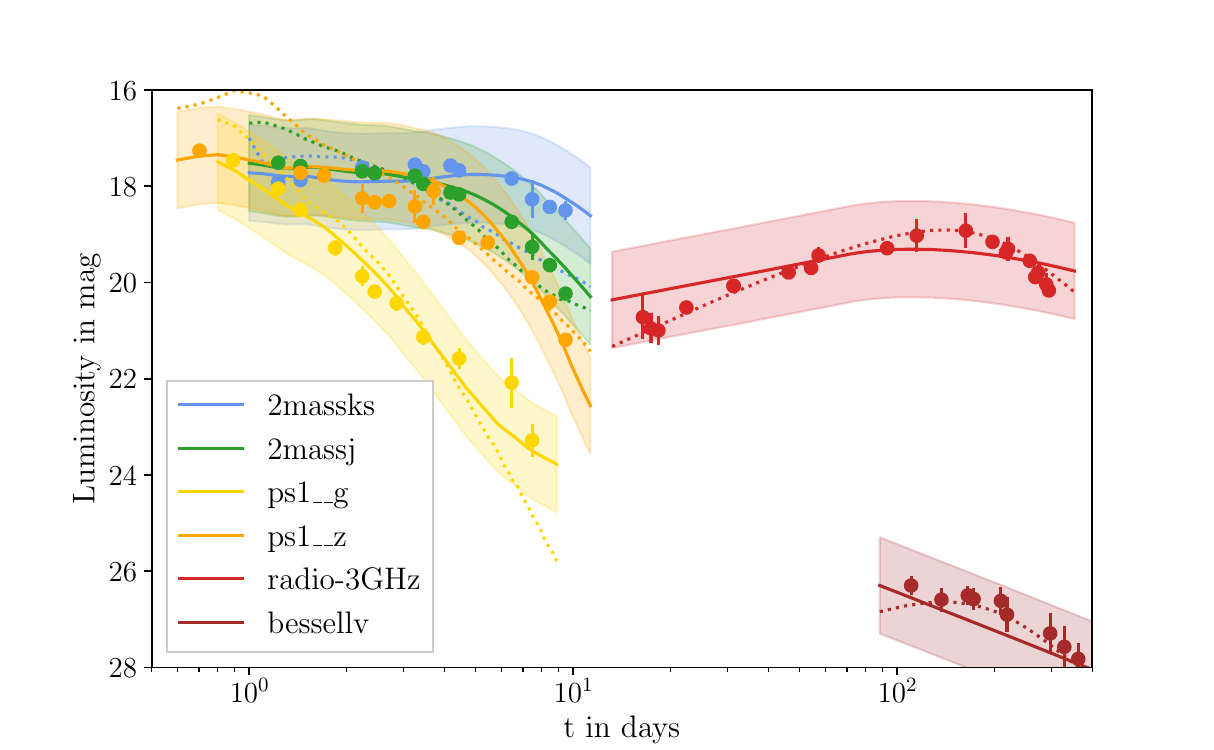}
    \caption{Best-fit light curves of the joint GW+KN+GRB inference with the \mbox{Bu2019} (dotted lines) and \mbox{Bu2023} model (solid lines) for selected filters. The different filters are color coded according to the legend, data points of the respected filters are displayed as circles with uncertainty bands. The colored bands indicate the systematic uncertainty of \qty{1}{mag} around the best-fit light curves of the \mbox{Bu2023} model.}
    \label{fig:fit_lightcurves}
\end{figure}

\begin{figure*}[t!]
    \centering
    \includegraphics[width = \linewidth]{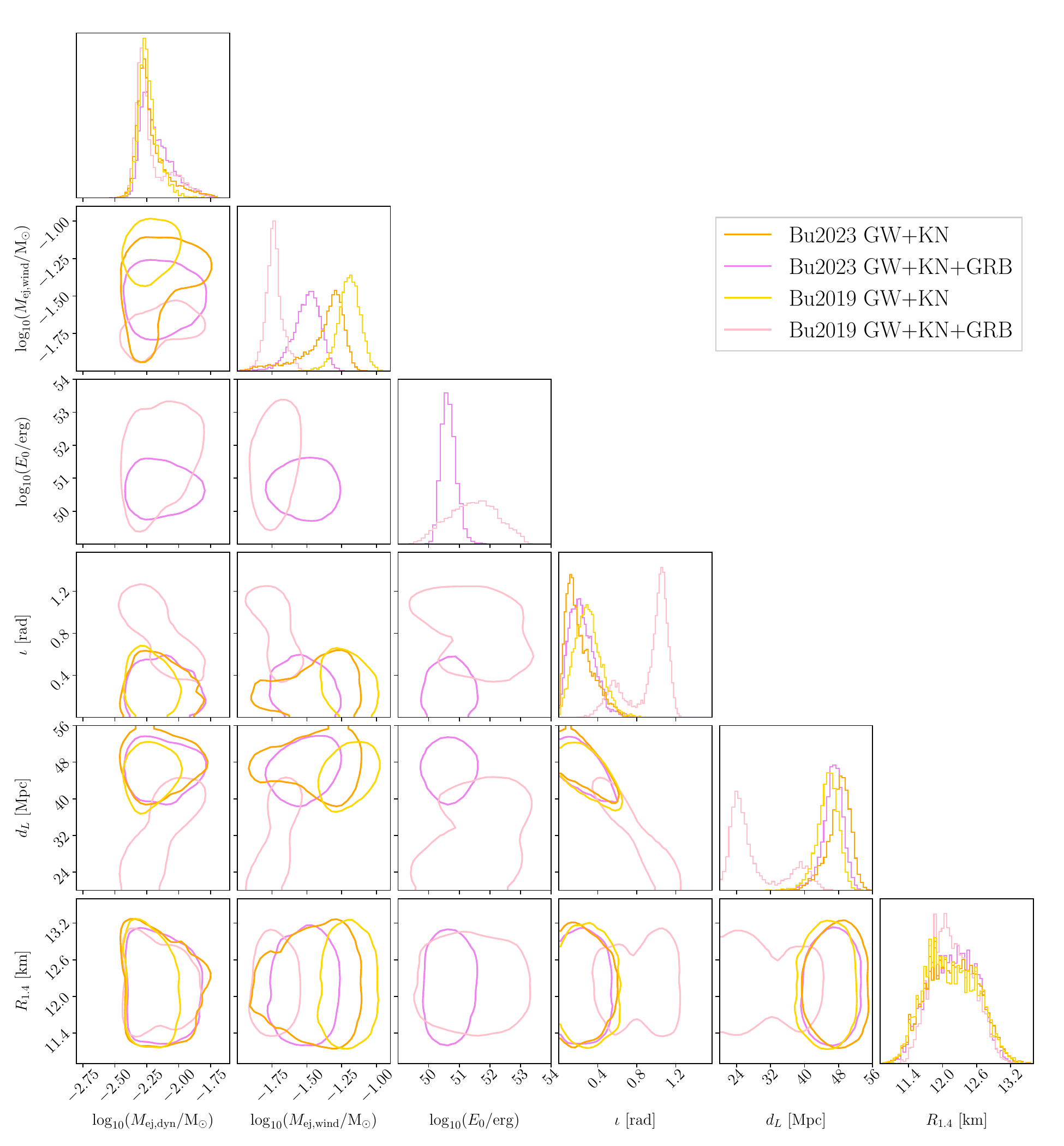}
    \caption{Comparison of the posterior estimates for selected parameters from inferences of GW170817 with electromagnetic counterparts. The contours show the 95\% credible regions of the respective parameters. Color coding refers to runs performed with either only the GW170817 and KN light-curve data, or GW170817, KN, and GRB afterglow light-curve data combined.}
    \label{fig:model_comp}
\end{figure*}
Our inferences from the electromagnetic counterpart of GW170817 deliver different results, depending on which of the two KN models from Sec.~\ref{sec:KN+GRB} is used and whether the GRB afterglow is included. We show corner plots for the most important electromagnetic parameters in Fig.~\ref{fig:model_comp}. GW parameters, in contrast, do not vary significantly across the separate inferences. While the differences with respect to the posterior distribution of the EOS are small, the estimates specifically for the wind ejecta masses show some discrepancy. Differences may be attributed to different physics prescriptions in the \mbox{Bu2019} and \mbox{Bu2023} model, however, adding the GRB afterglow also lowers the estimates for this parameter, since the GRB places some limits on the inclination and subsequently the disk mass and wind ejecta. Furthermore, it is apparent that during the joint GW+KN+GRB inference with the \mbox{Bu2019} model, the sampling of the luminosity distance is starkly biased toward distances of $\sim$\qty{24}{Mpc}. While such low values for the luminosity distances are technically reconcilable with the GW data alone given the degeneracy between inclination and distance, the identification of the electromagnetic counterpart in the galaxy NGC~4993 \citep{Coulter_2017} sets the luminosity distance to \qty{40.4}{Mpc}~\citep{Hjorth_2017}. Combined with the bimodal structure in the posterior, we interpret this as an error during the nested sampling process. The specific reason for this is uncertain, but may be related to artifacts in the neural network interpolation of the \textsc{possis} grid or some hitherto unknown degeneracy between the GRB parameters and ejecta properties. Because of the large computational cost for the joint inferences of GW, KN, and GRB afterglow data, we have to postpone a detailed investigation on this issue to a later date. The underestimated luminosity distance also correlates with the lower wind ejecta masses and broadens the estimate for the isotropic energy equivalent. We also note that the \mbox{Bu2023} tends to slightly overestimate the true value of the luminosity distance.

As mentioned, \mbox{Bu2019} is statistically preferred with $\ln \mathcal{B}^{\text{Bu2019}}_{\text{Bu2023}}=12.73$ for GW+KN only and $\ln \mathcal{B}^{\text{Bu2019}}_{\text{Bu2023}}=6.12$ when the GRB afterglow is added. However, given the aforementioned flaw in the sampling of observational parameters, we decided to quote our main results with respect to the newer \mbox{Bu2023} model. Indeed, both models fit the light-curve data reasonably well, as displayed in Fig.~\ref{fig:fit_lightcurves}. In the joint GW+KN+GRB inferences, the KN data are more appropriately fitted with \mbox{Bu2023}, though the run with \mbox{Bu2019} fits the data of the GRB afterglow better. The reduced $\chi^2$ value with the best-fit parameters $\vec{\theta_b}$,
\begin{align}
    \chi^2 = \frac{1}{\text{no. datapoints}}\sum_{f,j} \frac{(m^{f}(t_{j}, \vec{\theta_b}) - m^f(t_{j},d))^2}{(\sigma_{\text{syst}})^2+(\sigma^{f}_{\text{stat}}(t_{j}))^2}
\end{align}
reads 0.31 for the \mbox{Bu2019} run and 0.36 for the run with \mbox{Bu2023}. 
We point out that the best-fit light curve of the former assumes a luminosity distance of \qty{23.3}{Mpc}, so when we instead consider the best-fit parameter in the posterior region of $d_L \approx 40$\,Mpc, we obtain a $\chi^2$ value of 0.36 with \mbox{Bu2019} as well. 
The best-fit $\chi^2$ values for the GW+KN only inferences are 0.25 and 0.31, respectively.

\section{DETERMINING THE KEPLER MASS LIMIT WITH QUASIUNIVERSAL RELATIONS}
\label{app:Kepler_limit}
In Sec.~\ref{subsec:postmerger} we determined the baryonic Kepler limit for each of our EOS candidates to infer an upper bound on $\mtov$ from GW170817's postmerger fate. 
The calculation of $M_{\text{Kep},b}$ through the \textsc{rns} code carries some numerical error that could potentially impact the conclusions drawn. Even though we account for this uncertainty during sampling, in this appendix we briefly compare the result from \textsc{rns} to those obtained when employing the quasi-universal relations 
\begin{align}
    M_{\text{Kep}} &= \mathcal{R}\ M_{\text{TOV}}\,, \label{eq:m_max} \\
    M_{\text{Kep}, b} &= (1+\delta) \left(M_{\text{Kep}} + \frac{0.78 \text{\ km}}{R_{1.4}} M_{\text{Kep}}^2\right)
\label{eq:m_max_b}
\end{align}
from Refs.~\citep{Musolino_2023, Gao_2020}. 
There, the numerical values are $\mathcal{R} = 1.255^{+0.047}_{-0.040}$ at 95\% confidence and the relative error $\delta$ is at most 1.3\%. This agrees well to the fit for the (gravitational) Kepler mass from our own \textsc{rns} samples:
\begin{align}
    M_{\text{Kep}} &= 1.310\ M_{\text{TOV}} -0.278\,.
\end{align}
We implement the quasiuniversal relations Eqs.~(\ref{eq:m_max}) and (\ref{eq:m_max_b}) in our resampling procedure for the joint GW+KN+GRB posterior, and additionally sample $\mathcal{R}$ from a Gaussian prior with mean $1.255$ and standard deviation $0.024$, and sample uniformly over $\delta$ from $-1.3$\% to 1.3\%. The resulting posterior distribution on $\mtov$ is very similar to the one obtained in Sec.~\ref{subsec:postmerger}, although the EOS distribution here is slightly broader with $\mtov = 2.03^{+0.08}_{-0.06}\,\msun$ compared to the original $\mtov = 2.04^{+0.06}_{-0.06}\,\msun$ from above. The similarity between these results confirms that both methods, \textsc{rns} and quasi-universal relations, are valid methods in determining $M_{\text{Kep},b}$ for our purposes.

\vfill
\newpage
\bibliographystyle{prx.bst}
\bibliography{astro_bib, nuclear_bib, multimessenger_bib}
\end{document}